\documentclass[prb,notitlepage,reprint,superscriptaddress]{revtex4-2}
\usepackage{graphicx} % Required for inserting images
\usepackage{amsmath}
\usepackage{amssymb}
\usepackage{bm}
\usepackage{tikz}
\usepackage{pgfplots}
\usepackage{textcomp}

\pgfplotsset{compat=1.18}
\usepgfplotslibrary{groupplots}
\usepgfplotslibrary{external}
\newcommand{\change}[1]{{\color{black} #1}}

\begin{document}

\title{Measurement-driven Langevin modeling of superparamagnetic tunnel junctions}

\author{Liam A. Pocher}
\affiliation{Institute for Research in Electronics and Applied Physics, University of Maryland, College Park, MD, USA}
\affiliation{Associate, Physical Measurement Laboratory, National Institute of Standards and Technology, Gaithersburg, MD, USA}
\author{Temitayo N. Adeyeye}
\affiliation{Institute for Research in Electronics and Applied Physics, University of Maryland, College Park, MD, USA}
\affiliation{Associate, Physical Measurement Laboratory, National Institute of Standards and Technology, Gaithersburg, MD, USA}
\author{Sidra Gibeault}
\affiliation{Institute for Research in Electronics and Applied Physics, University of Maryland, College Park, MD, USA}
\affiliation{Associate, Physical Measurement Laboratory, National Institute of Standards and Technology, Gaithersburg, MD, USA}
\author{Philippe Talatchian}
\affiliation{Univ.~Grenoble Alpes, CEA, CNRS, Grenoble INP, SPINTEC, 38000 Grenoble, France}
\author{Ursula Ebels}
\affiliation{Univ.~Grenoble Alpes, CEA, CNRS, Grenoble INP, SPINTEC, 38000 Grenoble, France}
\author{Daniel P. Lathrop}
\affiliation{Institute for Research in Electronics and Applied Physics, University of Maryland, College Park, MD, USA}
\author{Jabez J. McClelland}
\affiliation{Physical Measurement Laboratory, National Institute of Standards and Technology, Gaithersburg, MD, USA}
\author{Mark D. Stiles}
\affiliation{Physical Measurement Laboratory, National Institute of Standards and Technology, Gaithersburg, MD, USA}
\author{Advait Madhavan}
\affiliation{Institute for Research in Electronics and Applied Physics, University of Maryland, College Park, MD, USA}
\affiliation{Associate, Physical Measurement Laboratory, National Institute of Standards and Technology, Gaithersburg, MD, USA}
\author{Matthew W. Daniels}
\affiliation{Physical Measurement Laboratory, National Institute of Standards and Technology, Gaithersburg, MD, USA}

\begin{abstract}
  Superparamagnetic tunnel junctions are important devices for a range of emerging technologies, but most existing compact models capture only their mean switching rates. Capturing qualitatively accurate analog dynamics of these devices will be important as the technology scales up. Here we present results using a one-dimensional overdamped Langevin equation that captures statistical properties of measured time traces, including voltage histograms, drift and diffusion characteristics as measured with Kramers-Moyal coefficients, and dwell times distributions. While common macrospin models are more physically-motivated magnetic models than the Langevin model, we show that for the device measured here, they capture even fewer of the measured experimental behaviors. 
\end{abstract}

\maketitle

\section{Introduction}
\label{sec:intro}

Magnetic tunnel junctions are versatile devices with many modes of operation, several of which are now being recognized for their value to neuromorphic and alternative computing schemes~\cite{grollierSpintronicNanodevicesBioinspired2016,roy2018perspective,grollierNeuromorphicSpintronics2020,blachowicz2020magnetic,zink2022review}. Their superparamagnetic regime, in particular, holds great promise not only for the acceleration of traditional scientific computing workloads~\cite{fukushima2014spin,vodenicarevicLowEnergyTrulyRandom2017,parks2018superparamagnetic,lv2022bipolar} but also for stochastic computing~\cite{danielsEnergyEfficientStochasticComputing2020,zink2023stochastic,cai2023unconventional}, other neural networks~\cite{mizrahiNeurallikeComputingPopulations2018,mizrahiOvercomingDeviceUnreliability2018,rajasekharan2021scanet,PhysRevApplied.8.064017,PhysRevApplied.11.034015}, and combinatorial optimization accelerators~\cite{camsari2017implementing,bordersIntegerFactorizationUsing2019,lv2019experimental,gibeault2023programmable}. Their technological relevance, and the attendant need to design scaled-up circuits that include these devices, has made the task of finding appropriate dynamical models an engineering priority. 

Currently, three classes of models are used for different levels of granularity in representing superparamagnetic tunnel junction (SMTJ) physics. The simplest is the N\'eel-Brown model~\cite{brownThermalFluctuationsSingleDomain1963,mizrahi2015magnetic,becle2021fast}, a two-state Markov model where the transition rates are exponential functions of the current and field applied to a device. These models fit transition rates of the Markov model to experiment~\cite{talatchianMutualControlStochastic2021}, and can sometimes be aligned with the measurable parameters of the system in a physics-driven way~\cite{carboniPhysicsBasedCompactModel2019,yangUniversalCompactModel2022}. Their discrete state spaces, however, hide any intermediate analog dynamics, and experiments to date indicate that this analog magnetic behavior is visibly non-negligible~\cite{schnitzspanNanosecondSuperparamagneticTunnel,talatchianMutualControlStochastic2021,safranskiDemonstrationNanosecondOperation2021,parks2018superparamagnetic,bapna2017current,hayakawa2021nanosecond}. N\'eel-Brown models implicitly assume an exponential distribution of the dwell times.

The most detailed form of modeling is achieved through micromagnetic simulations. Some micromagnetics packages can explicitly solve the stochastic Landau-Lifshitz-Gilbert (sLLG) equation over large magnetic domains using finite-element methods~\cite{leliaertAdaptivelyTimeStepping2017}. The simulation of devices in this manner can lead to physically realistic results, but the required computational timescale renders such models inappropriate for single elements in large circuit simulations. To wit, the authors of Ref.~\cite{endean2014tunable} show that micromagnetic simulations qualitatively reproduce measured two-level fluctuations but are not able to collect enough statistics with the full simulations to make a robust statistical comparison.  

The intermediate level between these approaches is that of analog compact modeling. Compact models -- relatively low-dimensional differential equations that capture the essential analog behaviors of a device without full physical realism -- have been developed for many nanodevices~\cite{shin2010compact,ding2022review,jimenez2021compact,araujo2022data}, including magnetic tunnel junctions~\cite{faber2009dynamic,guo2010spice,torunbalciModularCompactModeling2018}. \change{When SMTJs are incorporated into integrated circuits, modeling the interaction between the SMTJs and the transistors will require a statistically accurate description of the SMTJ dynamics on the time scale of switching events in both, but that description cannot be significantly slower than the modeling of the integrated circuit. Those restrictions force the development and use of a compact model that accurately describes the transition dynamics. A Markov model, while fast, does not accurately capture the dynamics of the transition, whereas a micromagnetic model would capture that dynamics, but is far too slow.}

Ideas for computing systems that use SMTJs are now being proposed that assume the existence and even interaction of thousands to millions of individual devices~\cite{mizrahiNeurallikeComputingPopulations2018,danielsEnergyEfficientStochasticComputing2020,chowdhuryFullstackViewProbabilistic2023,bunaiyan2023heisenberg}. The engineering work of design, test, and verification of large scale designs like those will require mature models that faithfully capture device physics. Such models already exist for memory-class MTJs used in magnetic random access memory (MRAM) designs and have enabled extraordinary progress in that field~\cite{jungCrossbarArrayMagnetoresistive2022}. Our present goal, then, is to facilitate a move beyond demonstrating the mere viability of stochastic and probabilistic computing proposals based on SMTJs and toward full-stack-engineered system design based on SMTJ device physics. 

For SMTJs in particular, a common compact modeling methodology in the literature is the use of a macrospin (single-domain) model~\cite{camsariImplementingPbitsEmbedded2017}. Macrospin modeling has been applied as a theoretical approach to understanding the physics of magnetic systems~\cite{xiao2005macrospin} and has been used to demonstrate the viability of SMTJ-based alternative computing schemes~\cite{danielsEnergyEfficientStochasticComputing2020,camsari2017implementing}. Yet it is generally understood that real magnetic devices rarely operate in the single domain regime, calling into question whether macrospin models provide an appropriate simulation framework for the engineering context. Even in devices that are expected to be near-single-domain, the authors of Ref.~\cite{KanaiTBP} show explicitly that a simple macrospin model cannot reproduce their experimental results because details of the dynamics change in non-trivial ways during the reversal process. They theorize that the dissipative mechanisms of the model behave outside of what can be captured by the usual sLLG equation, which has a diffusive part that is pre-determined~\cite{berkov2002thermally} and cannot be tuned except in overall amplitude (through the Gilbert damping). The authors of Ref.~\cite{desplat_2020} show that entropic effects due to degrees of freedom neglected in the macrospin approximation significantly affect both the energetics and the dynamics of the reversal process.

From a practical standpoint, numerical integration of the sLLG equation often requires careful attention~\cite{cimrak2007survey}; spherical coordinate approaches can depend in subtle ways on the integration scheme~\cite{aron2014magnetization}, and integrating the more stable Euclidean equations requires renormalization of the magnetization~\cite{PhysRevE.90.023203,PhysRevB.58.14937}\change{, which slows down the integration,}  or carefully chosen solvers~\cite{ament2016solving} that an end-user \change{may not be able to access} in an engineering context (\textit{e.g.}~when using a commercial circuit simulator).

Facing these limitations of simple macrospin models, we propose a data-driven method to capture the dynamics of interest in circuit simulations. We turn to generic overdamped 1D Langevin models for the voltage across the device and fit such models to experimental measurements. Similar data-driven approaches have been applied in other fields of physics to learn the underlying physics of stochastic processes through a variety of methods~\cite{boninsegna2018sparse,callaham2021nonlinear,gorjao2019analysis}. We show that our approach leads to high-fidelity matching of the voltage histograms and the drift and diffusion coefficients of the device between model and experiment while maintaining the analog nature of macrospin approaches. We also show that our model correctly predicts the drift dynamics and dwell-time distributions of the experiment without explicitly encoding these in the simulation, confirming a level of self-consistency between the model and the underlying physics.

We organize the remainder of the paper as follows. We give an overview of the approach we use to fit a time trace measured for an SMTJ in Sec.~\ref{sec:roadmap} so that we can reproduce a statistically identical time trace in simulation. In Sec.~\ref{sec:data}  we describe the different statistical measures of such voltage-time traces that provide reductions of the data to relevant dynamical quantities. Section~\ref{sec:Langevin} introduces an overdamped, 1D Langevin model and describes the method that we use to fit the experimental data. Section~\ref{sec:results} compares the Langevin modeling to experiment and shows that the model reproduces the experimental data used to determine the model and exhibits self-consistency with the assumed Fokker-Planck dynamics. In Sec.~\ref{sec:macrospin}, we analyze standard macrospin models and use a comparison of their allowed probability distributions and drift-diffusion statistics with those of the experimental data to show that they cannot reasonably provide a quantitative model for the data. Finally, in Sec.~\ref{sec:discuss}, we discuss possible extensions to our compact modeling approach, orienting future research in ways we view as most useful for supporting the device-circuit-system codesign practices needed to realize large-scale computational systems based on SMTJs. 

Experimental details are given in the appendices. Appendix~\ref{sec:samples} briefly introduces the fabrication of our experimental device. Appendix~\ref{app:Exp_DC_charac} describes the direct current (DC) characterization of the measured device. Appendix~\ref{app:Elec_Meas_Circuit} describes the circuitry used to make the high-speed measurements of the superparamagnetic behavior. Appendix~\ref{app:fitting} describes some alternate fitting schemes. Appendix~\ref{app:dwell-times} describes the extraction of mean dwell times from time traces. 

\section{Model Construction Roadmap}
\label{sec:roadmap}

Our goal is to develop a data-driven, compact, semi-analytic model that produces dynamics that are statistically identical to those of the voltage-time trace. We refer to the model as data-driven because we use some of the statistical properties of the data to produce the model. In contrast to other, physically motivated models, such as a macrospin approximation, our models are not derived from approximations based on device physics. Not being tied to approximations of the physics can be an advantage, as it allows the model to capture behavior that is neglected by physically-motivated models that may be  oversimplified compared to experimental reality.

\begin{figure*}
    \centering
    \includegraphics[width=0.8\textwidth]{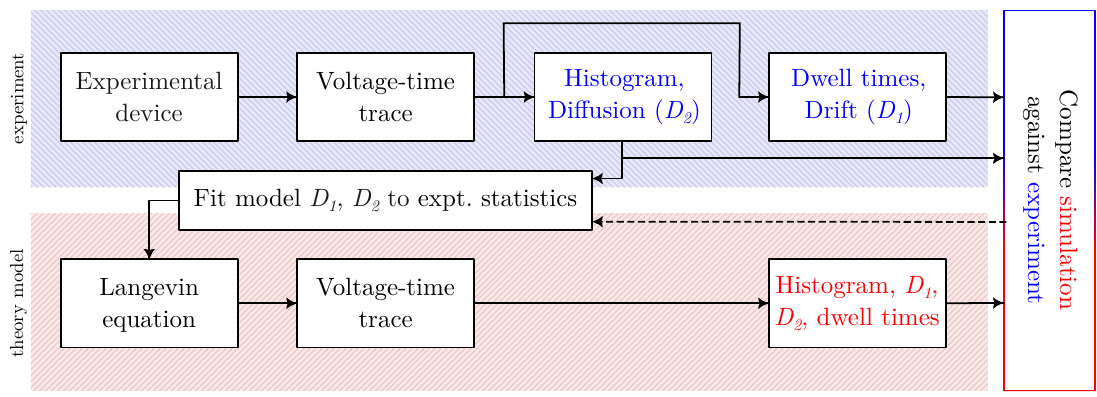}
    \caption{Visual outline of the model generation process developed in this paper. Top row: we start with an experimental device from which a voltage-time trace is extracted under fixed experimental conditions. The histogram and the voltage-dependent diffusion coefficient (second Kramers-Moyal coefficient $D_2$) are extracted numerically from the time trace. The drift coefficient and dwell-time distributions are also extracted for later use. Middle row: We fit drift and diffusion characteristics for our model to the histogram and diffusion characteristics found in experiment, in a way that compensates for the high-frequency cutoff of the experimental data. Bottom row: we then use these fitted drift and diffusion characteristics to simulate a Langevin equation from which we extract a voltage-time trace and its attendant statistics. Far right: the suite of statistics from the model is validated against the experiment. In future work with more complex fits, one may need to introduce a self-consistent fitting procedure that uses observed error between theory and experiment to inform iterative refinements of the model (dashed line).}
    \label{fig:model-roadmap}
\end{figure*}

Figure~\ref{fig:model-roadmap} shows the modeling flowchart for this paper. From the experiment with the SMTJ device we obtain a voltage-time trace. The statistics of that data are used to determine our model, a one-dimensional equation of motion (directed, overdamped Brownian motion). This equation of motion describes the state of the system with drift (deterministic) and diffusion (stochastic) terms, with parameterizations that can be calculated directly from the experimental voltage-time trace. We integrate the model forward in time to produce a simulated time trace, do the same statistical analysis that we had performed for the experiment, and  compare the experimental and simulated statistics.

% evolution of an ensemble of systems with mixed deterministic-stochastic properties is described with the Fokker-Plank equation, which evolves according to drift and diffusion terms given by the Krmaers-Moyal coefficients~\cite{risken1996fokker,honisch2011estimation,friedrich2000quantify}, which can be measured directly from the voltage-time trace of our experiment. 
% The Langevin equation's drift and diffusion is related to the Kramers-Moyal coefficients. These coefficients encode short time dynamics into our model. The $D_1$ coefficients captures deterministic (drift) behavior of the system while $D_2$ captures stochastic (diffusive) behavior. We perform data-analysis on the voltage-time trace to obtain statistics to seed our model.
% histogram, diffusion $(D_2)$ coefficient, and the dwell time distribution. . 
% We then integrate the model forward in time to produce a simulated time trace which we then do the same statistical analysis that we had done for the experiment. Finally, we compare the experimental and simulated data metrics. 

\begin{figure}
    \centering
    \includegraphics[width=\columnwidth]{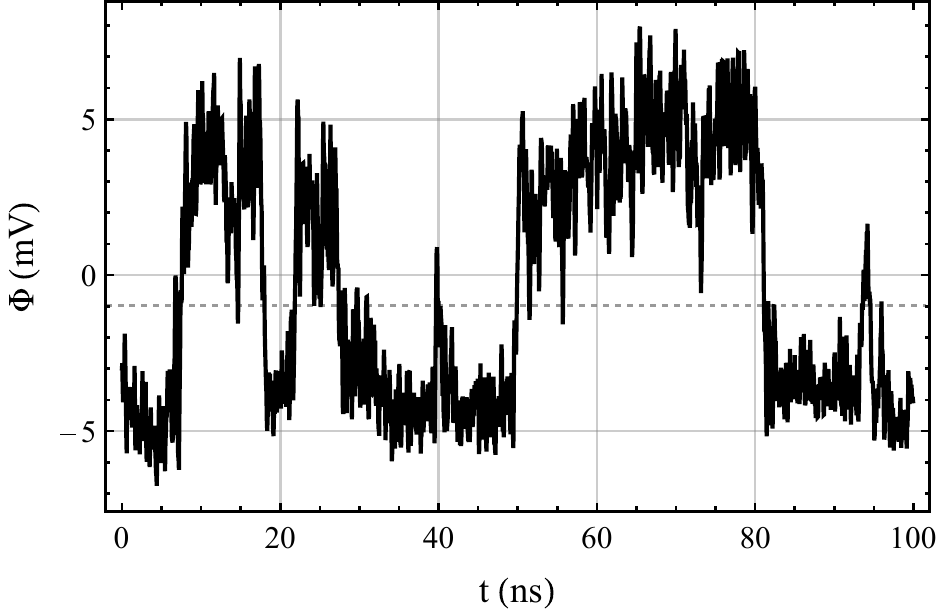}
    \caption{Typical bi-stable voltage fluctuations of an SMTJ in our experimental setup, measured at a sampling rate of $\tau_{\text{exp}}=50$~ps. Shown is a small subset -- a 100~ns window -- of the full-time 1~ms trace captured in the experiment. The total number of transitions between the two states in the full voltage-time trace is on the order of $3\times 10^5$. \change{The dashed line indicates the boundary between the parallel and antiparallel wells. That value is chosen from the local minimum in the distribution of measured voltages given in Fig.~\ref{fig:histogram}. }}
    \label{fig:experiment-trace}
\end{figure}

Figure~\ref{fig:experiment-trace} shows a representative segment of a voltage-time trace of our experimental device described in App.~\ref{sec:samples}. The models we use to reproduce this behavior are one-dimensional stochastic differential equations whose drift and diffusion parts are determined by both short-time statistics and long-time statistics. The resulting equation is the Langevin equation
\begin{equation} \label{eq:Langevin-roadmap}
   \dot \Phi = \underbrace{f(\Phi)}_{\text{drift}} + \underbrace{g(\Phi) \; \eta_\Phi(t)}_{\text{diffusion}},
\end{equation}
where $\Phi$ is the dependant variable (voltage), $t$ the independent variable (time), $f(\Phi)$ the function characterizing the drift, $g(\Phi)$ the function characterizing the diffusion, and $\eta_{\Phi}(t)$ a Gaussian white noise term~\footnote{\change{We note that in reality the driving noise terms likely possess additional structure beyond mere white noise. However, as we will show in Sec.~\ref{sec:results}, the white noise assumption does very well in reproducing the statistics observed in experiment. Physically, the choice of white noise is the unique choice that ensures the dissipative mechanisms of the system are local in time~\cite{kubo1966fluctuation}.}} with the properties $\langle \eta_\Phi(t) \eta_\Phi(t') \rangle_t = 2 \delta(t' - t)$ and $\langle \eta_\Phi(t) \rangle_t = 0$. The drift and diffusion terms of the Langevin equation are related to each other through the voltage-time trace histogram, which gives the long-time average of the stationary state of the system for a fixed current flowing through the SMTJ; see Sec.~\ref{sec:Langevin} for further details. 

The one-dimensional Langevin models are completely determined by fitting the histogram and either the first, $n=1$, or second, $n=2$, Kramers-Moyal coefficient 
\begin{equation} \label{eq:KM-coef-D2}
    D_{n}(\Phi) = \frac{1}{n!} \lim_{\tau \rightarrow 0} \frac{\left\langle \left[\Phi(t+\tau) - \Phi(t)\right]^n\right\rangle_t}{\tau},
\end{equation}
with $\tau$  the sampling time, the angle brackets denoting a time average over the voltage-time trace, and $\left[\Phi(t + \tau) - \Phi(t)\right]$ a $\tau$-delayed difference in voltage. The first and second Kramers-Moyal coefficients encode short-time dynamics into our model. The $D_1$ coefficients capture the drift behavior of the system while $D_2$ captures the diffusive behavior. Fitting the histogram and one of the Kramers-Moyal coefficients then determines the functional coefficients in the Langevin model Eq.~\eqref{eq:Langevin-roadmap}, fully characterizing it and allowing us to perform simulations that mimic the experimental measurement. 

As will be shown in Sec.~\ref{sec:results} and discussed in Sec.~\ref{sec:discuss}, this data-driven approach captures multiple statistical and dynamical metrics with high fidelity. Specifically, it captures the Kramers-Moyal coefficients, the histogram, the dwell time distributions, and the power spectral density (discussed in App.~\ref{app:Elec_Meas_Circuit}). The agreement for the dwell time distributions is perhaps the most significant validation of the model because the experimental input to the model does not directly encode these time scales. This ability to recover higher-order statistics is necessary so that we can use the model to design scaled-up circuits in engineering applications. 

\section{Statistics of Time Traces}
\label{sec:data}

In this section, we describe in more detail the statistical properties that we use to set up the Langevin model. These properties and others are compared to those of simulations of the resulting model in Sec.~\ref{sec:results}. 

\begin{figure}
    \centering\includegraphics[width=\columnwidth]{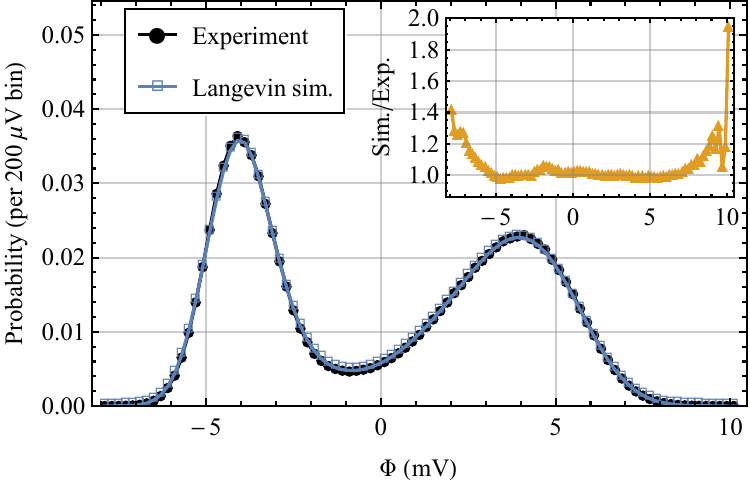}
    \caption{Histogram of SMTJ state. The black circles give the experimental resultsfor the entire voltage-time trace of 160 ms, part of which is shown in Fig.~\ref{fig:experiment-trace}, using bins 200~{\textmu}V wide. The blue \change{rectangles} give the results of a simulation, discussed in Sec.~\ref{sec:results} using a data-driven Langevin model. Note the excellent recovery of the experimental histogram's values in the simulation, demonstrating the capacity of the first-order Langevin model to capture coarse-grained statistics. Statistical error bars are smaller than the symbols.  The inset plot shows the ratio between the simulation and experiment histogram values. Note the near unity agreement for the entire domain save where both histogram's values approach zero. \change{The local minimum near $-$1~mV is taken as the boundary between the parallel and antiparallel wells.} }
    \label{fig:histogram}
\end{figure}

The first statistical reduction of the voltage-time trace we use is the histogram of the device state over the entire 1~ms measurement window, see Fig.~\ref{fig:histogram}. We assume that the system giving rise to the voltage-time trace is in quasiequilibrium, so the histogram primarily depends on the effective energy and entropy of the system at each voltage and less on the short-time dynamics of the system. The binning resolution used for the histogram and other statistical properties described below along the voltage axis impacts the fit quality and the observed characteristics. If the bins are too small, the statistical uncertainty prevents good fits; if they are too large, the details of the dynamics are obscured. The bin size we use for the histogram is a compromise between these factors at approximately 200~{\textmu}V. Each bin captures approximately $50$ quantized voltage signals out of the approximately $2^{12}$ unique levels reported by the oscilloscope.

Though the histograms we observed above can (together with an assumption of Boltzmann statistics) tell us about the effective energy landscape of the system, the detailed short-time dynamics generated by thermally induced magnetic fluctuations -- which are by construction separate from the conservative forces on the system -- remain hidden. To determine their behaviors, we compute conditional moments~\footnote{These conditional moments are reminiscent of Kolmogorov's structure functions in turbulence. However, structure functions are conditional moments that use a spatial separation and spatial ensemble average to obtain a scalar value for fluid velocity, whereas the Kramers-Moyal coefficients here are evaluated from a voltage-time trace (which by the ergodic hypothesis is equivalent to an ensemble average) that integrates solely over the probability distribution.} of the experimental voltage-time trace. The $n^\text{th}$ order conditional moment $M_n$ is
\begin{equation}
\begin{aligned} \label{eq:KM-coef-M}
    M_{n}(\Phi,\tau) &= \left\langle \left[\Phi(t+\tau) - \Phi(t)\right]^n\right\rangle_t,
\end{aligned}
\end{equation}
with the angle brackets denoting a time average over an entire trajectory, and $\Phi(t + \tau) - \Phi(t)$ a $\tau$-delayed difference in the system's state. We capture these conditional moments with the same bins as we use for the histogram.

At very small $\tau$, Eq.~\eqref{eq:KM-coef-M} can be used to approximate the $n^\text{th}$ order Kramers-Moyal coefficient~\cite{risken1996fokker,honisch2011estimation,friedrich2000quantify} of the system. Formally, these Kramers-Moyal coefficients are connected to the conditional moments as
\begin{equation} \label{eq:KM-coef-D}
    D_{n}(\Phi) = \frac{1}{n!} \lim_{\tau \rightarrow 0} \frac{M_n (\Phi,\tau)}{\tau}.
\end{equation}
The $D_1$ and $D_2$ terms, which we refer to as the drift and diffusion terms, are the specific Kramers-Moyal coefficients used in our first-order Langevin model and are the only nonzero Kramers-Moyal coefficients required to describe systems obeying the Fokker-Planck equation~\cite{PhysRev.162.186}. 

Due to the finite resolution inherent in experimental data, we cannot immediately take the limit as $\tau \rightarrow 0$ in our calculation of the Kramers-Moyal coefficients. We also find in practice that our experimental timestep was not quite small enough to approximate this limit; subsampling our data to compute the $M_n$ at slightly longer timesteps indicates that we are not in the converged regime, and strongly so for $M_2$. As the $\tau$ increases from zero, however, finite-time corrections can be used to relate the conditional, time-delayed moments $M_n$ to the underlying Kramers-Moyal coefficients that are needed to construct an analytic Langevin model~\cite{gottschall2008definition,rydin2021arbitrary}. These corrections up to second order in the time delay $\tau$ are
\begin{align} 
    M_1 &= \tau D_1 + \frac{\tau^2}{2} \left( D_1 D_1' + D_2 D_1'' \right) + \mathcal{O}(\tau^3)\label{eq:KM-correctionM1} \\
    \text{and}\;\; M_2 &= 2 \tau D_2 + \tau^2 \big( D_1^2 + D_1 D_2'\nonumber\\ 
    &\quad\quad+ D_2 D_2'' + 2 D_2 D_1' \big) + \mathcal{O}(\tau^3), \label{eq:KM-correctionM2}
\end{align}
with a prime denoting differentiation with respect to $\Phi$, and each Kramers-Moyal coefficient and time-delayed moment understood to be a function of $\Phi$. Equations \eqref{eq:KM-correctionM1} and \eqref{eq:KM-correctionM2} describe a second-order differential system for the Kramers-Moyal coefficients in terms of calculated $M_1$ and $M_2$; these equations cannot be solved analytically in general. Our approach to extracting the Kramers-Moyal coefficients, which we require to parameterize our Langevin model, is to choose a parameterized \textit{ansatz} for $D_2$; that is, we choose some functional form with a number of free parameters. Combining this \textit{ansatz} with an analytic fit to the histogram induces an \textit{ansatz} on $D_1$ via the stationary solution to the Fokker-Planck equation. We then choose the free parameters (see Eq.~\eqref{eq:D2ansatz} below) by fitting the right-hand-sides of Eqs.~\eqref{eq:KM-correctionM2} (using our \textit{ansatz} on the right-hand sides of Eq.~\eqref{eq:KM-correctionM1} and Eq.~\eqref{eq:KM-correctionM2}) to the $M_2$ extracted from our experimentally measured voltage-time trace. We could fit both $M_1$ and $M_2$ simultaneously, but finding agreement between $M_1$ and the predictions of the model when only $M_2$ is fit argues for the appropriateness of using the one-dimensional Fokker-Planck equation.

\section{One-dimensional Langevin model\label{sec:Langevin}}

We note in Sec.~\ref{sec:capturing-macrospin-Langevin} that macrospin models fail to capture qualitative metrics -- let alone statistical metrics -- associated with the experimental data. Yet models that capture the statistics of experimental devices will be required for high-fidelity circuit simulations of SMTJ devices as their attendant technological applications scale. To address this discrepancy, we introduce a first-order Langevin model inspired by existing works on the modeling of fluctuating bi-stable processes~\cite{sicard2021position,vercauteren2005numerical}. This approach may not be capable of predicting the behavior of uncharacterized devices, but we anticipate it could be used to characterize the devices from a particular manufacturing process and those fit results could be used for circuit simulations of those devices.

\begin{figure}
    \centering
    \includegraphics[width=\columnwidth]{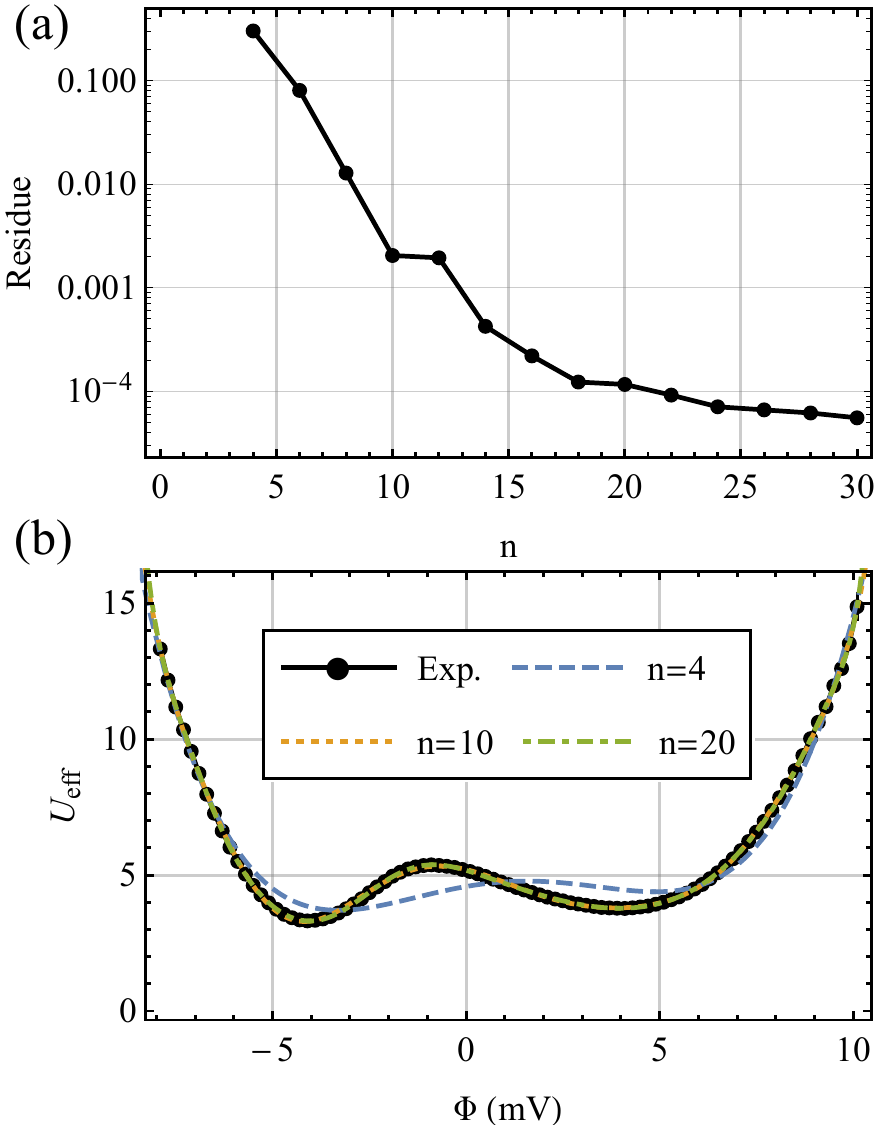}
    \caption{Fit to the effective energy. (a) Residues (normalized $l^2$ norm) between the Chebyshev polynomial fits to the data as a function of even order $n$ and (b) sample fits---$n=4,10,20$--- to the experimental effective energy $U_{\text{eff}}$. Error bars on experimental data indicate statistical uncertainties due to the number of counts per histogram bin; however, these uncertainties are smaller than the symbols.}
    \label{fig:Ueff}
\end{figure}

We take a data-driven approach to determining the details of the one-dimensional Langevin model. The first step is to find an analytic expression for the experimentally determined probability density $\rho_0$. We require an analytic expression because we will need to take its derivative to infer the deterministic forces in the system.  To guarantee a positive-definite fit, we assume a Boltzmann distribution and then fit not to the histogram itself but to a dimensionless effective energy $U_\text{eff}(\Phi)$ defined so that $\rho_0(\Phi)=(1/Z)\exp(-U_{\text{eff}}(\Phi))$, where $Z$ ensures that $\int d\Phi\rho_0(\Phi)=1$. A judicious choice of basis function is required to capture the distribution with high fidelity; we scale our data appropriately and then use the Chebvyshev polynomials of the first kind $T_n(\Phi)$ which form an orthonormal, complete basis on $[-1,1]$ and are each bounded between $-1$ and 1 over this range~\footnote{Legrendre polynomials were also considered because of their similar properties, but we chose Chebyshev polynomials because they obtained more accurate results with fewer total coefficients.}. Figure~\ref{fig:Ueff} shows progressively higher order fits to $U_{\text{eff}}$; for the rest of the work presented in this paper, we use the $n=20$ fit for the effective energy and the stationary distribution. For practical applications the choice of the fit order would be a balance between fidelity to the data and speed of calculation.

The effective energy in Fig.~\ref{fig:Ueff}, derived from the stationary distribution for the experimental data, exhibits several features that are inconsistent with a macrospin model as we show later in Sec.~\ref{sec:macrospin}. These include asymmetry for the interior/exterior sides of the P/AP states, disparate P/AP well widths, and exterior boundaries that are significantly rounded compared to the macrospin results. We believe that these discrepancies may be related to a common cause: the way the macrospin model neglects many degrees of freedom that may play an important role in the dynamics. The asymmetry between the parallel and antiparallel alignments could be due to asymmetry between the configurations due to the non-uniform fringing fields from the fixed layer and synthetic antiferromagnet. The tails in the distribution could have several explanations. At a basic level, it could be that the states with the minimum and maximum resistances are not the lowest energy states. One may also note that for an open system (the SMTJ) in contact with a thermal reservoir it is the free energy, rather than internal energy, that is the relevant quantity~\cite{desplat_2020} for making thermodynamic predictions like the probability distributions. The measured distribution integrates over the unmeasured degrees of freedom (those that do not lie along the fixed layer magnetization) so, even if the strictly parallel and strictly antiparallel states \emph{were} energy minima, slightly higher energy states may be significantly more prolific so that the peaks in the marginal distrubtion along the voltage, $\Phi$, may be shifted from the energy minima.

The desire to capture these features of the histograms and other statistical measures are what motivates us to develop a compact model beyond the macrospin approach. As we have elaborated on above, our approach is to assume a 1D overdamped Langevin equation driven by white noise. The simplicity of this model makes our task tractable, and in fact, we will show in Sec.~\ref{sec:results} that the model gives very good agreement with experiment. 

To determine the parameters for the Langevin model, we start with the Fokker-Plank equation, which describes the evolution of the probability density $\rho(\Phi,t)$ as
\begin{equation} \label{eq:fokker-planck}
    \frac{\partial \rho}{\partial t} = \frac{\partial }{\partial \Phi} \left[ -D_1(\Phi) \rho(\Phi,t) + \frac{\partial}{\partial \Phi} D_2(\Phi) \rho(\Phi,t) \right],
\end{equation}
with $\Phi$ a generic state variable (voltage, in the present paper), $t$ the time, $D_1(\Phi,t)$ the drift, and $D_2(\Phi,t)$ the diffusion. The terms $D_1(\Phi,t)$ and $D_2(\Phi,t)$ are precisely the first- and second-order Kramers-Moyal coefficients from Eq.~\eqref{eq:KM-coef-D}. The Kramers-Moyal coefficients may be functions of time in general, but the measured probability distribution $\rho_0$ is stationary on the timescale of the experiment and thus we also assume $D_n(\Phi,t)$ are not functions of time. In the steady-state limit $\partial \rho/\partial t= 0$ we have
\begin{equation} \label{eq:D1ansatz}
    D_1(\Phi) = D_2'(\Phi) - D_2(\Phi) \frac{dU_\text{eff}(\Phi)}{d\Phi} ,
\end{equation}
so that $D_1$ is uniquely determined given the diffusion coefficient and the steady-state distribution of the system. In passing from Eq.~\eqref{eq:fokker-planck} to Eq.~\eqref{eq:D1ansatz}, we used the fact that the right-hand side of Eq.~\eqref{eq:fokker-planck} is the divergence of the probability current $J$. In the steady state limit, $\partial_\Phi J= 0$, and thus the stationary current $J_0$ is constant. For the system to be physically bounded, the constant value of $J_0$ must in fact be zero, which allows us to uniquely solve for $D_1(\Phi)$ as a function of $D_2(\Phi)$ and $\rho_0(\Phi)$.

In a system with additive noise -- that is, $D_2'(\Phi)=0$ -- the drift term would be determined through the derivative of the effective energy, in which case $D_1$ coincides with a typical conservative force, up to prefactors. A system with multiplicative noise -- where $D_2'(\Phi) \neq 0$ -- exhibits mixing between the stochastic terms and the deterministic part; here the nonzero diffusion gradient introduces a stochastic drift term.

The Langevin equation derived from the Fokker-Plank equation is
\begin{equation} \label{eq:Langevin}
   \dot \Phi = f(\Phi) + g(\Phi) \; \eta_\Phi(t)
\end{equation}
with $\Phi$ the voltage state, $f(\Phi)$ the deterministic drift term, $g(\Phi)$ the stochastic diffusion term, and $\eta_\Phi$ is white noise with $\langle \eta_\Phi(t) \eta_\Phi(t') \rangle_t = 2 \delta(t' - t)$ and $\langle \eta_\Phi(t) \rangle_t = 0$.
We interpret this equation in the Stratonovich sense~\cite{van1981ito}, as one would for the typical construction of the sLLG equation. In the Stratonovich interpretation, the Langevin equation is related to the Fokker-Planck equation's Kramers-Moyal coefficients by $f(\Phi) = D_1(\Phi) - D_2'(\Phi)/2$ and $g(\Phi) = \sqrt{D_2(\Phi)}$. The second term on the right-hand side of $f(\Phi)$ comes from gradients in the diffusion which give rise to an effective drift term independent of the usual drift coefficient $D_1(\Phi)$.

%\subsection{Constructing the model}
%\label{sec:ansatz}

\begin{figure}
    \centering
    \includegraphics[width=\columnwidth]{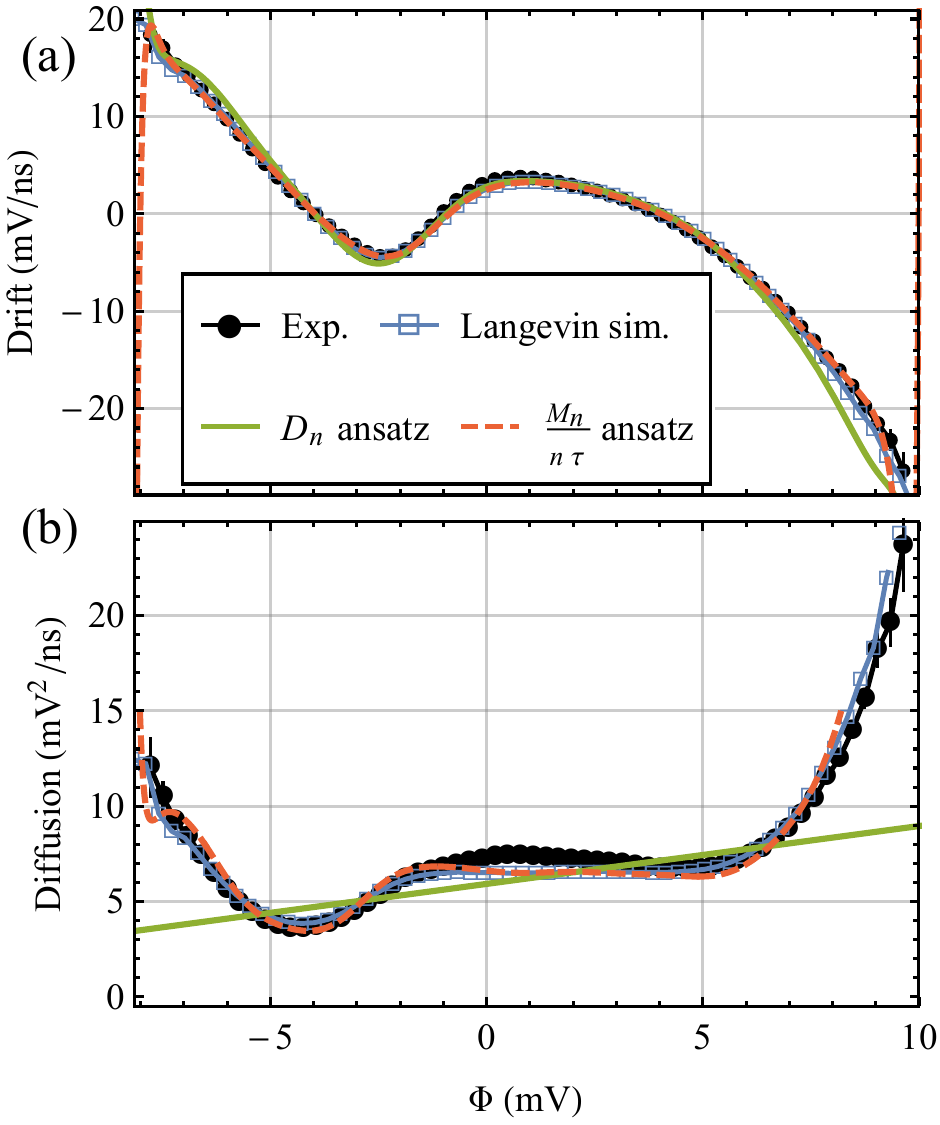}
    \caption{Time-delayed drift coefficient (a) and diffusion coefficient (b), showing experiment (black circles), Langevin simulation (blue squares), computed at a sampling time $\tau_{\text{samp}} = 50$~ps equivalent to the experimental measurement time, and the analytic, fitted results that were inserted into the model, $D_n$ (solid green line) and $M_n/n\tau$ (dashed red line), where $n=1$ for the drift term and $n=2$  for the diffusion term. Error bars on the experimental indicate single standard deviation uncertainties in the mean. They are smaller than the plot symbols except at the extreme voltages.}
    \label{fig:KMcomparison}
\end{figure}

The first step in determining a Langevin model that can reproduce aspects of the experimental voltage-time trace in Fig.~\ref{fig:experiment-trace} is to compute the conditional moments in Eq.~\eqref{eq:KM-coef-M}. These moments are shown in Fig.~\ref{fig:KMcomparison}.
With these experimental conditional moments, we determine the Kramers-Moyal coefficients by making an \textit{ansatz} $\widetilde{D}_2$ for the diffusion part of the Fokker-Planck equation, which when combined with the stationary distribution is used to obtain an induced form for $D_1$ via Eq.~\eqref{eq:D1ansatz}. 
Measurements of $M_2$, shown in Fig.~\ref{fig:KMcomparison}, inform our choice for $\widetilde{D}_2$. The observation that the data show larger diffusion in the AP well than in the P well leads us to propose
\begin{equation} \label{eq:D2ansatz}
    \widetilde{D}_2(\Phi;\bm{\mu}) = m \Phi + b,
\end{equation}
with $\bm{\mu}=(m,b)$ the fit parameters. In Appendix~\ref{app:fitting} we show that fitting with a constant $\widetilde{D}_2$ also provides an adequate fit, and introducing more fitting parameters can give an even better fit. To ensure that $D_2$ is positive definite we take $D_2 = \lambda\log(1 + \text{exp}(\widetilde{D}_2)/\lambda)$, where we take $\lambda=1$~V$^2$/s, which does not significantly affect $D_2>0$. Combining this \textit{ansatz} for $D_2$ with the analytical expressions for $D_1$ and the stationary distribution $\rho_0$ completely specifies the model. In the next section, we use this model to run stochastic Langevin equation simulations, that is, numerical integrations of Eq.~\eqref{eq:Langevin}.

We fit the parameters of the \textit{ansatz} in Eq.~\eqref{eq:D2ansatz} following the procedure outlined in Sec.~\ref{sec:Langevin}.  The fit parameters in Eq.~\eqref{eq:D2ansatz} are determined by first computing an analytic form for $D_1$ from the \textit{ansatz} for $D_2$, the fit to the histogram, and Eq.~\eqref{eq:D1ansatz}. Then an analytic form for $M_2$ is determined from Eq.~\eqref{eq:KM-correctionM2}. The parameters of this last expression are adjusted to fit the experimental data. The agreement between the model and the experimental data is shown in Fig.~\ref{fig:KMcomparison}. The error in the drift and diffusion coefficients were calculated from the standard deviation $\sigma_i/\sqrt{N_i}$ of the underlying $\delta \Phi^{n}(\Phi)$ distribution as a function for the binned $\Phi_i$ levels shown in the figure, where $\sigma$ is the standard deviation and $N_i$ is the number of counts within the $i^\text{th}$ $\Phi_i$ level. The binning level used for $\Phi_i$ here was approximately 295~{\textmu}V to have exactly 64 total bins over the sample. The maximum error shown in the plots is observed on the exterior well boundaries where the number of $N_i$ counts is on the order of hundreds of data points, instead of millions nearer the well's interior.

Both drift and diffusion coefficients (that is, the induced \emph{ansatzes} for $M_n$), agree well with the conditional moments extracted from experiment except at the extreme voltage values where the statistical certainty is poor.  Figure~\ref{fig:KMcomparison} shows the functional forms determined for the model. The model is constructed so that the $M_2/2\tau$ is fit to the experimental data; the consequent agreement between $M_1/\tau$ and the experimental data speaks to the appropriateness of the underlying Fokker-Planck model as a description of this system. 

\section{Results}
\label{sec:results}

Our ultimate goal is to develop a stochastic differential equation of motion for SMTJs for use in circuit simulators. Though our current model is restricted to a particular biasing condition, we can still explore how well our model matches the device. We simulate the Langevin equation (Eq.~\eqref{eq:Langevin}) and compare simulation results both to the experimental data used to fit the model and to selected statistical metrics of the experimental data.

After obtaining the Kramers-Moyal coefficients $D_1$ and $D_2$, we insert them into Eq.~\eqref{eq:Langevin}. We set the integration timestep to a fifth of the experimental sampling rate, $dt=\tau_{\text{exp}}/5 = 10$~ps. Integration is performed using the Euler-Maruyama method. Simulation histogram results are compared to the experimental histogram in Fig.~\ref{fig:histogram}, confirming that the first-order, data-driven Langevin model captures this aspect of the experiment well. The histogram displays asymmetrical energy-well probabilities and widths, greater-than-exponential decay for the probability density at exterior data boundaries, and a large energy barrier between the two well peaks. We assert that the reproduction of statistical features such as these is key to developing an engineering-appropriate model for the device. For purely analog circuits in particular -- such as those proposed in Refs.~\cite{daniels2023neural} or \cite{bunaiyan2023heisenberg} -- correct modeling of these distributions may be crucial for capturing emergent statistics of scaled-up circuitry based on superparamagnetic tunnel junctions. 

Comparing the conditional moment extracted from experiment and simulation demonstrates that the first-order Langevin model gives dynamics similar to those seen in experiment. Figure~\ref{fig:KMcomparison} confirms excellent agreement for $M_1(\Phi,\tau)$ and $M_2(\Phi,\tau)$ calculated for a time delay equal to the experimental sampling time ($\tau_{\text{samp}}=\tau_{\text{exp}} = 50$~ps).  

The statistics that have received perhaps the most attention in the literature are the mean dwell times of the parallel and antiparallel states. These are captured directly by construction in N\'eel-Brown models, and macrospin models are generally fit to ensure these dwell times are accurate~\cite{torunbalciModularCompactModeling2018}. In circuits and systems where the SMTJ state is binarized into a two-level telegraph signal, the mean dwell times together with the assumption that the switching events are independent of the past history -- the Markov property -- completely determine the circuit output. With a voltage threshold chosen to demark the boundary between the P and AP states, the dwell times imbue the data we have already captured in the histogram with a characteristic time scale. Crucially, this timescale information is carried into the model through our fitted \textit{ansatz} on $D_2$.
For the purposes of this work we consider a transition between the P and AP states to occur when the voltage crosses the barrier location determined by the local maximum of the energy landscape in Fig.~\ref{fig:Ueff}. We define a dwell time as the amount of time the system spends in the positive or negative half of state space before crossing that $\partial_\Phi U_\text{eff} = 0$ threshold into the other half of state space. 

%Among other potential complications, then, higher order corrections (i.e. $\mathcal{O}(\tau^3)$ and beyond) in the expressions for $M_1$ and $M_2$ begin to have larger effects in Eqs.~\eqref{eq:KM-correctionM1} and \eqref{eq:KM-correctionM2} in relating these subsampled data to the underlying Kramers-Moyal coefficients, leading to a failure of the underpinning assumptions of the Langevin model.

\begin{figure}
    \centering\includegraphics[width=\columnwidth]{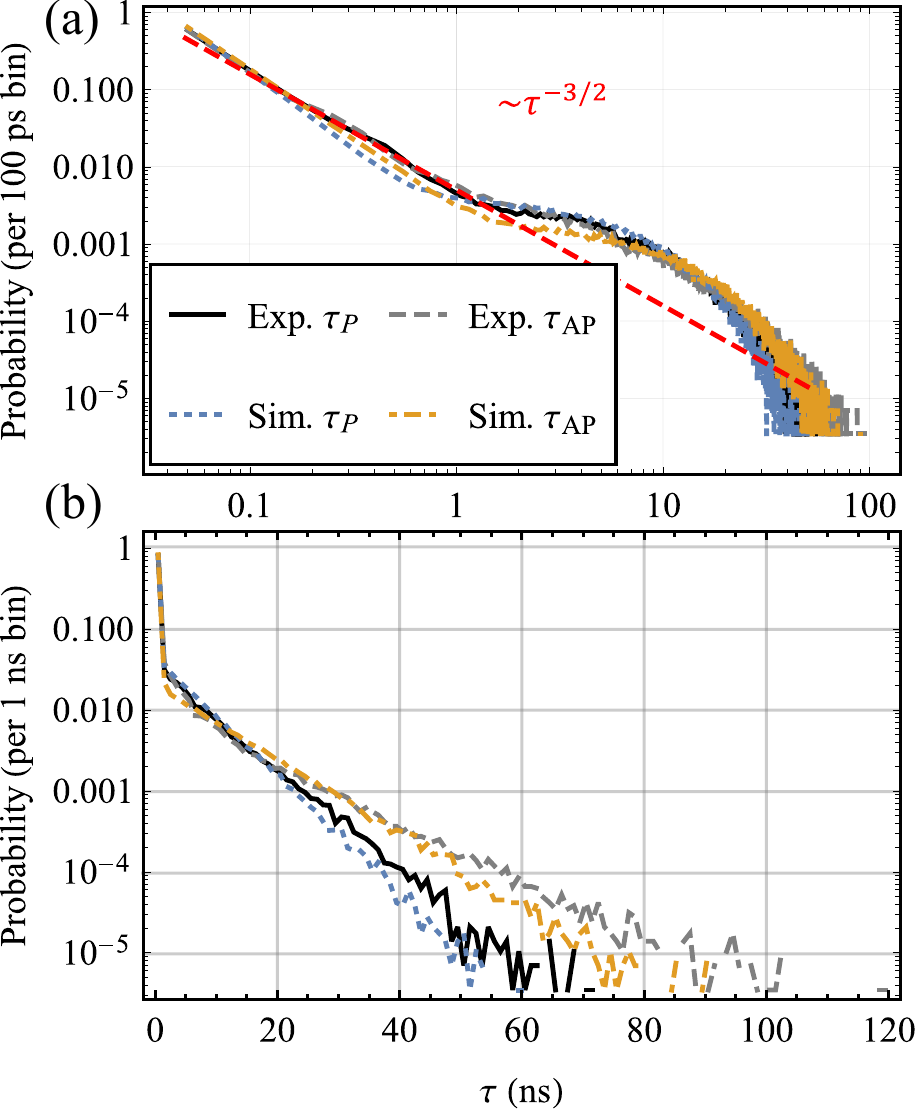}
    \caption{Dwell time distributions as extracted from experiment (black) and  simulation (blue) in (a) a log-log plot (a) and (b) a semi-log plot. The log-log plot uses a sampling time bin size of 100~ps, while the semi-log plot uses a sampling time bin size of 1~ns.  The experimental data and simulation both have on the order of 300,000 transitions. The straight-line behavior of both the simulation and experiment for short times in the upper panel indicates a power-law distribution and for long times in the lower panel an exponential distribution. }
    \label{fig:dwell-times}
\end{figure}

%Even in the converged limit where are mean dwell times to capture the experimental dwell times, further complications arise in considering the dwell time \emph{distributions}.
The main panels in Fig.~\ref{fig:dwell-times} show simulation and experimental dwell times for a sampling rate of $\tau_{\text{samp}} = \tau_{\text{exp}}$ on log-log and semi-log plots for the two distinct P and AP states. These figures show that the simulation captures the full distribution of dwell times for both states. This agreement for the dwell time distributions is perhaps the most significant validation of the model because the experimental input to the model does not directly encode these time scales. The model uses just the short-time dynamics captured by the Kramers-Moyal coefficients and the stationary limit of the time-averaged time series in the form of the histogram. 

One unusual feature of both experimental and simulated distributions is the crossover from the familiar exponential behavior to $\sim t^{-3/2}$ power-law behavior at short times. We attribute this behavior to fluctuations around the threshold we defined between the states. Recall that we chose the threshold between states to correspond to the local maximum of the effective energy landscape; linearizing about this threshold therefore reveals a locally flat energy landscape where the dynamics are driven entirely by stochastic fluctuations. In other words, the dynamics in this small neighborhood around the threshold are given by a random walk. It is well-known that the dwell time (return time) distribution of a random walk is proportional to $\sqrt{\tau/t^3}$, where $\tau$ is the timestep of the walk~\cite{kostinski2016elementary}. As $\tau$ decreases, more and more probability mass accumulates at smaller and smaller $t$, dominating the dwell time distribution. This suggests that the transition between the power-law and exponentially distributed dwell times is controlled by the curvature of the energy barrier: the flatter the energy near the threshold, the longer the system can behave as an unbiased random walk. 

\change{
\begin{figure}
    \centering
    \includegraphics[width=\columnwidth]{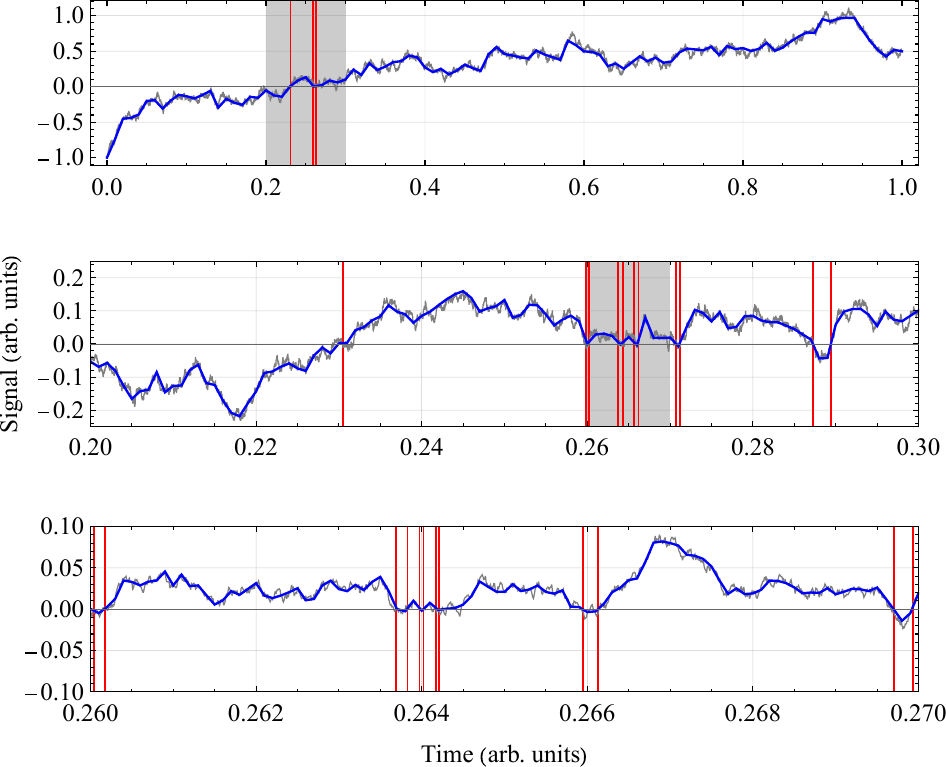}
    \caption{Progressively increased zoom and sampling rates on a random trajectory generated by an arbitrarily simulated Ornstein-Uhlenbeck process. Top panel: the full-resolution trajectory which has $10^7$ sampling points is given in gray. The blue trajectory, subsampled to have only $10^2$ points, represents what might one collect experimentally. Red vertical bars denote zero-crossings of the blue trajectory. We zoom in on the gray region of the top panel in the center panel. Center panel: here we locally resample with $10^2$ points, an order of magnitude faster sampling than in the top panel. The resampling reveals more zero-crossings than were present at that larger scale. Bottom panel: zooming and resampling by factors of 10 again reveals yet more crossings. Note the qualitative self-similarity of the blue curves in each panel, suggesting that (with infinite underlying resolution) we would continue to reveal more and more crossings with higher and higher sampling rates.}
    \label{fig:fractal-zoom}
\end{figure}
}
Intuitively, we can imagine zooming in to a threshold-crossing trajectory sampled at some frequency. \change{If we ``zoom in'' by increasing our sampling frequency on the same underlying trajectory, the newly sampled points have the potential to introduce additional threshold crossings.} Since the spectrum of the white noise process contains all frequencies, we expect to introduce more and more crossing events as we sample at higher and higher frequencies; as the sampling time goes to zero, the number of crossings will diverge. 

\change{To see this visually, consider Fig.~\ref{fig:fractal-zoom}. In the top panel, we plot in gray a simulated trajectory of an Ornstein-Uhlenbeck process with unit mean and volatility, spring constant $\theta = 2$, and initial condition $y=-1$. As it traverses from the initial condition toward its energy minimum at $y=1$, it crosses the $y=0$ line, which we regard as a threshold. The question is how many times it crosses this threshold. This question is easily addressed if we have the full trajectory data. But suppose the number of points we have access to on the trajectory is much smaller than the ``true'' number of points. In Figs.~\ref{fig:fractal-zoom}, we denote subsampled trajectories with a blue line.

In the top panel of Fig.~\ref{fig:fractal-zoom}, we observe three crossings of the subsampled trajectory across the threshold. But if we zoom in on the gray region and resample the underlying trajectory there (middle panel), we find that far more crossings become apparent, approximately 11. We zoom in and resample once again on a subdomain of this trajectory to find an even finer resolution (bottom panel), which has yet more transitions. Note that in each case, the size scale of the flucutations (i.e. the range of the vertical axis) gets smaller and smaller. But the thresholding process, by definition, ignores this amplitude information, and since the white noise driving the system has support at all frequencies, we will always find more transitions by going to higher and higher sampling rates.}

In the physical SMTJ system, we expect some physical mechanism to impose an ultraviolet cutoff---but whatever this cutoff may be clearly resides at a higher frequency than our experiment can access. Since this power-law behavior is simply a property of stochastic dynamics around a threshold, we expect it to be found in other stochastic magnetic systems as well. In Sec.~\ref{sec:macrospin}, we will find that the behavior is also reproduced in the macrospin model, but note that it is impossible to capture in a simple N\'eel-Brown model, which is based on a two-state Markov model that assumes exponential distributions of dwell times; such a system has no notion of dynamics in the barrier, which is the essential cause of the power-law effect.

Traditionally, mean dwell times are extracted from experimental dwell time histograms by examining the slope of the exponential dwell time distributions on a semilog scale (wherein the distribution appears linear, as in Fig.~\ref{fig:easy-axis-macrospin-long}). Because of the power-law behavior at small times in our data, these inverse slopes -- which we will call \emph{characteristic dwell times} -- are clearly very different than the literal mean dwell times. Yet merely isolating the slope of the curve where it does appear exponential in Fig.~\ref{fig:dwell-times} does not lead to a physically invariant characteristic dwell time; this slope can change significantly under minor subsampling of the signal, because this changes the probability-dominating power-law part. We elaborate more on the topic of characteristic dwell times in Appendix~\ref{app:dwell-times}.

%The overdamped, first-order Langevin model we use throughout this paper assumes the underlying dynamics are Markovian.  
%We can test the validity of this assumption by constructing a rough metric describing the memorylessness of the observed trajectories. For each time $t$, we compute the probability of the system to increase (go up in voltage) or decrease (go down in voltage) at time $t+\tau$ conditioned on whether it increased or decreased from time $t-\tau$ to $t$. From these observed quantities we can compute four distinct probabilities: up after up $P_{\uparrow \uparrow}$; down after down $P_{\downarrow \downarrow}$; up after down $P_{\uparrow \downarrow}$; and down after up $P_{\downarrow \uparrow}$; in these expressions, the left arrow shows the second change and the right arrow shows the first change. The sum of $P_{\uparrow \uparrow}$ and $P_{\downarrow \uparrow}$ equals one since the system must either increase or decrease in value. Likewise the sum of $P_{\downarrow \downarrow}$ and $P_{\uparrow \downarrow}$ also equals one.
%If the system is Markovian, the probabilities that have the same second-step direction will be independent of the direction of their previous steps, and only dependent on the current system state (i.e. $P_{\uparrow \downarrow}(\Phi) = P_{\uparrow \uparrow}(\Phi)$ and $P_{\downarrow \downarrow}(\Phi) = P_{\downarrow \uparrow}(\Phi)$). 

Finally, we compare the power spectral densities of the experimental signal to those produced by the model. Accurate power spectra are crucial from an electrical engineering context. Unlike the diffusion characteristics and histograms, power spectra are known to exhibit the correct qualitative (Lorentzian) form in macrospin models~\cite{leliaertAdaptivelyTimeStepping2017} and in N\'eel-Brown models~\cite{fitzhughStatisticalPropertiesAsymmetric1983}. In Appendix~\ref{app:Elec_Meas_Circuit}, we show that the power spectral density of the simulation agrees with that of the experiment and both follow the expected behavior at high frequencies.

\section{Analysis of macrospin modeling\label{sec:macrospin}}

In the previous sections, we described the modeling of an experimental dataset collected from a device in the laboratory. In service of our engineering-focused modeling goals, our objective was to create compact models that faithfully capture the relevant physics of \emph{that particular device}. A common approach in the literature to modeling SMTJs is through the use of a macrospin model. \change{In the present section, we consider the macrospin model for SMTJs in two separate contexts. First, we show that it cannot reproduce important aspects of the measured properties of our device. Second, use the macrospin model as experimental data to test the generality of our approach using the Langevin model to fit data. In this context, we show that our Langevin modeling approach can capture the dynamics of an easy-axis macrospin. However, for other devices of current interest, referred to as low-barrier devices~\cite{hayakawa2021nanosecond,schnitzspanNanosecondSuperparamagneticTunnel}, in which there is a strong easy-plane anisotropy axis and a weaker in-plane easy axes, the Langevin approach can model the drift and diffusion characteristics but is unable to capture the correct dwell-time distributions of the device. This failure is due to the inertial dynamics that are neglected in the overdamped Langevin approach. Treating such systems will require additional developments as discussed at the end of this section.}
%In the present section we consider the macrospin model for SMTJs, confirming some of this model's problems for modeling our device. We also show that while our Langevin modeling approach can capture the dynamics of an easy-axis macrospin, adding an additional strong anisotropy axis (as in so-called low-barrier devices~\cite{hayakawa2021nanosecond,schnitzspanNanosecondSuperparamagneticTunnel}) introduces enough inertial dynamics that the overdamped Langevin approach is unable to capture the correct dwell-time distributions of the device -- though it can still model the drift and diffusion characteristics.

\subsection{Macrospin model}
\begin{figure}
    \centering
    \includegraphics[width=\columnwidth]{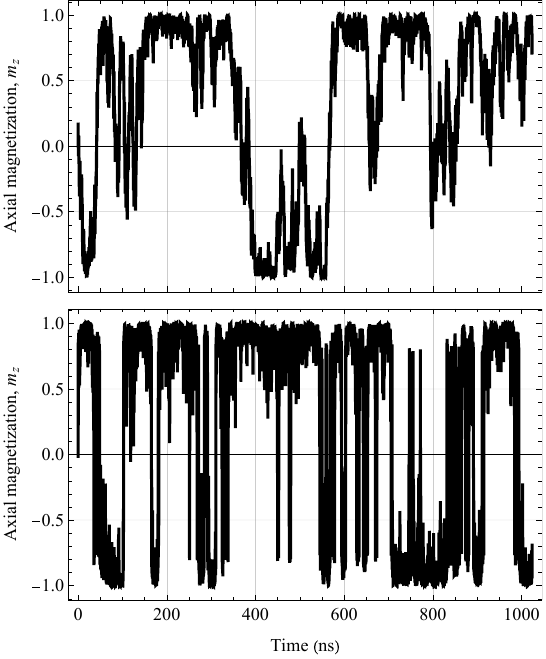}
    \caption{Typical magnetization-time traces from the sLLG simulation of the easy-axis (top) and hard-axis with in-plane easy axis (bottom) macrospin model. In both panels, the simulation was conducted with a timestep of $10$~fs, but the trajectories are plotted by subsampling the raw data with a $5$~ps timestep (corresponding to the subsampling used in the experimental section). Compare with Fig.~\ref{fig:experiment-trace}: we have harder walls at the extremal states in the vertical direction.}
    \label{fig:simulation-trace}
\end{figure}

We consider here a typical macrospin model with anisotropy. The dynamics are generated by the stochastic Landau-Lifshitz-Gilbert (sLLG) equation for dimensionless magnetization $\bm m$,
\begin{equation}
    \dot{\bm m} = -\tilde\gamma_0\bm m\times\bm{h}_\text{eff} - \lambda \bm m \times \left(\bm m \times \bm{h}_\text{eff} \right),
    \label{eq:sllg-full}
\end{equation}
with $\bm{h}_\text{eff} = \bm{\xi} - \partial_{\bm m}\mathcal{E}/(\mu_0M_s)$ the effective field with thermal part $\bm \xi$ and $M_s$ the saturation magnetization, $\tilde\gamma_0 = \gamma_0/(1+\alpha^2)$ the renormalized gyromagnetic ratio, $\lambda = \gamma_0\alpha/(1+\alpha^2)$ with $\alpha$ the Gilbert damping, and $\gamma_0 = \mu_0 g\mu_B/\hbar$, with $\mu_0$ the vacuum magnetic permeability, $g$ the electron $g$-factor, $\mu_B$ the Bohr magneton, and $\hbar$ the reduced Planck constant. We take $\bm \xi$ to be a spherically symmetric Gaussian noise term with variance $2\Gamma = 2\alpha kT/(\mu_0 \gamma_0 M_s \mathcal{V})$~\cite{xiao2005macrospin}; with $\mathcal{V}$ the volume, $k$ the Boltzmann constant, and $T$ the temperature. Finally, $E =  -K_z m_z^2 -K_x m_x^2 $ is the energy density. The first term, with $K_z>0$, is an easy-axis anisotropy along the direction of the fixed layer magnetization, \textit{i.e.} the component of the magnetization that determines the magnetoresistance. We take $m_z$ as a proxy for the voltage signal across an MTJ at fixed bias; in Fig.~\ref{fig:simulation-trace}, for instance, $m_z$ is the vertical axis (compare to Fig.~\ref{fig:experiment-trace}). The second term with $K_x\leq 0$ is a (potential) easy-plane anisotropy.  Other contributions, like external fields, other anisotropies, or current-driven torques could be included but would not change the conclusions we draw below. We assume the Stratonovich interpretation here as we have for all other stochastic differential equations throughout the paper~\cite{van1981ito}.

\subsection{Capturing macrospin physics in a first-order Langevin model \label{sec:capturing-macrospin-Langevin}}
Figure~\ref{fig:simulation-trace} shows typical time traces for a macrospin where $K_z = 5\,kT$; in the top panel, $K_x = 0$, and in the bottom panel, $K_x = -10\,kT$. Compare with Fig.~\ref{fig:experiment-trace} (or experimentally measured voltage-time trace signals); macrospin simulations show much harder walls at the $m_z = \pm 1$ state than we see in experiment. We will soon see that the diffusion characteristics also look quite different than the experiment. Nevertheless, in this section, we attempt to capture the physics of a macrospin within an overdamped Langevin model with a single degree of freedom, $m_z$. 
\begin{figure}
    \centering
    \includegraphics[width=\columnwidth]{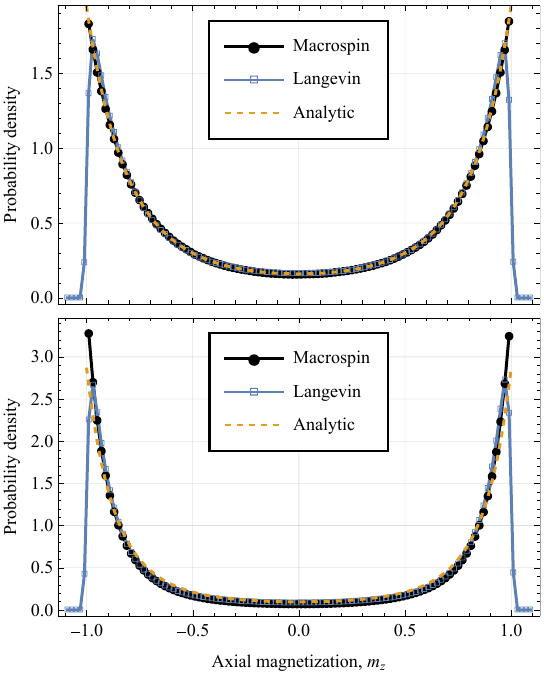}
        \caption{Comparison of histograms between our sLLG simulation of the easy-axis macrospin model and the first-order Langevin model that was fit to the sLLG simulation statistics, together with the analytic solution for these distributions (dashed line). Top panel shows the easy-axis macrospin;  bottom, a hard-axis macrospin with an in-plane easy axis.}
    \label{fig:easy-axis-macrospin}
\end{figure}

\begin{figure}
    \centering
    \includegraphics[width=\columnwidth]{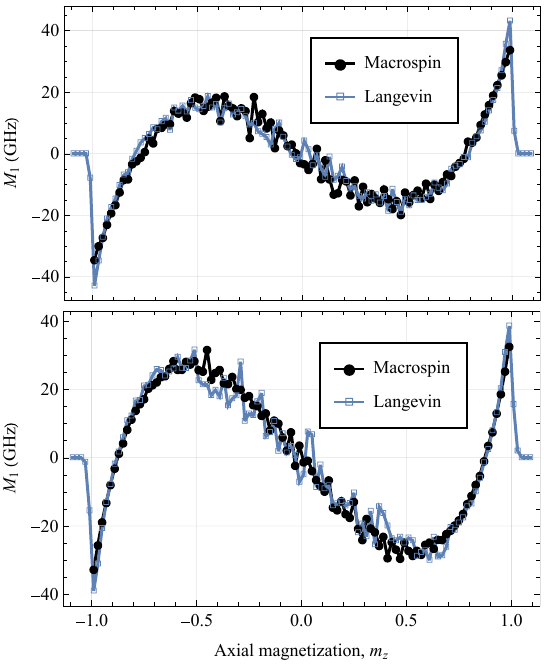}
    \caption{Comparisons of $M_1(\tau)/\tau$ between our sLLG simulation of the easy-axis macrospin model and the first-order Langevin model that was fit to the sLLG simulation statistics. Top panel shows the easy-axis macrospin; bottom, a hard-axis macrospin with an in-plane easy axis. In both cases, $\tau=10$~fs.}
    \label{fig:easy-axis-macrospin-m1}
\end{figure}

Our experimental results indicate that our real device has two relatively well-defined states, which suggests an easy-axis description in the macrospin limit. If the system has only easy-axis anisotropy in the energy density, we can isolate the equations of motion for the $z$-component of a macrospin. Rewriting Eq.~\eqref{eq:sllg-full} in terms of $m_z$ and $\phi$ gives an equation for $\dot\phi$ that depends on $m_z$ and one for $\dot m_z$ that is independent of $\phi$
\begin{equation}
    \dot m_z = \frac{\lambda K_z}{\mu_0 M_s \mathcal{V}} (m_z-m_z^3) - \tilde\gamma_0(L_\phi + \alpha L_\theta)\sqrt{1-m_z^2},
\end{equation}
where the  two independent white noise processes,
$L_\phi$ and $L_\theta$ each have variance $2\Gamma$ and produce thermal fields in the $\hat\phi$ and $\hat\theta$ directions. We combine these and regard them as a combined white noise process $L$ with variance $2\tilde\Gamma = 2\Gamma(1+\alpha^2) = 2\lambda kT/(\mu_0M_s\mathcal{V})$. Then it is straightforward to derive the drift and diffusion coefficients,
\begin{subequations}
\label{eq:d1-d2-macrospin}
\begin{align}
    D_1(m_z) &= \frac{\lambda m_z}{\mu_0 M_s V}\left( K_z (1 - m_z^2) - kT \right)\\
    \text{and}\;\;D_2(m_z) &= \frac{\lambda kT}{\mu_0 M_sV}(1-m_z^2).\label{eq:d2-macrospin}
\end{align}
\end{subequations}
The critical point $K_z=kT$ is a pitchfork bifurcation in the drift coefficient where $m_z=0$ changes stability and two additional zeros of $D_1(m_z)$ appear. Note that the value of $D_1(\pm 1)$ is exactly $\mp \tilde\Gamma$; the conservative forces do not contribute there.
%We plot these for a few sample values of $K$ in Fig.~\ref{fig:macrospin-d1-d2}.

%\begin{figure}
    %\centering
    %\includegraphics[width=\columnwidth]{macrospin-d1-d2.pdf}
    %\caption{Top: behavior of the drift coefficient $D_1(m_z)$ for the easy-axis macrospin model, for varying strengths of the easy axis anistropy $K$. Bottom: behavior of the diffusion coefficient $D_2(m_z)$. Note that }
    %\label{fig:macrospin-d1-d2}
%\end{figure}

\begin{figure}
    \centering
    \includegraphics[width=\columnwidth]{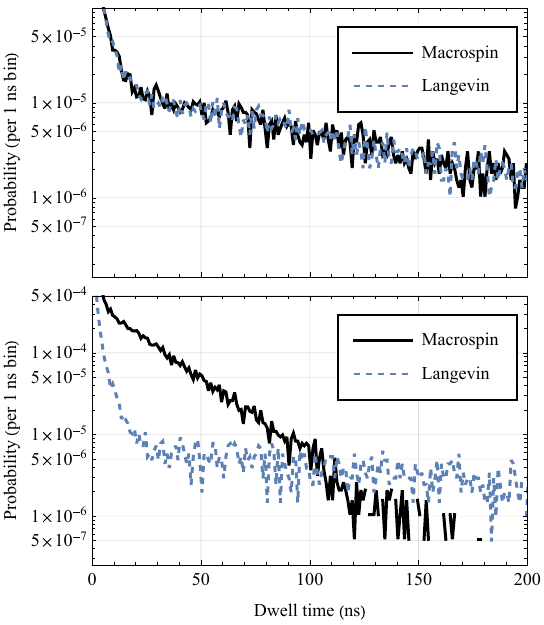}
    \caption{Long-time behavior of the dwell time histograms for the easy-axis macrospin (top) and hard-axis macrospin with an in-plane easy axis (bottom). As in Fig.~\ref{fig:easy-axis-macrospin-short}, the model fit to the hard axis macrospin has poor agreement, but the easy axis system matches well. Note that this data has been rebinned to a 1~GHz sampling frequency compared to Fig.~\ref{fig:easy-axis-macrospin-short}; on this time scale, the coherent switching dynamics of the easy-plane oscillator are washed out.}
    \label{fig:easy-axis-macrospin-long}
\end{figure}
\begin{figure}
    \centering
    \includegraphics[width=\columnwidth]{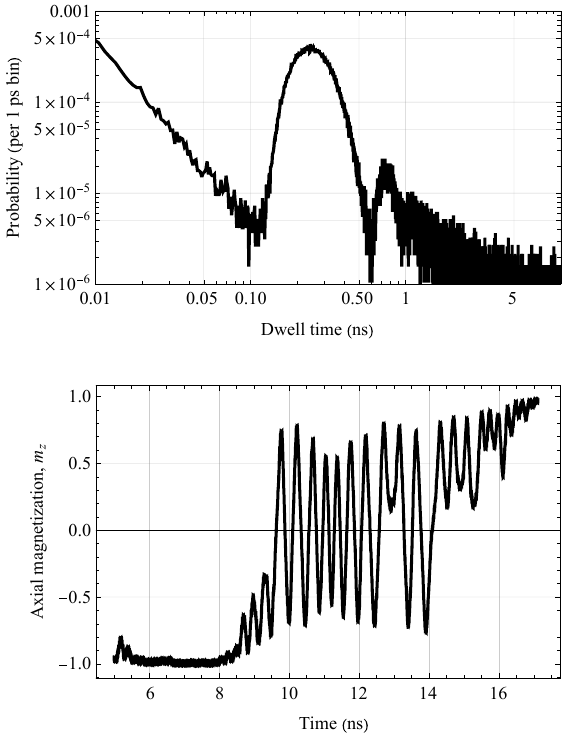}
    \caption{Top: short-time behavior of the dwell time histogram for the hard-axis macrospin with an in-plane easy axis. The central peak in the distribution is not captured by the fitted Langevin model; abnormally frequent dwell times around $250$~ps (and, to a lesser degree, around 750~ps) are attributed to coherent in-plane switching cycles. Bottom: a high resolution (10~fs timestep) time trace of a transition in the hard-axis macrospin model that exhibits coherent ringing effects on subnanosecond timescales leading to $\approx 0.2$~ns dwell events, commensurate with the central peak in the top panel.}
    \label{fig:easy-axis-macrospin-short}
\end{figure}
Equipped with the drift and diffusion coefficients of this process in $m_z$ coordinates, it would be theoretically valid to write down a Langevin equation
\begin{equation}
    dm_z = \left[D_1(m_z) - \frac{1}{2}D_2'(m_z)\right] dt+\sqrt{2 D_2(m_z)} dW,\label{eq:dmz-macrospin}
\end{equation}
where $W$ is a Wiener process~\cite{lemons2002introduction} with unit variance and zero mean. Given Eqs.~\eqref{eq:d1-d2-macrospin}, however, it is clear that this equation would be difficult to numerically simulate. Any finite timestep integration of this process is liable to step beyond the valid domain $m_z\in[-1,1]$, leading both to imaginary diffusion and to a drift term that would push the system even further away from the valid regime.

There may well be methods to properly integrate Eq.~\eqref{eq:dmz-macrospin}, but for the purposes of this paper, we will instead consider our data-driven approach. First, we simulate the system according to the original description Eq.~\eqref{eq:sllg-full}. By examination of the conditional moments $M_1 = \langle m_z(t+\tau) - m_z(t)\rangle$ and $M_2 = \langle [m_z(t+\tau) - m_z(t)]^2\rangle$ of that ``experimental data'', we once again choose an \textit{ansatz} for the $D_2$ of our synthetic model and fit it to the data via Eq.~\eqref{eq:D2ansatz}. The histogram of the data again defines an energy function $E(m_z) = -\log \rho(m_z)$ which, together with $D_2$, determines $D_1$ through the equilibrium condition of the Fokker-Planck equation,
\begin{equation}
    D_1(m_z) = D_2'(m_z) - D_2(m_z) E'(m_z),
\end{equation}
just as in Sec.~\ref{sec:Langevin}. Then we simulate Eq.~\eqref{eq:dmz-macrospin} using these fitted $D_1$, $D_2$ which will be chosen to have good behavior at the boundaries.

We numerically integrate the sLLG equation that will provide our dataset in two different ways: first, using only an easy axis anisotropy $K_z = 5\,kT$; second, using a hard axis anisotropy $K_x = -10\,kT$ with a weak easy axis $K_z = 5\,kT$. In both cases, we take $\alpha = 0.01$, $M_s = 1.5$~MA/m, and $T = 300$~K. We assume a free layer thickness of 1~nm and a diameter of 20~nm, which approximates our experimental device. 

We then take these data and apply our procedure from Sec.~\ref{sec:Langevin} to extract an effective first-order Langevin system. As the structure of the macrospin dynamics is much simpler than the experimental device, we use only a fourth-order polynomial to fit the effective energy function, though we also include two sharp exponential functions $A_\pm \exp[\pm100(x\mp1)]$ in order to capture the hard-wall boundaries at $m_z = \pm 1$. We then simulate this effective model and compare its histogram to the macrospin simulation data in Fig.~\ref{fig:easy-axis-macrospin}. Both simulations used a timestep of 10~fs and ran for $10^{11}$ timesteps (a total simulated time of 1~ms). While the agreement is good in the easy-axis simulation, our models naturally exhibit ``leak'' of probability beyond $[-1,1]$. Both models have analytically tractable histograms with which the macrospin and Langevin simulations well agree.

Theoretically, the structure of the multiplicative noise in the sLLG equation (which manifests here as $D_2$) arises entirely from the geometric considerations of $\bm{m}$'s normalized amplitude~\cite{berkov2002thermally}; the thermally driven part of the sLLG equation in Cartesian coordinates has coefficients $\epsilon_{ijk}m_k + \alpha(\delta_{ij}-m_im_j)$ irrespective of the effective field. The agreement between $M_2$ from all simulations with the known $1-m_z^2$ form of $D_2$ in the macrospin model is excellent; they are essentially indistinguishable except for very tiny deviations at $m_z=\pm1$.   We conclude that a simple macrospin model cannot capture the complex diffusion characteristics such as those observed in Fig.~\ref{fig:KMcomparison}(b), since the functional form of $D_2$ is fixed by construction. Fig.~\ref{fig:easy-axis-macrospin-m1} shows the first conditional moments for these sLLG simulations compared to their fitted Langevin models. These also have nice agreement, though the convergence of the noise is slower than for $M_2$ and the boundary conditions more artificially imposed.

In both the easy axis and hard axis models, the \emph{literal} mean dwell times agree very well between macrospin and Langevin models: the easy-axis mean dwell times are $0.624$~ns (macrospin) and $0.623$~ns (Langevin); the hard-axis mean dwell times are $0.730$~ns (macrospin) and $0.731$~ns (Langevin). We also consider the distribution of dwell times, that is, the probability that any particular dwell event has a given dwell time. These agree very well between macrospin and Langevin models in the easy-axis system (Fig.~\ref{fig:easy-axis-macrospin-long}, top panel). Although we have elaborated above and in App.~\ref{app:dwell-times} on the potential unphysicality of characteristic dwell times extracted from the slopes of these curves, they can still offer a comparison between macrospin and Langevin models. Using a linear fit to extract the slope of the exponential distribution from the semi-log plot, we obtain ``characteristic dwell times'' of 121~ns and 120~ns for both the macrospin and Langevin simulations, respectively. In the hard-axis system, despite the remarkable agreement of the mean dwell times, the dwell time distributions are considerably different (bottom panel). The ``characteristic dwell time'' of the Langevin model is 293~ns, while the ``characteristic dwell time'' of the macrospin model is only 27~ns---an order of magnitude difference.

The top panel of Fig.~\ref{fig:easy-axis-macrospin-short} zooms into the dwell time distribution of the hard-axis macrospin simulation for very short times. Both the easy- and hard-axis macrospin models exhibit the same $t^{-3/2}$ power law behavior~\footnote{We remark that, unlike what we would expect in a real magnetic tunnel junction, there is no physically imposed ultraviolet cutoff for macrospin dynamics. This is not particularly problematic, though, since any good circuit simulation involving such a device would impose its own high-frequency cutoffs through the finite bandwidth of realistic circuit components.} at the smallest time steps (not shown) as was observed in the experimental device~(Fig.~\ref{fig:dwell-times}(a)). At intermediate but subnanosecond dwell times, however, the hard-axis system shows a distinctive set of peaks. We associate these peaks with coherent precession events around multiple cycles of the easy-plane. The bottom panel of Fig.~\ref{fig:easy-axis-macrospin-short} shows an example of a ``transition'' we found in a trajectory generated from the hard-axis macrospin simulations; the ringing here artificially creates ``switching events'' around twice some resonant frequency of the macrospin. Such dynamics rely crucially on second-order behavior, with energy being stored in the out-of-plane angle serving as a source of inertia. Since these dynamics strongly affect the switching times but cannot possibly be modeled in an 1D overdamped Langevin equation, it is unsurprising that our model fails to capture these particular statistics.

The inclusion of memory and higher-order differential terms would, therefore, seem a necessary next step to capture higher fidelity models, especially given the high interest in low-barrier easy-plane devices. Representation of such systems as multiple coupled first-order Langevin models seems the natural direction of study, but this will come hand-in-hand with a quadratic increase in the number of free parameters. \change{Fitting the increased set of parameters will require extracting more statistical information from the time traces. In the simple case we consider here, the first-order Kramers-Moyal coefficient is not needed nor are the dwell time distributions. In addition to these, higher-order or time-lagged Kramers-Moyal coefficients could be extracted and used. It is not implausible that the inverse problem (that is, divining the model directly from the experimental data without having to run simulations to compare) may become intractable. One solution might be to use an iterative, self-consistent forward-fitting approach, where trial models are refined by comparing the statistics extracted from simulations to those of experiments and refining the parameters until the comparison is satisfactory. While finding the appropriate model will become much more complicated, using the resulting model will not.} This type of approach is represented by the dashed arrow in Fig.~\ref{fig:model-roadmap}, and may be a topic of future research.

\section{Discussion}
\label{sec:discuss}

While the sLLG equation for a macropsin is physically motivated and contains terms directly identifiable with distinct physics, it cannot capture certain statistical metrics associated with our experimental device. This is partly due to the sLLG model's drift and diffusion coefficients; fitting the former to the observed drift would require including sufficiently ``data-driven'' terms as to lose the simple plausibility of the model, while directly fitting the diffusion is simply impossible in general. Indeed, even single-domain MTJs may show diffusion properties that do not generally agree with the macrospin. Theory indicates that the Gilbert damping (an overall prefactor for $D_2$) in common magnetic materials has strong directional dependence~\cite{PhysRevB.81.174414}, while recent experiments on SMTJs switching suggests that the Suhl instability may create dynamically variable damping mechanisms during single-domain reversal~\cite{KanaiTBP}. These considerations suggest that we must move beyond the simple macrospin model to accurately capture SMTJ physics in compact models in general, and the data-driven approach explored in this paper is our initial attempt at this enterprise. 

We show that our data-driven approach captures multiple statistical and dynamical metrics with high fidelity in an experimentally measured SMTJ device with perpendicular anisotropy. Specifically, we were able to capture the Kramers-Moyal coefficients (Fig.~\ref{fig:KMcomparison}), the stationary distribution $\rho_0$ by construction (Fig.~\ref{fig:histogram}), the dwell time distributions (Fig.~\ref{fig:dwell-times}). %The model's ability to recover higher-order statistics is required to satisfy the sufficiency condition so that we can use the model to design scaled-up circuits in engineering applications. 
%It is an open question whether this memory arises from the underlying magnetic physics or represents an experimental artifact. One distinct possibility is that the 1.8~MHz response frequency of the current source, which will be strongly relevant during barrier crossings but not while the system explores a single well, contributes to the apparent inertia. However, this mechanism does not explain the ``negative inertia'' found in the wells, where the parameteric plots of Fig.~\ref{fig:Memory} falls below the the equiprobability line. In any case, the system being modeled is ultimately a joint circuit-device system, which must be accounted for if the compact model we develop is to be used in an engineering context as part of a larger architecture.
The first-order Langevin model outperforms traditional sLLG macrospin simulations in capturing coarse-grained statistics of the experimental data. \change{We note that perpendicular devices of this type are of strong applied interest, as the fabrication of their thermally stable kin are already integrated into commercial foundry processes, and challenges with device-to-device variation are expected to be easy to address in this geometry compared to in-plane magnets (so-called low-barrier magnets)~\cite{soumah2024nanosecond}.}

\change{We have also simulated macrospin models to demonstrate that the behaviors they exhibit are not commensurate with the experimental observations that we capture with our Langevin model.  Of course, the macrospin model \emph{is}, ultimately, a Langevin equation. So long as one is willing to accept a macrospin's finite domain, it should be possible in principle to deduce complicated, potentially pathological forms of the effective field and (state-dependent) Gilbert damping functions that would indeed mimic the behavior we observe in experiment. But to do so would be to lose the physical motivation of the macrospin model; we contend that achieving this would amount to using a data-driven approach similar to what we present here, simply in spherical rather than Euclidean geometry.

There are several refinements of our approach that are necessary to successfully design scaled-up circuits that operate over ranges of input parameters. One, which is referred to at the end of the previous section, is to extend the model to include memory effects and incorporate additional statistical properties of the device into the fitting procedure. A second is to extend the model to varying currents through the device. In the present results, the current through the device was kept constant, but for a model to be useful in simulation it must extrapolate to an extensive current range. Additionally, it is likely that the devices will have a range of properties and cannot all be described with a single model. If the distributions of device properties are wide enough, it will be necessary to fit a distribution of device properties with a distribution of model parameters. }
Ultimately, the needs of compact models will be dictated by advancements in circuits and systems that use SMTJs as critical computational elements. Progress on the front of applied physics and engineering in this regard must advance in tandem with physics-level device modeling to ensure that large-scale circuit simulation remains rooted in physical reality.

\section*{Acknowledgements}
The authors are indebted to Di Xiao for his useful notes on stochastic processes and to Steve Moxim for his help with the experimental setup. We also thank William Rippard and Robert McMichael for their invaluable comments on the manuscript, and we again thank Robert McMichael for his advice on microwave theory. This work was funded by the National Institute of Standards and Technology, National Science Foundation, and Agence nationale de la recherche. SG, TA, LP, DPL, and AM acknowledge support under NSF grant No.~CCF-CISE-ANR-FET-2121957. AM acknowledges support under the NIST Cooperative Research Agreement Award No.~70NANB14H209 through the University of Maryland. PT and UE acknowledge support under the ANR StochNet Project award No.~ANR-21-CE94-0002-01. The authors acknowledge J. Langer and J. Wrona from Singulus Technologies for the MTJ stack deposition and N. Lamard, R. Sousa, L. Prejbeanu, and the Upstream Technological platform PTA, Grenoble, France for the device nanofabrication.

\appendix

\section{Experimental samples}
\label{sec:samples}

% Add discussion of the ~10ns RC timescale found in SPICE simulations
To serve as a basis for our modeling efforts we collected experimental time-voltage traces from an SMTJ. The device's material stack is Si (base) / SiO$_2$ / TaN / [Co(0.5)/Pt(0.2)]$_6$ / Ru(0.8)/ [Co(0.6)/Pt(0.2)]$_3$ / Ta(0.2) / Co(0.9) / W(0.25) / CoFeB(1) / MgO(0.8) / CoFeB(1.4)/ W(0.3) / CoFeB(0.5) / MgO(0.75) / Ta(150) / Ru(8), where the numbers in parentheses refer to layer thicknesses in nanometers and the subscripts on square brackets show bilayer repetitions. The device is roughly circular in cross-section and has a diameter of about 20~nm~\cite{chavent2020multifunctional, ma2021microwave}. The materials in this stack produce a perpendicular anisotropy~\cite{dieny2017perpendicular} that largely cancels the demagnetization field, leaving only a small energy barrier in the final device. The device thereby shows superparamagnetic behavior at a small range of applied currents and applied field. The biasing scheme in this paper is described in Appendix~\ref{app:Elec_Meas_Circuit}. 
An external magnetic field of 124.5~mT was applied at an 18$^\circ$ angle to the device's normal axis. This tilt accelerates the occurrence of stochastic reversals and allows for operation at lower bias voltages and currents.

%Field-free stochastic magnetization fluctuations have been observed in these perpendicular MTJ devices for a large range of diameters, for details see Ref.~\cite{el2022radiofrequency}.

% I just start to draft the DC charachterization part, here (following the discussion with Bill)
\section{Experimental DC characterization}
\label{app:Exp_DC_charac}

We first characterized the DC properties of the SMTJ by performing static DC measurements using a varied  DC applied voltage and magnetic field. This initial measurement provides the operating voltage and field conditions needed to find superparamagnetic regimes in the fabricated devices.

To accurately measure the DC resistance of the fabricated SMTJs, a constant DC voltage $V_\mathrm{0}$ was applied to the SMTJ through a static series resistance $R_\mathrm{0}=3$~k$\Omega$ For real-time monitoring and precise measurement of the device's DC resistance, an oscilloscope with an internal impedance of $Z_\mathrm{0} = 1\;\text{M}\Omega$ was connected in parallel to the SMTJ. The oscilloscope, used as a DC voltmeter, outputs an average voltage that is used to evaluate the DC resistance of the SMTJ.
At the $V_0=505$~mV operating voltage used to measure the time trace analyzed in this paper, the parallel (P) and antiparallel (AP) resistances are 2500~$\Omega$ and 3400~$\Omega$ respectively.

\begin{figure}
    \centering
    \includegraphics[width=\columnwidth]{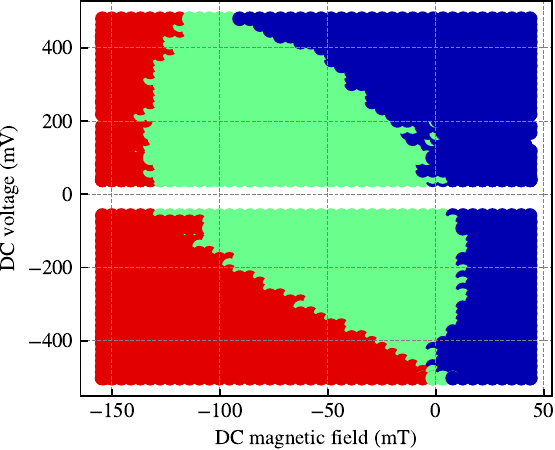}
    \caption{DC Resistance phase diagram of the investigated SMTJ device. The chart summarizes the DC resistance versus field (x-axis) curves obtained at different SMTJ voltages (y-axis). The red (blue) region corresponds to AP (P) configurations. The central green region illustrates a hysteretic bistable configuration where both AP and P states are possible depending on the variation direction of the applied field. A reduction in the dimension of the bistable region at large voltages coincides with the emergence of a superparamagnetic regime in the device.}
    \label{fig:DC_phase_diagram}
\end{figure}

We measured the evolution of the DC resistance in response to a perpendicular magnetic field, which was swept back and forth while maintaining the constant applied voltage during the field sweep. The obtained magnetoresistance loops (resistance versus field curves) are summarized in Fig.~\ref{fig:DC_phase_diagram}, presenting them as a DC resistance phase diagram.

As shown in the diagram, positive (negative) fields tend to stabilize the P (AP) state. Similarly, positive (negative) voltages increase (decrease) the stability range of the P (AP) state. For an intermediate set of fields and voltages, a ``bistable'' region corresponding to configurations where both AP and P states can be stabilized is depicted in green. In a traditional resistance versus field loop, that region can be understood as the opening of the hysteresis. For small voltage values, roughly close to zero, the bistable region occupies a roughly $-150$~mT hysteresis opening, which coincides with the behavior of a thermally stable MTJ, typically suitable for nonvolatile memory applications [e.g. magnetic random access memory (MRAM)].

The size of the bistable region can be reduced by applying positive (negative) voltages, which in our case is interpreted as a stabilization of the P (AP) state through spin-transfer torque. In this line of thinking, the slope of the top-right and bottom-left boundaries of the bistable region can be interpreted as the efficiency of the spin-transfer torque. For large voltages, the boundaries present a curvature (quadratic or higher order contribution in voltage), which could be a partial consequence of Joule heating effects in perpendicular MTJs~\cite{strelkov2018impact}.

For conditions close to the zero field, the bistable region seems to progressively disappear with voltage, at least for negative voltages larger than $-400$~mV, leading to a superparamagnetic regime. In our case, such a set of DC field and voltage conditions is typically used to observe superparamagnetic behavior, which is confirmed by voltage-time trace measurements in real-time presenting random telegraph noise.

\section{High bandwidth measurement circuit}
\label{app:Elec_Meas_Circuit}
\begin{figure}
    \centering
    \includegraphics[width=\columnwidth]{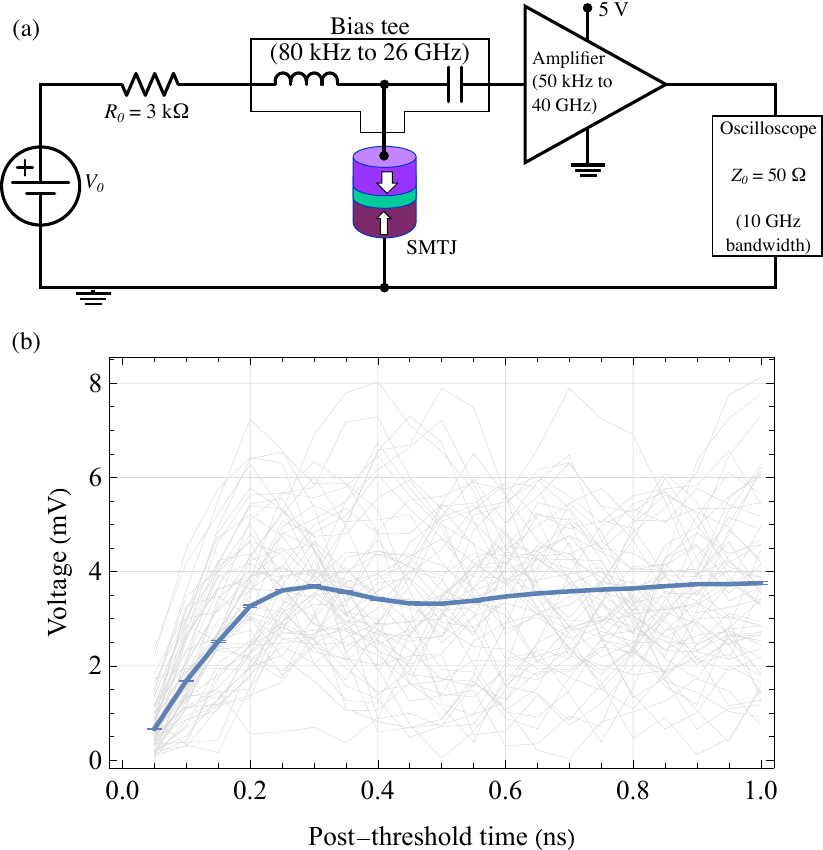}
    \caption{
(a) A schematic diagram of the electrical circuit designed for the detection of resistance fluctuations between parallel (P) and antiparallel (AP) states on sub-microsecond timescales. This setup is specifically tailored for measuring the nanosecond-scale random telegraph noise generated by SMTJs. (b) Schematic illustration of the trajectory-averaging process we use to analyze transition dynamics. The gray lines represent a small subset of (the initial timesteps of) the positive voltage subtrajectories. Error bars on the blue mean curve indicate a standard deviation of uncertainty on the mean.}
    \label{fig:AC_circuit}
\end{figure}

To precisely capture random telegraph noise on the nanosecond scale, we engineered a radio frequency measurement setup optimized for identifying transitions with dwell times below $10$~{\textmu}s as shown in Fig.~\ref{fig:AC_circuit}(a). The setup uses a wide-band bias-tee in order to isolate the high-frequency noise from other electrical components interfering with our signal while allowing the fast, alternating current (AC) fluctuations of our devices to be captured effectively. It is specified to have an operating frequency range from 80~kHz to 26~GHz. The DC-only terminal of the bias-tee is connected to a voltage source through a 3~k$\Omega$ source resistor, while the RF+DC terminal is connected to the SMTJ. The AC-only terminal is connected to a 5~cm long coaxial cable which is terminated into a wide bandwidth amplifier.  The characteristic impedance of the cable and the termination into the amplifier is 50~$\Omega$. The amplifier is specified to have a gain of 4 and an operating frequency range from 50~kHz to 40~GHz, and is powered by a filtered source to minimize the effects of noise. The output of the amplifier is impedance-matched to a 50~$\Omega$ coaxial cable which terminates into the 50~$\Omega$ input of the oscilloscope. The oscilloscope has a frequency range of 10~GHz and a sampling rate of 20 billion samples per second.

The RF path of the measurement setup has a nominal complex impedance of 50~$\Omega$, which helps in achieving impedance matching with the oscilloscope, thus ensuring the efficient transmission of RF signals from the SMTJ. This is effective even when the cumulative impedance of the SMTJ and the static resistor diverge from the 50~$\Omega$ target. Within such a setup, SMTJ resistance changes lead to a fluctuating current, which acts as the primary stimulus for the entire RF circuitry. Nonetheless, this configuration significantly reduces the voltage amplitude of the outgoing RF signal observed on the oscilloscope. 

\subsection{Temporal sensitivity}
\label{app:temporal}

After implementing the measurement circuit outlined above, the effective  $RC$ time constant of the circuit is decreased to below 0.5~ns, as illustrated in Fig.~\ref{fig:AC_circuit}(b). This adjustment facilitates the electrical detection of phenomena within the SMTJs occurring on timescales from submicroseconds to nanoseconds.
Of particular interest is how the electronic setup affects our measurement of the transitions between P and AP states. This is difficult to interrogate for single trajectories due to the intrinsic noise of the system, so we use the following procedure to compute averaged switching events. We consider all positive-voltage (negative-voltage) subtrajectories of our time trace, representing stretches of time that the device spent purely in the AP (P) state. We discard subtrajectories shorter than 20~ns to ensure that we are studying events that reliably transition, sit in a metastable state, and then transition back. This filtering leaves us with 3263 positive-voltage subtrajectories and 5550 negative-voltage subtrajectories.

From each of these subtrajectories, we extract their initial (final) 10 ns stretches and average these across all subtrajectories, creating averaged trajectories into (out of) the two states. The averaging process on the initial-stretch trajectories for the P and AP states is depicted in Fig.~\ref{fig:AC_circuit}(b) to give a sense of the underlying distributions. We repeat this entire averaging procedure for the voltage-time trace generated by our Langevin simulation from Sec.~\ref{sec:results} so that we may compare theory and experiment.

\begin{figure}
    \centering
    \includegraphics[width=\columnwidth]{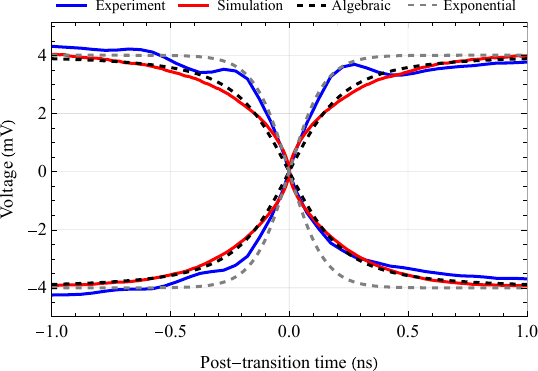}
    \caption{Blue curves show the experimental subtrajectory averages around a transition; red shows the simulation subtrajectory averages. Note that at 1~ns the simulation curve is larger in amplitude than the experimental curve, but by the end of the trajectory (\emph{i.e.} at $-1$~ns) the experimental amplitude exceeds the simulated amplitude. The dashed black line, which approximately captures the simulation behavior, is $V_0 t/\sqrt{t^2+(0.25\;\text{ns})^2}$ for $V_0 = 4$~mV. The early rise times and asympototic behavior of the experiment is captured by the dashed gray line, $V_0\tanh[t/(0.15\;\text{ns})]$.}
    \label{fig:transition-detail}
\end{figure}
We glue the averaged state-entering and state-exiting trajectories together to get a sense of average transition events. We zoom in on the 2~ns window around barrier crossings to examine crossing dynamics in Fig.~\ref{fig:transition-detail}. There are some apparently coherent oscillations in the experimental trajectories before and after barrier crossings. These are not captured by the Langevin model in simulation -- an unsurprising result given that the Langevin model is interia-free by construction. It is not straightforward at this point to tell whether these oscillations represent magnetic physics or artifacts of the experimental setup. We note that the simulation curve was difficult to fit with functions of exponential character (like hyperbolic tangent); here we fit it with a curve of the form $t/\sqrt{\tau^2+t^2}$. We speculate that, since the Langevin model is given by diffusion motion over the barrier, there is no characteristic time scale \emph{per se} available to make a sensible exponential-type fit, but again, more investigation is needed. The timescales reported in Fig.~\ref{fig:transition-detail} are close to the limits of the oscilloscope. Accurately observing even faster dynamics necessitates further enhancements in the electrical circuit design, including the use of oscilloscopes and circuits that support a broader frequency bandwidth.

\begin{figure}
    \centering
    \includegraphics{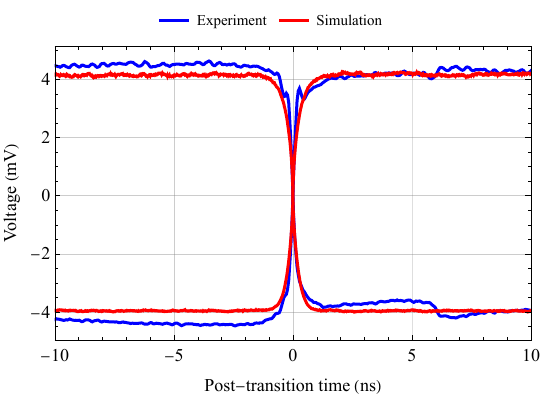}
    \caption{Zoomed-out version of Fig.~\ref{fig:easy-axis-macrospin-short}. A notable artifact in the signal is present at 6~ns post-transition. Note that at long times (\emph{i.e.} the left half of the plot) the simulated signal stays flat at around $\pm4$~mV while the experiment signal rises to almost $\pm4.5$~mV.}
    \label{fig:transition-long-time}
\end{figure}
Finally, in Fig.~\ref{fig:transition-long-time}, we zoom out and show these averaged trajectories for a span of 20~ns. We note that at about 6~ns post-transition in the experimental data, there is a notable voltage shift in both the P and AP wells (though it is more visible in the P well). Since these curves are averaged over thousands of subtrajectories, it seems unlikely that these correspond to device physics. We attribute them to signal reflection in the cable connecting our oscilloscope to the amplifier. 

\subsection{Frequency response}
\label{app:frequency}

\begin{figure}
    \centering
    \includegraphics[width=\columnwidth]{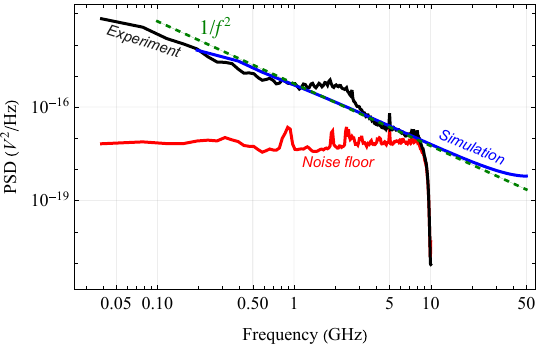}
    \caption{Estimated~power spectral density (PSD) of the experimental time trace compared to the noise floor of the measurement apparatus and the output of the simulated Langevin model described in Sec.~\ref{sec:results}.  The dashed line shows the slope of $1/f^2$ behavior, which would be typical of the spectral density expected from asymmetric random telegraph noise (\textit{i.e.}~a N\'eel-Brown model) beyond the its Lorentzian low-frequency cutoff.}
    \label{fig:psd-comparisons}
\end{figure}

This paper is focused on a temporal analysis of the behavior of the device. Analysis in the frequency domain is complicated by the dependence of the behavior on the state of the SMTJ as it crosses the barrier. Nonetheless, it can be useful to look in the frequency domain in the additive noise limit (where no state dependence of the diffusion processes exists) to understand the effect of measurement noise on the analysis. 
First, we note that measurements of electronic noise in this measurement circuit indicate that it only makes minor contributions to the voltage-space histograms. The electrical noise is characterized by replacing the SMTJ with a 3~k$\Omega$ resistor giving measured noise with a standard deviation of $(256 \pm 1)$~{\textmu}V. Meanwhile, the full-widths at half-maximum of the P and AP wells found in our data are approximately 2~mV and 4~mV, respectively. Since these widths add in quadrature, we expect the effects of electrical noise to contribute to less than 5~\% of the distributional full-widths at half-maximum around each well. In what follows, we use a voltage-time trace collected with this 3~k$\Omega$ resistor to represent the effective noise floor of the measurement apparatus.
%Note that the voltage data was taken with a 50~ps timestep, which is a 20~GHz sampling rate, giving the physical bandwidth of 10 GHz. 

In Fig.~\ref{fig:psd-comparisons}, we note that power spectral density~\footnote{Numerical experimental power spectral densities were estimated using Welch's method~\cite{welch1967use}, with a Hann window.}  of our model simulations are well-matched to the experiment to which it was fit. These spectra are consistent with the expected Lorentzian behavior of an asymmetric random telegraph noise model~\cite{fitzhughStatisticalPropertiesAsymmetric1983}. Fig.~\ref{fig:psd-comparisons} shows the 10~GHz cutoff frequency of the oscilloscope. Note that the slight upward curvature of the simulation curve at high frequency is an expected deviation from the continuous time behavior of any discrete timestep simulation of the Langevin equation (\textit{i.e.} any autoregressive process of order one).

The power spectral densities make two things clear. First, although we do measure slightly above the cutoff frequency of the scope, the loss of total signal power resulting from this choice is minimal. Because of the $1/f^2$ behavior, the spectral contributions here would be negligible compared to where we do have meaningful signal. Estimating from the plot (and assuming a continuation of the Lorentzian tail), the total missing signal power from 10 GHz out to infinity is $\approx3\times10^{-8}~\text{V}^2$. The integrated power of such a Lorentzian integrated from zero frequency up to the cutoff is three orders of magnitude larger, $\approx 2\times 10^{-5}\;\text{V}^2$.

\begin{figure}
    \centering
    \includegraphics[width=\columnwidth]{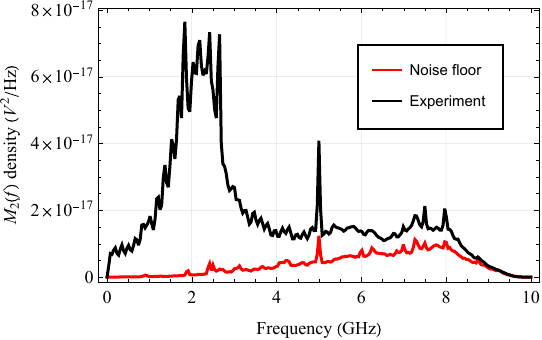}
    \caption{The integrand of Eq.~\eqref{eq:m2-freq} for the experimental data and for the noise analysis data of the measurement setup without the SMTJ. At all frequencies, the experimental curves equals or exceeds the diffusion density found in the noise floor characterization.}
    \label{fig:m2-integrand}
\end{figure}
Of course, the quantity we extract from the data that is most sensitive to this noise power is $M_2$, which is our direct characterization of the noise from experiment. Together with the histograms, $M_2$ is a central quantity on which all our fitting is based, so its fidelity is crucial to our process. One may ask how not only the high-frequency truncation but also the \emph{low-frequency} cutoff from the bias-tee may affect this characterization. As a first approximation, we can consider the case of additive noise (in Appendix~\ref{app:fitting} we show that, though modeling our device with additive noise loses some of the features present in the data, it still does a reasonably good job). In this case, one can expand the signal $\Phi(t)$ in the definition of $M_2$ [Eq.~\eqref{eq:KM-coef-M}] in its Fourier components, and after some simplification one finds that
\begin{equation}
\label{eq:m2-freq}
    M_2 = \frac{2}{\pi}\int_0^{\pi/\tau} S(\omega)[1-\cos(\omega \tau)]\,d\omega,
\end{equation}
where $S(\omega)$ is the power spectral density of the signal measured with timestep $\tau$; note that the upper integration limit is the (angular) Nyquist frequency. This expression of $M_2$ suggests that while the maximal contribution is nominally from the Nyquist frequency itself, there are still strong contributions at half the Nyquist frequency (where $\cos\omega\tau = 0$) and lower. 

The infrared cutoff of about $0.1$~MHz to 1~MHz imposed by the bias tee corresponds to $(1-\cos\omega\tau)$ of about $5\times10^{-11}$ to $5\times 10^{-9}$, respectively. This suppression is sufficiently strong that any signal power lost from the sub-megahertz range would never meaningfully contribute to $M_2$ even if we could capture it.

More problematic, however, is the ultraviolet cutoff. Fig.~\ref{fig:m2-integrand} plots the integrand of Eq.~\eqref{eq:m2-freq} for the experimental data and the noise floor characterization. Because of the convergence of the noise floor with the signal at the cutoff frequency (Fig.~\ref{fig:psd-comparisons}), the contribution of the electrical noise to $M_2$ becomes significant there; all in all, the total $M_2$ from the noise floor derived from integrating Eq.~\eqref{eq:m2-freq} amounts to 20~\% of the same integral for the experiment. Unlike the histograms, these quantities do not add in quadrature. While the majority of our $M_2$ signal is indeed device physics, it is also clear that the effect of the electrical noise is far from negligable.

From a theoretical standpoint, however, accounting for the electrical noise may be straightforward. Assuming the magnetic fluctuations have zero mean and are independent of the electrical noise, Eq.~\eqref{eq:KM-coef-M} factors to give $M_2 = M_2^\text{device} + M_2^\text{electrical}$. The measured $M_2$ for the noise characterization could then be subtracted out directly before beginning the fitting process for the theoretical model. Verifying this model against the experiment would then involve the additional step of injecting modeled electrical noise on top of the signal from the Langevin simulation in order to make a fair comparison. Though this procedure could be used to mitigate the effect of experimental artifacts on the model, we omit such a procedure from the main text in the interest of clarity.

\section{Other parameterizations of $\tilde{D}_2$}
\label{app:fitting}

\begin{figure}
    \centering
    \includegraphics[width=\columnwidth]{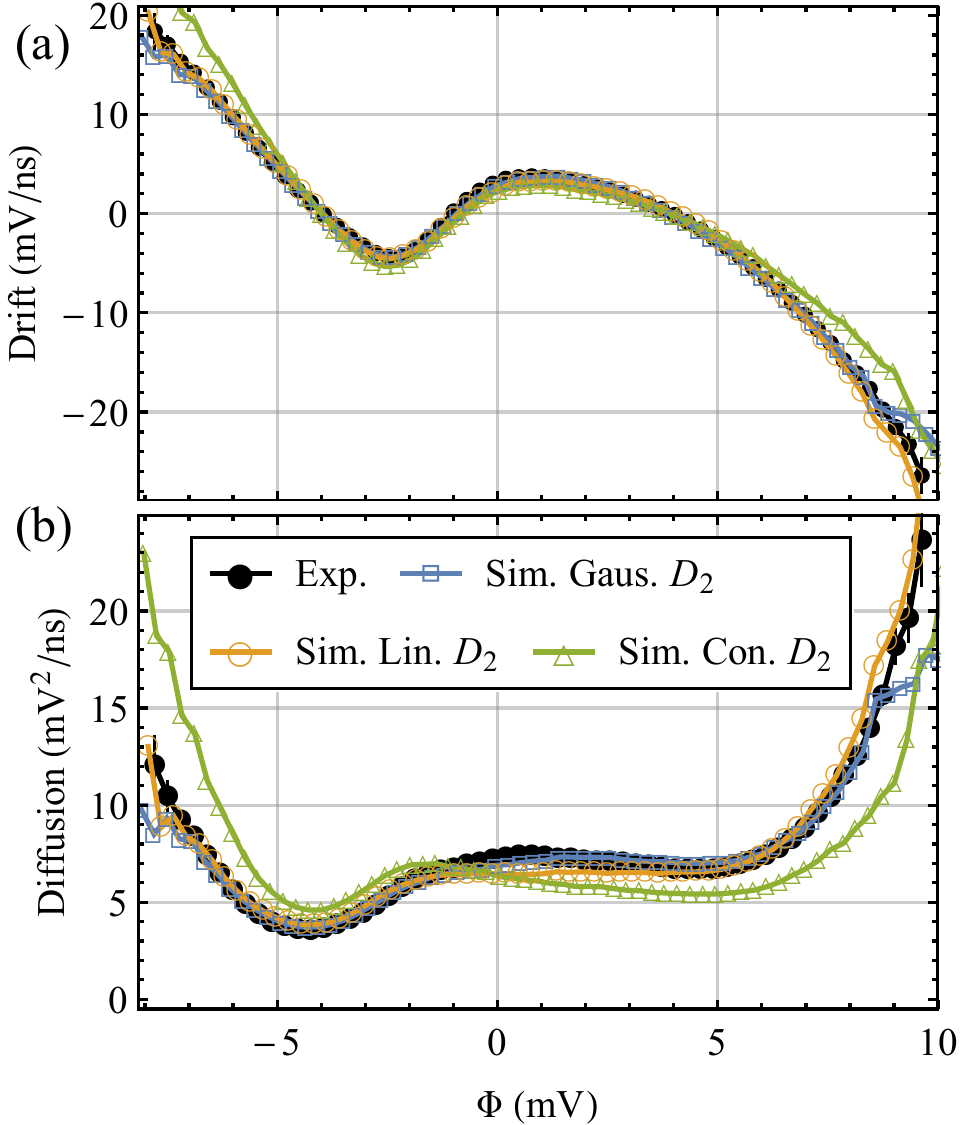}
    \caption{Experimental and Langevin simulation results for the time-delayed, drift panel (a) and diffusion panel (b) coefficients computed at a sampling time $\tau_{\text{samp}} = 50 $ ps equivalent to the experimental measurement time. The various lines in the figures are as follows: black circles show experimental data; blue squares show first-order, data-driven Langevin simulations with a Gaussian plus linear $\tilde{D}_2$; yellow circles show first-order, data-driven Langevin simulations with a linear $\tilde{D}_2$ as reproduced from the main paper body; green triangles show first-order, data-driven Langevin simulations with a constant $\tilde{D}_2$.}
    \label{fig:KMcomparisonAll}
\end{figure}

\begin{figure}
    \centering
    \includegraphics[width=\columnwidth]{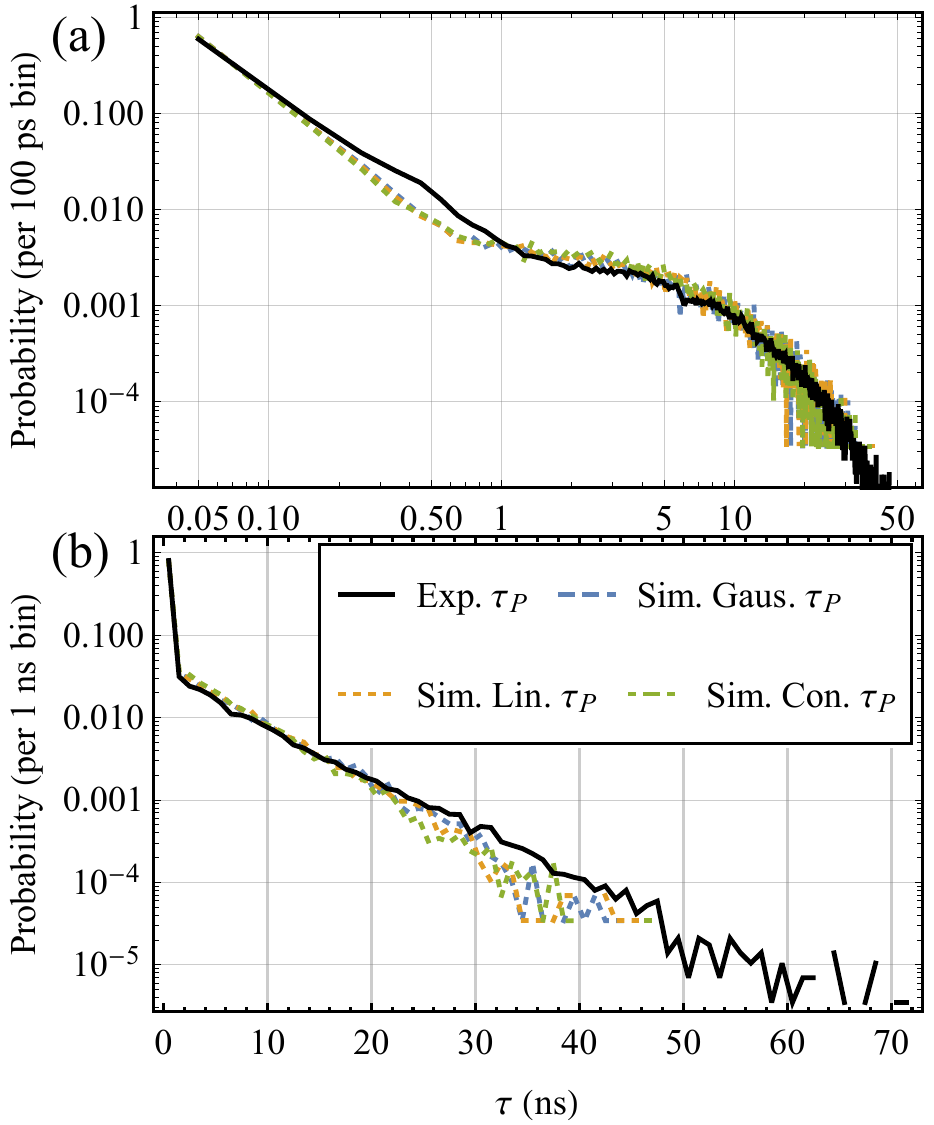}
    \caption{Log-log panel (a) and semi-log panel (b) dwell time distribution for the P state $\tau_{p}$ as calculated in experiment and produced by simulation: black circles show experimental data; the blue, dashed line shows the first-order, data-driven Langevin simulations with a Gaussian plus linear $\tilde{D}_2$; the yellow, dotted line shows the first-order, data-driven Langevin simulations with a linear $\tilde{D}_2$ as reproduced from the main paper body; the green, dot-dashed line shows the first-order, data-driven Langevin simulations with a constant $\tilde{D}_2$. The line-like behavior of the simulation and experiment in the upper plot at short times indicated a power-law dwell time distribution in that regime. In the lower plot, the line-like behavior of the simulation and experiment for longer times indicates an exponential dwell time distribution in that regime. The same sampling time $\tau_{\text{samp}}$ is used as in Fig.~\ref{fig:dwell-times}, but the overall dwell time distribution is only measured over the first 2,000,000 measurement points of the experiment.}
    \label{fig:dwell-timesAll}
\end{figure}

The ansatz chosen in Sec.~\ref{sec:Langevin} only has two fit parameters that can be optimized to fit the experiment's calculated $M_2/2\tau$ Kramers-Moyal coefficient. This section describes the results using a higher and lower number of fitting parameters for the $\tilde{D}_2$ ansatz.

The higher order fit differs from the linear fit by including a Gaussian that seeks to capture the difference between the Langevin model's calculated $M_2/2 \tau$ and the experiment's calculated $M_2/2\tau$ seen in Fig.~\ref{fig:KMcomparison}. The analytic form for this fit $\tilde{D}_2(\Phi;\bm{\mu}) = m \Phi + b + a \, e^{(\Phi-\mu_0)/2\sigma^2}$. Note we ensure that $D_2$ remains positive, as discussed in the main text. The lower order fit assumes a constant $D_2 \rightarrow D_2(\Phi;\bm{\mu}) = b$. 

Figure~\ref{fig:KMcomparisonAll} compares the calculated $M_2/2\tau$ coefficients from experiment and simulations using the Gaussian plus linear fit, the linear fit, and the constant fit for $\tilde{D}_2$. With progressively higher-order fits, the simulation fidelity increases, indicating that the actual noise term in the SMTJ device has a more complicated structure than these simple analytic models. However, model fidelity to the empirical data is remarkable and demonstrates the flexibility and robustness of our data-driven method to capture gross statistics of the STMJ device.

In addition to comparing calculated Kramers-Moyal coefficients, we also compare the dwell time distributions of each model. Figure~\ref{fig:dwell-timesAll} shows the dwell time distributions in the P state for the experiment and simulations measured using $2\times 10^6$ points. This number of points is fewer than used in Fig.~\ref{fig:dwell-times}, which is why there is a difference between the experimental curves in the two figures. This was done to ensure magnitude agreement in the dwell time distribution between the experiment and simulations, which were limited in total length because of computational expense. The dwell time distribution is unchanged for these different assumed fits for $D_2$. 

%The slope of the semi-log plots in panel (b) Fig.~\ref{fig:dwell-timesAll} is shown in Tab.~\ref{tab:dwell-time-all}. As the assumed fit for $\tilde{D}_2$ becomes simpler, the $\tau_{\text{P}}$ exponential distribution slope diverges from the experiment, whereas the slope of the $\tau_{\text{AP}}$ exponential distribution approaches the experimental value. 

\section{Characteristic dwell times}
\label{app:dwell-times}

In the discussion of Sec.~\ref{sec:results}, we described how the short-time power law behavior of the dwell time histograms in both experiment its fitted simulations [Fig.~\ref{fig:dwell-times}] and, later, in simulations of the macrospin model [Fig.~\ref{fig:easy-axis-macrospin-long}] can be attributed to the scale-free, fractal-like nature of random walks around the switching threshold at high sampling frequencies. They are not mere artifacts of the experiment, but additional structure that is usually neglected in dwell time analysis of magnetic devices, and structure that we expect to similarly arise in any magnetic nanodevice.

%An obvious culprit leading to this behavior is the choice to define our P and AP states strictly according to a threshold. Other authors have made other choices. For instance, Ref.~\cite{hayakawa2021nanosecond} separates the P and AP states by a ``barrier'' state that is then discarded f

\begin{figure}
    \centering
    \includegraphics[width=\columnwidth]{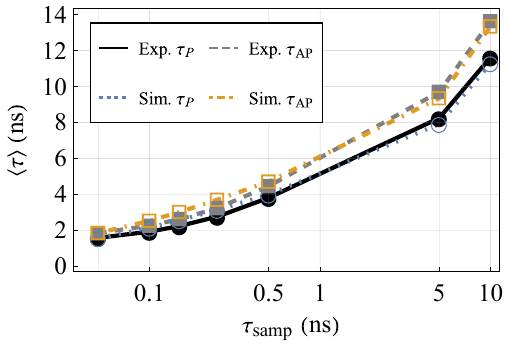}
    \caption{Mean dwell times of experiment and simulation as a function of sampling time.}
    \label{fig:dwell-times-inset}
\end{figure}
Calculating the mean dwell times from the entire distributions gives very small dwell times dominated by these short-time power-law behaviors. Figure~\ref{fig:dwell-times-inset}
shows good agreement between the mean dwell times $\langle \tau \rangle$ extracted from the simulation and from the experiment. Even when subsampled, the simulation and experimental average dwell times agree moderately well. Recall that $D_2$, which is fit to $M_2$ via finite-time corrections, holds all this timescale information in the model. The mean dwell times increases as the subsampling time increases because more and more short-time events are eliminated from the distribution.

Generally speaking, one expects that if a dwell time is relevant for a particular application, it is likely the characteristic dwell time of the exponential distribution that is most meaningful. If we ignore the power-law behavior of the very short dwell times, we can extract these from the slope of the exponential in the semi-log plot, panel (b) in Fig.~\ref{fig:dwell-times} The resulting fits give $\tau_{\text{P,exp}} = 7.10$~ns; $\tau_{\text{AP,exp}} = 11.5$~ns; $\tau_{\text{P,sim}} = 5.95$~ns; $\tau_{\text{AP,sim}} = 9.51$~ns. The one standard deviation ranges for these slopes are $\tau_{\text{P,exp}} \pm \sigma = [6.82,7.39]$~ns;  $\tau_{\text{AP,exp}} \pm \sigma = [11.23,11.85]$~ns; $\tau_{\text{P,sim}} \pm \sigma = [5.79,6.12]$~ns; and $\tau_{\text{AP,sim}} \pm \sigma = [9.33,9.69]$~ns. The agreement is good even though both simulation dwell time slopes slightly underestimate the corresponding experimental values. 

It is important to note that different fitting schemes will give different results that lie outside these ranges, which are strictly determined by the statistics of the points that have been fit. 
Indeed, it is not clear that these extracted times represent anything of experimental relevance.  By a fine-tuned subsampling or $RC$-filtering of the data, we can artificially flatten the power law behavior so that the resulting distribution is exponential. Using this procedure to obtain a plausibly exponential distribution gives $\tau_{P,\text{sim}} \approx 25$~ns and $\tau_{AP,\text{sim}}\approx 30$~ns.  
Going back to Fig.~\ref{fig:experiment-trace} qualitatively illustrates this effect. Keeping all short-time events gives on the order of 35 crossings of the zero line, which would correspond to mean dwell times of approximately 3~ns. On the other hand, if the rapid fluctuations were filtered out, there would be approximately six transitions for a mean dwell time of about 17~ns.

\bibliography{altcomp}

%apsrev4-2.bst 2019-01-14 (MD) hand-edited version of apsrev4-1.bst
%Control: key (0)
%Control: author (8) initials jnrlst
%Control: editor formatted (1) identically to author
%Control: production of article title (0) allowed
%Control: page (0) single
%Control: year (1) truncated
%Control: production of eprint (0) enabled
\begin{thebibliography}{82}%
\makeatletter
\providecommand \@ifxundefined [1]{%
 \@ifx{#1\undefined}
}%
\providecommand \@ifnum [1]{%
 \ifnum #1\expandafter \@firstoftwo
 \else \expandafter \@secondoftwo
 \fi
}%
\providecommand \@ifx [1]{%
 \ifx #1\expandafter \@firstoftwo
 \else \expandafter \@secondoftwo
 \fi
}%
\providecommand \natexlab [1]{#1}%
\providecommand \enquote  [1]{``#1''}%
\providecommand \bibnamefont  [1]{#1}%
\providecommand \bibfnamefont [1]{#1}%
\providecommand \citenamefont [1]{#1}%
\providecommand \href@noop [0]{\@secondoftwo}%
\providecommand \href [0]{\begingroup \@sanitize@url \@href}%
\providecommand \@href[1]{\@@startlink{#1}\@@href}%
\providecommand \@@href[1]{\endgroup#1\@@endlink}%
\providecommand \@sanitize@url [0]{\catcode `\\12\catcode `\$12\catcode `\&12\catcode `\#12\catcode `\^12\catcode `\_12\catcode `\%12\relax}%
\providecommand \@@startlink[1]{}%
\providecommand \@@endlink[0]{}%
\providecommand \url  [0]{\begingroup\@sanitize@url \@url }%
\providecommand \@url [1]{\endgroup\@href {#1}{\urlprefix }}%
\providecommand \urlprefix  [0]{URL }%
\providecommand \Eprint [0]{\href }%
\providecommand \doibase [0]{https://doi.org/}%
\providecommand \selectlanguage [0]{\@gobble}%
\providecommand \bibinfo  [0]{\@secondoftwo}%
\providecommand \bibfield  [0]{\@secondoftwo}%
\providecommand \translation [1]{[#1]}%
\providecommand \BibitemOpen [0]{}%
\providecommand \bibitemStop [0]{}%
\providecommand \bibitemNoStop [0]{.\EOS\space}%
\providecommand \EOS [0]{\spacefactor3000\relax}%
\providecommand \BibitemShut  [1]{\csname bibitem#1\endcsname}%
\let\auto@bib@innerbib\@empty
%</preamble>
\bibitem [{\citenamefont {Grollier}\ \emph {et~al.}(2016)\citenamefont {Grollier}, \citenamefont {Querlioz},\ and\ \citenamefont {Stiles}}]{grollierSpintronicNanodevicesBioinspired2016}%
  \BibitemOpen
  \bibfield  {author} {\bibinfo {author} {\bibfnamefont {J.}~\bibnamefont {Grollier}}, \bibinfo {author} {\bibfnamefont {D.}~\bibnamefont {Querlioz}},\ and\ \bibinfo {author} {\bibfnamefont {M.~D.}\ \bibnamefont {Stiles}},\ }\bibfield  {title} {\bibinfo {title} {Spintronic {{Nanodevices}} for {{Bioinspired Computing}}},\ }\href {https://doi.org/10.1109/JPROC.2016.2597152} {\bibfield  {journal} {\bibinfo  {journal} {Proceedings of the IEEE}\ }\textbf {\bibinfo {volume} {104}},\ \bibinfo {pages} {2024} (\bibinfo {year} {2016})}\BibitemShut {NoStop}%
\bibitem [{\citenamefont {Roy}\ \emph {et~al.}(2018)\citenamefont {Roy}, \citenamefont {Sengupta},\ and\ \citenamefont {Shim}}]{roy2018perspective}%
  \BibitemOpen
  \bibfield  {author} {\bibinfo {author} {\bibfnamefont {K.}~\bibnamefont {Roy}}, \bibinfo {author} {\bibfnamefont {A.}~\bibnamefont {Sengupta}},\ and\ \bibinfo {author} {\bibfnamefont {Y.}~\bibnamefont {Shim}},\ }\bibfield  {title} {\bibinfo {title} {Perspective: Stochastic magnetic devices for cognitive computing},\ }\href@noop {} {\bibfield  {journal} {\bibinfo  {journal} {Journal of Applied Physics}\ }\textbf {\bibinfo {volume} {123}} (\bibinfo {year} {2018})}\BibitemShut {NoStop}%
\bibitem [{\citenamefont {Grollier}\ \emph {et~al.}(2020)\citenamefont {Grollier}, \citenamefont {Querlioz}, \citenamefont {Camsari}, \citenamefont {{Everschor-Sitte}}, \citenamefont {Fukami},\ and\ \citenamefont {Stiles}}]{grollierNeuromorphicSpintronics2020}%
  \BibitemOpen
  \bibfield  {author} {\bibinfo {author} {\bibfnamefont {J.}~\bibnamefont {Grollier}}, \bibinfo {author} {\bibfnamefont {D.}~\bibnamefont {Querlioz}}, \bibinfo {author} {\bibfnamefont {K.~Y.}\ \bibnamefont {Camsari}}, \bibinfo {author} {\bibfnamefont {K.}~\bibnamefont {{Everschor-Sitte}}}, \bibinfo {author} {\bibfnamefont {S.}~\bibnamefont {Fukami}},\ and\ \bibinfo {author} {\bibfnamefont {M.~D.}\ \bibnamefont {Stiles}},\ }\bibfield  {title} {\bibinfo {title} {Neuromorphic spintronics},\ }\href {https://doi.org/10.1038/s41928-019-0360-9} {\bibfield  {journal} {\bibinfo  {journal} {Nature Electronics}\ }\textbf {\bibinfo {volume} {3}},\ \bibinfo {pages} {360} (\bibinfo {year} {2020})}\BibitemShut {NoStop}%
\bibitem [{\citenamefont {Blachowicz}\ and\ \citenamefont {Ehrmann}(2020)}]{blachowicz2020magnetic}%
  \BibitemOpen
  \bibfield  {author} {\bibinfo {author} {\bibfnamefont {T.}~\bibnamefont {Blachowicz}}\ and\ \bibinfo {author} {\bibfnamefont {A.}~\bibnamefont {Ehrmann}},\ }\bibfield  {title} {\bibinfo {title} {Magnetic elements for neuromorphic computing},\ }\href@noop {} {\bibfield  {journal} {\bibinfo  {journal} {Molecules}\ }\textbf {\bibinfo {volume} {25}},\ \bibinfo {pages} {2550} (\bibinfo {year} {2020})}\BibitemShut {NoStop}%
\bibitem [{\citenamefont {Zink}\ \emph {et~al.}(2022)\citenamefont {Zink}, \citenamefont {Lv},\ and\ \citenamefont {Wang}}]{zink2022review}%
  \BibitemOpen
  \bibfield  {author} {\bibinfo {author} {\bibfnamefont {B.~R.}\ \bibnamefont {Zink}}, \bibinfo {author} {\bibfnamefont {Y.}~\bibnamefont {Lv}},\ and\ \bibinfo {author} {\bibfnamefont {J.-P.}\ \bibnamefont {Wang}},\ }\bibfield  {title} {\bibinfo {title} {Review of magnetic tunnel junctions for stochastic computing},\ }\href@noop {} {\bibfield  {journal} {\bibinfo  {journal} {IEEE Journal on Exploratory Solid-State Computational Devices and Circuits}\ }\textbf {\bibinfo {volume} {8}},\ \bibinfo {pages} {173} (\bibinfo {year} {2022})}\BibitemShut {NoStop}%
\bibitem [{\citenamefont {Fukushima}\ \emph {et~al.}(2014)\citenamefont {Fukushima}, \citenamefont {Seki}, \citenamefont {Yakushiji}, \citenamefont {Kubota}, \citenamefont {Imamura}, \citenamefont {Yuasa},\ and\ \citenamefont {Ando}}]{fukushima2014spin}%
  \BibitemOpen
  \bibfield  {author} {\bibinfo {author} {\bibfnamefont {A.}~\bibnamefont {Fukushima}}, \bibinfo {author} {\bibfnamefont {T.}~\bibnamefont {Seki}}, \bibinfo {author} {\bibfnamefont {K.}~\bibnamefont {Yakushiji}}, \bibinfo {author} {\bibfnamefont {H.}~\bibnamefont {Kubota}}, \bibinfo {author} {\bibfnamefont {H.}~\bibnamefont {Imamura}}, \bibinfo {author} {\bibfnamefont {S.}~\bibnamefont {Yuasa}},\ and\ \bibinfo {author} {\bibfnamefont {K.}~\bibnamefont {Ando}},\ }\bibfield  {title} {\bibinfo {title} {Spin dice: A scalable truly random number generator based on spintronics},\ }\href@noop {} {\bibfield  {journal} {\bibinfo  {journal} {Applied Physics Express}\ }\textbf {\bibinfo {volume} {7}},\ \bibinfo {pages} {083001} (\bibinfo {year} {2014})}\BibitemShut {NoStop}%
\bibitem [{\citenamefont {Vodenicarevic}\ \emph {et~al.}(2017)\citenamefont {Vodenicarevic}, \citenamefont {Locatelli}, \citenamefont {Mizrahi}, \citenamefont {Friedman}, \citenamefont {Vincent}, \citenamefont {Romera}, \citenamefont {Fukushima}, \citenamefont {Yakushiji}, \citenamefont {Kubota}, \citenamefont {Yuasa}, \citenamefont {Tiwari}, \citenamefont {Grollier},\ and\ \citenamefont {Querlioz}}]{vodenicarevicLowEnergyTrulyRandom2017}%
  \BibitemOpen
  \bibfield  {author} {\bibinfo {author} {\bibfnamefont {D.}~\bibnamefont {Vodenicarevic}}, \bibinfo {author} {\bibfnamefont {N.}~\bibnamefont {Locatelli}}, \bibinfo {author} {\bibfnamefont {A.}~\bibnamefont {Mizrahi}}, \bibinfo {author} {\bibfnamefont {J.~S.}\ \bibnamefont {Friedman}}, \bibinfo {author} {\bibfnamefont {A.~F.}\ \bibnamefont {Vincent}}, \bibinfo {author} {\bibfnamefont {M.}~\bibnamefont {Romera}}, \bibinfo {author} {\bibfnamefont {A.}~\bibnamefont {Fukushima}}, \bibinfo {author} {\bibfnamefont {K.}~\bibnamefont {Yakushiji}}, \bibinfo {author} {\bibfnamefont {H.}~\bibnamefont {Kubota}}, \bibinfo {author} {\bibfnamefont {S.}~\bibnamefont {Yuasa}}, \bibinfo {author} {\bibfnamefont {S.}~\bibnamefont {Tiwari}}, \bibinfo {author} {\bibfnamefont {J.}~\bibnamefont {Grollier}},\ and\ \bibinfo {author} {\bibfnamefont {D.}~\bibnamefont {Querlioz}},\ }\bibfield  {title} {\bibinfo {title} {Low-{{Energy Truly Random Number Generation}} with {{Superparamagnetic Tunnel Junctions}} for {{Unconventional
  Computing}}},\ }\href {https://doi.org/10.1103/PhysRevApplied.8.054045} {\bibfield  {journal} {\bibinfo  {journal} {Physical Review Applied}\ }\textbf {\bibinfo {volume} {8}},\ \bibinfo {pages} {054045} (\bibinfo {year} {2017})}\BibitemShut {NoStop}%
\bibitem [{\citenamefont {Parks}\ \emph {et~al.}(2018)\citenamefont {Parks}, \citenamefont {Bapna}, \citenamefont {Igbokwe}, \citenamefont {Almasi}, \citenamefont {Wang},\ and\ \citenamefont {Majetich}}]{parks2018superparamagnetic}%
  \BibitemOpen
  \bibfield  {author} {\bibinfo {author} {\bibfnamefont {B.}~\bibnamefont {Parks}}, \bibinfo {author} {\bibfnamefont {M.}~\bibnamefont {Bapna}}, \bibinfo {author} {\bibfnamefont {J.}~\bibnamefont {Igbokwe}}, \bibinfo {author} {\bibfnamefont {H.}~\bibnamefont {Almasi}}, \bibinfo {author} {\bibfnamefont {W.}~\bibnamefont {Wang}},\ and\ \bibinfo {author} {\bibfnamefont {S.~A.}\ \bibnamefont {Majetich}},\ }\bibfield  {title} {\bibinfo {title} {Superparamagnetic perpendicular magnetic tunnel junctions for true random number generators},\ }\href@noop {} {\bibfield  {journal} {\bibinfo  {journal} {AIP Advances}\ }\textbf {\bibinfo {volume} {8}} (\bibinfo {year} {2018})}\BibitemShut {NoStop}%
\bibitem [{\citenamefont {Lv}\ \emph {et~al.}(2022)\citenamefont {Lv}, \citenamefont {Zink},\ and\ \citenamefont {Wang}}]{lv2022bipolar}%
  \BibitemOpen
  \bibfield  {author} {\bibinfo {author} {\bibfnamefont {Y.}~\bibnamefont {Lv}}, \bibinfo {author} {\bibfnamefont {B.~R.}\ \bibnamefont {Zink}},\ and\ \bibinfo {author} {\bibfnamefont {J.-P.}\ \bibnamefont {Wang}},\ }\bibfield  {title} {\bibinfo {title} {Bipolar random spike and bipolar random number generation by two magnetic tunnel junctions},\ }\href@noop {} {\bibfield  {journal} {\bibinfo  {journal} {IEEE Transactions on Electron Devices}\ }\textbf {\bibinfo {volume} {69}},\ \bibinfo {pages} {1582} (\bibinfo {year} {2022})}\BibitemShut {NoStop}%
\bibitem [{\citenamefont {Daniels}\ \emph {et~al.}(2020)\citenamefont {Daniels}, \citenamefont {Madhavan}, \citenamefont {Talatchian}, \citenamefont {Mizrahi},\ and\ \citenamefont {Stiles}}]{danielsEnergyEfficientStochasticComputing2020}%
  \BibitemOpen
  \bibfield  {author} {\bibinfo {author} {\bibfnamefont {M.~W.}\ \bibnamefont {Daniels}}, \bibinfo {author} {\bibfnamefont {A.}~\bibnamefont {Madhavan}}, \bibinfo {author} {\bibfnamefont {P.}~\bibnamefont {Talatchian}}, \bibinfo {author} {\bibfnamefont {A.}~\bibnamefont {Mizrahi}},\ and\ \bibinfo {author} {\bibfnamefont {M.~D.}\ \bibnamefont {Stiles}},\ }\bibfield  {title} {\bibinfo {title} {Energy-{{Efficient Stochastic Computing}} with {{Superparamagnetic Tunnel Junctions}}},\ }\href {https://doi.org/10.1103/PhysRevApplied.13.034016} {\bibfield  {journal} {\bibinfo  {journal} {Physical Review Applied}\ }\textbf {\bibinfo {volume} {13}},\ \bibinfo {pages} {034016} (\bibinfo {year} {2020})}\BibitemShut {NoStop}%
\bibitem [{\citenamefont {Zink}\ \emph {et~al.}(2023)\citenamefont {Zink}, \citenamefont {Lv}, \citenamefont {Zabihi}, \citenamefont {Cilasun}, \citenamefont {Sapatnekar}, \citenamefont {Karpuzcu}, \citenamefont {Riedel},\ and\ \citenamefont {Wang}}]{zink2023stochastic}%
  \BibitemOpen
  \bibfield  {author} {\bibinfo {author} {\bibfnamefont {B.~R.}\ \bibnamefont {Zink}}, \bibinfo {author} {\bibfnamefont {Y.}~\bibnamefont {Lv}}, \bibinfo {author} {\bibfnamefont {M.}~\bibnamefont {Zabihi}}, \bibinfo {author} {\bibfnamefont {H.}~\bibnamefont {Cilasun}}, \bibinfo {author} {\bibfnamefont {S.~S.}\ \bibnamefont {Sapatnekar}}, \bibinfo {author} {\bibfnamefont {U.~R.}\ \bibnamefont {Karpuzcu}}, \bibinfo {author} {\bibfnamefont {M.~D.}\ \bibnamefont {Riedel}},\ and\ \bibinfo {author} {\bibfnamefont {J.-P.}\ \bibnamefont {Wang}},\ }\bibfield  {title} {\bibinfo {title} {A stochastic computing scheme of embedding random bit generation and processing in computational random access memory (sc-cram)},\ }\href@noop {} {\bibfield  {journal} {\bibinfo  {journal} {IEEE Journal on Exploratory Solid-State Computational Devices and Circuits}\ } (\bibinfo {year} {2023})}\BibitemShut {NoStop}%
\bibitem [{\citenamefont {Cai}\ \emph {et~al.}(2023)\citenamefont {Cai}, \citenamefont {He}, \citenamefont {Xin}, \citenamefont {Yuan}, \citenamefont {Zhang}, \citenamefont {Zhu},\ and\ \citenamefont {Liang}}]{cai2023unconventional}%
  \BibitemOpen
  \bibfield  {author} {\bibinfo {author} {\bibfnamefont {B.}~\bibnamefont {Cai}}, \bibinfo {author} {\bibfnamefont {Y.}~\bibnamefont {He}}, \bibinfo {author} {\bibfnamefont {Y.}~\bibnamefont {Xin}}, \bibinfo {author} {\bibfnamefont {Z.}~\bibnamefont {Yuan}}, \bibinfo {author} {\bibfnamefont {X.}~\bibnamefont {Zhang}}, \bibinfo {author} {\bibfnamefont {Z.}~\bibnamefont {Zhu}},\ and\ \bibinfo {author} {\bibfnamefont {G.}~\bibnamefont {Liang}},\ }\bibfield  {title} {\bibinfo {title} {Unconventional computing based on magnetic tunnel junction},\ }\href@noop {} {\bibfield  {journal} {\bibinfo  {journal} {Applied Physics A}\ }\textbf {\bibinfo {volume} {129}},\ \bibinfo {pages} {236} (\bibinfo {year} {2023})}\BibitemShut {NoStop}%
\bibitem [{\citenamefont {Mizrahi}\ \emph {et~al.}(2018{\natexlab{a}})\citenamefont {Mizrahi}, \citenamefont {Hirtzlin}, \citenamefont {Fukushima}, \citenamefont {Kubota}, \citenamefont {Yuasa}, \citenamefont {Grollier},\ and\ \citenamefont {Querlioz}}]{mizrahiNeurallikeComputingPopulations2018}%
  \BibitemOpen
  \bibfield  {author} {\bibinfo {author} {\bibfnamefont {A.}~\bibnamefont {Mizrahi}}, \bibinfo {author} {\bibfnamefont {T.}~\bibnamefont {Hirtzlin}}, \bibinfo {author} {\bibfnamefont {A.}~\bibnamefont {Fukushima}}, \bibinfo {author} {\bibfnamefont {H.}~\bibnamefont {Kubota}}, \bibinfo {author} {\bibfnamefont {S.}~\bibnamefont {Yuasa}}, \bibinfo {author} {\bibfnamefont {J.}~\bibnamefont {Grollier}},\ and\ \bibinfo {author} {\bibfnamefont {D.}~\bibnamefont {Querlioz}},\ }\bibfield  {title} {\bibinfo {title} {Neural-like computing with populations of superparamagnetic basis functions},\ }\href {https://doi.org/10.1038/s41467-018-03963-w} {\bibfield  {journal} {\bibinfo  {journal} {Nature Communications}\ }\textbf {\bibinfo {volume} {9}},\ \bibinfo {pages} {1533} (\bibinfo {year} {2018}{\natexlab{a}})}\BibitemShut {NoStop}%
\bibitem [{\citenamefont {Mizrahi}\ \emph {et~al.}(2018{\natexlab{b}})\citenamefont {Mizrahi}, \citenamefont {Grollier}, \citenamefont {Querlioz},\ and\ \citenamefont {Stiles}}]{mizrahiOvercomingDeviceUnreliability2018}%
  \BibitemOpen
  \bibfield  {author} {\bibinfo {author} {\bibfnamefont {A.}~\bibnamefont {Mizrahi}}, \bibinfo {author} {\bibfnamefont {J.}~\bibnamefont {Grollier}}, \bibinfo {author} {\bibfnamefont {D.}~\bibnamefont {Querlioz}},\ and\ \bibinfo {author} {\bibfnamefont {M.~D.}\ \bibnamefont {Stiles}},\ }\bibfield  {title} {\bibinfo {title} {Overcoming device unreliability with continuous learning in a population coding based computing system},\ }\href {https://doi.org/10.1063/1.5042250} {\bibfield  {journal} {\bibinfo  {journal} {Journal of Applied Physics}\ }\textbf {\bibinfo {volume} {124}},\ \bibinfo {pages} {152111} (\bibinfo {year} {2018}{\natexlab{b}})}\BibitemShut {NoStop}%
\bibitem [{\citenamefont {Rajasekharan}\ \emph {et~al.}(2021)\citenamefont {Rajasekharan}, \citenamefont {Rangarajan}, \citenamefont {Patnaik}, \citenamefont {Sinanoglu},\ and\ \citenamefont {Chauhan}}]{rajasekharan2021scanet}%
  \BibitemOpen
  \bibfield  {author} {\bibinfo {author} {\bibfnamefont {D.}~\bibnamefont {Rajasekharan}}, \bibinfo {author} {\bibfnamefont {N.}~\bibnamefont {Rangarajan}}, \bibinfo {author} {\bibfnamefont {S.}~\bibnamefont {Patnaik}}, \bibinfo {author} {\bibfnamefont {O.}~\bibnamefont {Sinanoglu}},\ and\ \bibinfo {author} {\bibfnamefont {Y.~S.}\ \bibnamefont {Chauhan}},\ }\bibfield  {title} {\bibinfo {title} {Scanet: Securing the weights with superparamagnetic-mtj crossbar array networks},\ }\href@noop {} {\bibfield  {journal} {\bibinfo  {journal} {IEEE transactions on neural networks and learning systems}\ } (\bibinfo {year} {2021})}\BibitemShut {NoStop}%
\bibitem [{\citenamefont {Liyanagedera}\ \emph {et~al.}(2017)\citenamefont {Liyanagedera}, \citenamefont {Sengupta}, \citenamefont {Jaiswal},\ and\ \citenamefont {Roy}}]{PhysRevApplied.8.064017}%
  \BibitemOpen
  \bibfield  {author} {\bibinfo {author} {\bibfnamefont {C.~M.}\ \bibnamefont {Liyanagedera}}, \bibinfo {author} {\bibfnamefont {A.}~\bibnamefont {Sengupta}}, \bibinfo {author} {\bibfnamefont {A.}~\bibnamefont {Jaiswal}},\ and\ \bibinfo {author} {\bibfnamefont {K.}~\bibnamefont {Roy}},\ }\bibfield  {title} {\bibinfo {title} {Stochastic spiking neural networks enabled by magnetic tunnel junctions: From nontelegraphic to telegraphic switching regimes},\ }\href {https://doi.org/10.1103/PhysRevApplied.8.064017} {\bibfield  {journal} {\bibinfo  {journal} {Phys. Rev. Appl.}\ }\textbf {\bibinfo {volume} {8}},\ \bibinfo {pages} {064017} (\bibinfo {year} {2017})}\BibitemShut {NoStop}%
\bibitem [{\citenamefont {Cai}\ \emph {et~al.}(2019)\citenamefont {Cai}, \citenamefont {Fang}, \citenamefont {Zhang}, \citenamefont {Lv}, \citenamefont {Zhang}, \citenamefont {Zhou}, \citenamefont {Finocchio},\ and\ \citenamefont {Zeng}}]{PhysRevApplied.11.034015}%
  \BibitemOpen
  \bibfield  {author} {\bibinfo {author} {\bibfnamefont {J.}~\bibnamefont {Cai}}, \bibinfo {author} {\bibfnamefont {B.}~\bibnamefont {Fang}}, \bibinfo {author} {\bibfnamefont {L.}~\bibnamefont {Zhang}}, \bibinfo {author} {\bibfnamefont {W.}~\bibnamefont {Lv}}, \bibinfo {author} {\bibfnamefont {B.}~\bibnamefont {Zhang}}, \bibinfo {author} {\bibfnamefont {T.}~\bibnamefont {Zhou}}, \bibinfo {author} {\bibfnamefont {G.}~\bibnamefont {Finocchio}},\ and\ \bibinfo {author} {\bibfnamefont {Z.}~\bibnamefont {Zeng}},\ }\bibfield  {title} {\bibinfo {title} {Voltage-controlled spintronic stochastic neuron based on a magnetic tunnel junction},\ }\href {https://doi.org/10.1103/PhysRevApplied.11.034015} {\bibfield  {journal} {\bibinfo  {journal} {Phys. Rev. Appl.}\ }\textbf {\bibinfo {volume} {11}},\ \bibinfo {pages} {034015} (\bibinfo {year} {2019})}\BibitemShut {NoStop}%
\bibitem [{\citenamefont {Camsari}\ \emph {et~al.}(2017{\natexlab{a}})\citenamefont {Camsari}, \citenamefont {Salahuddin},\ and\ \citenamefont {Datta}}]{camsari2017implementing}%
  \BibitemOpen
  \bibfield  {author} {\bibinfo {author} {\bibfnamefont {K.~Y.}\ \bibnamefont {Camsari}}, \bibinfo {author} {\bibfnamefont {S.}~\bibnamefont {Salahuddin}},\ and\ \bibinfo {author} {\bibfnamefont {S.}~\bibnamefont {Datta}},\ }\bibfield  {title} {\bibinfo {title} {Implementing p-bits with embedded mtj},\ }\href@noop {} {\bibfield  {journal} {\bibinfo  {journal} {IEEE Electron Device Letters}\ }\textbf {\bibinfo {volume} {38}},\ \bibinfo {pages} {1767} (\bibinfo {year} {2017}{\natexlab{a}})}\BibitemShut {NoStop}%
\bibitem [{\citenamefont {Borders}\ \emph {et~al.}(2019)\citenamefont {Borders}, \citenamefont {Pervaiz}, \citenamefont {Fukami}, \citenamefont {Camsari}, \citenamefont {Ohno},\ and\ \citenamefont {Datta}}]{bordersIntegerFactorizationUsing2019}%
  \BibitemOpen
  \bibfield  {author} {\bibinfo {author} {\bibfnamefont {W.~A.}\ \bibnamefont {Borders}}, \bibinfo {author} {\bibfnamefont {A.~Z.}\ \bibnamefont {Pervaiz}}, \bibinfo {author} {\bibfnamefont {S.}~\bibnamefont {Fukami}}, \bibinfo {author} {\bibfnamefont {K.~Y.}\ \bibnamefont {Camsari}}, \bibinfo {author} {\bibfnamefont {H.}~\bibnamefont {Ohno}},\ and\ \bibinfo {author} {\bibfnamefont {S.}~\bibnamefont {Datta}},\ }\bibfield  {title} {\bibinfo {title} {Integer factorization using stochastic magnetic tunnel junctions},\ }\href {https://doi.org/10.1038/s41586-019-1557-9} {\bibfield  {journal} {\bibinfo  {journal} {Nature}\ }\textbf {\bibinfo {volume} {573}},\ \bibinfo {pages} {390} (\bibinfo {year} {2019})}\BibitemShut {NoStop}%
\bibitem [{\citenamefont {Lv}\ \emph {et~al.}(2019)\citenamefont {Lv}, \citenamefont {Bloom},\ and\ \citenamefont {Wang}}]{lv2019experimental}%
  \BibitemOpen
  \bibfield  {author} {\bibinfo {author} {\bibfnamefont {Y.}~\bibnamefont {Lv}}, \bibinfo {author} {\bibfnamefont {R.~P.}\ \bibnamefont {Bloom}},\ and\ \bibinfo {author} {\bibfnamefont {J.-P.}\ \bibnamefont {Wang}},\ }\bibfield  {title} {\bibinfo {title} {Experimental demonstration of probabilistic spin logic by magnetic tunnel junctions},\ }\href@noop {} {\bibfield  {journal} {\bibinfo  {journal} {IEEE Magnetics Letters}\ }\textbf {\bibinfo {volume} {10}},\ \bibinfo {pages} {1} (\bibinfo {year} {2019})}\BibitemShut {NoStop}%
\bibitem [{\citenamefont {Gibeault}\ \emph {et~al.}(2024)\citenamefont {Gibeault}, \citenamefont {Adeyeye}, \citenamefont {Pocher}, \citenamefont {Lathrop}, \citenamefont {Daniels}, \citenamefont {Stiles}, \citenamefont {McClelland}, \citenamefont {Borders}, \citenamefont {Ryan}, \citenamefont {Talatchian} \emph {et~al.}}]{gibeault2023programmable}%
  \BibitemOpen
  \bibfield  {author} {\bibinfo {author} {\bibfnamefont {S.}~\bibnamefont {Gibeault}}, \bibinfo {author} {\bibfnamefont {T.~N.}\ \bibnamefont {Adeyeye}}, \bibinfo {author} {\bibfnamefont {L.~A.}\ \bibnamefont {Pocher}}, \bibinfo {author} {\bibfnamefont {D.~P.}\ \bibnamefont {Lathrop}}, \bibinfo {author} {\bibfnamefont {M.~W.}\ \bibnamefont {Daniels}}, \bibinfo {author} {\bibfnamefont {M.~D.}\ \bibnamefont {Stiles}}, \bibinfo {author} {\bibfnamefont {J.~J.}\ \bibnamefont {McClelland}}, \bibinfo {author} {\bibfnamefont {W.~A.}\ \bibnamefont {Borders}}, \bibinfo {author} {\bibfnamefont {J.~T.}\ \bibnamefont {Ryan}}, \bibinfo {author} {\bibfnamefont {P.}~\bibnamefont {Talatchian}}, \emph {et~al.},\ }\bibfield  {title} {\bibinfo {title} {Programmable electrical coupling between stochastic magnetic tunnel junctions},\ }\href@noop {} {\bibfield  {journal} {\bibinfo  {journal} {Physical Review Applied}\ }\textbf {\bibinfo {volume} {21}},\ \bibinfo {pages} {034064} (\bibinfo {year} {2024})}\BibitemShut {NoStop}%
\bibitem [{\citenamefont {Brown}(1963)}]{brownThermalFluctuationsSingleDomain1963}%
  \BibitemOpen
  \bibfield  {author} {\bibinfo {author} {\bibfnamefont {W.~F.}\ \bibnamefont {Brown}},\ }\bibfield  {title} {\bibinfo {title} {Thermal {{Fluctuations}} of a {{Single-Domain Particle}}},\ }\href {https://doi.org/10.1103/PhysRev.130.1677} {\bibfield  {journal} {\bibinfo  {journal} {Physical Review}\ }\textbf {\bibinfo {volume} {130}},\ \bibinfo {pages} {1677} (\bibinfo {year} {1963})}\BibitemShut {NoStop}%
\bibitem [{\citenamefont {Mizrahi}\ \emph {et~al.}(2015)\citenamefont {Mizrahi}, \citenamefont {Locatelli}, \citenamefont {Matsumoto}, \citenamefont {Fukushima}, \citenamefont {Kubota}, \citenamefont {Yuasa}, \citenamefont {Cros}, \citenamefont {Kim}, \citenamefont {Grollier},\ and\ \citenamefont {Querlioz}}]{mizrahi2015magnetic}%
  \BibitemOpen
  \bibfield  {author} {\bibinfo {author} {\bibfnamefont {A.}~\bibnamefont {Mizrahi}}, \bibinfo {author} {\bibfnamefont {N.}~\bibnamefont {Locatelli}}, \bibinfo {author} {\bibfnamefont {R.}~\bibnamefont {Matsumoto}}, \bibinfo {author} {\bibfnamefont {A.}~\bibnamefont {Fukushima}}, \bibinfo {author} {\bibfnamefont {H.}~\bibnamefont {Kubota}}, \bibinfo {author} {\bibfnamefont {S.}~\bibnamefont {Yuasa}}, \bibinfo {author} {\bibfnamefont {V.}~\bibnamefont {Cros}}, \bibinfo {author} {\bibfnamefont {J.-V.}\ \bibnamefont {Kim}}, \bibinfo {author} {\bibfnamefont {J.}~\bibnamefont {Grollier}},\ and\ \bibinfo {author} {\bibfnamefont {D.}~\bibnamefont {Querlioz}},\ }\bibfield  {title} {\bibinfo {title} {Magnetic stochastic oscillators: Noise-induced synchronization to underthreshold excitation and comprehensive compact model},\ }\href@noop {} {\bibfield  {journal} {\bibinfo  {journal} {IEEE Transactions on Magnetics}\ }\textbf {\bibinfo {volume} {51}},\ \bibinfo {pages} {1} (\bibinfo {year} {2015})}\BibitemShut {NoStop}%
\bibitem [{\citenamefont {Becle}\ \emph {et~al.}(2021)\citenamefont {Becle}, \citenamefont {Talatchian}, \citenamefont {Prenat}, \citenamefont {Anghel},\ and\ \citenamefont {Prejbeanu}}]{becle2021fast}%
  \BibitemOpen
  \bibfield  {author} {\bibinfo {author} {\bibfnamefont {E.}~\bibnamefont {Becle}}, \bibinfo {author} {\bibfnamefont {P.}~\bibnamefont {Talatchian}}, \bibinfo {author} {\bibfnamefont {G.}~\bibnamefont {Prenat}}, \bibinfo {author} {\bibfnamefont {L.}~\bibnamefont {Anghel}},\ and\ \bibinfo {author} {\bibfnamefont {I.-L.}\ \bibnamefont {Prejbeanu}},\ }\bibfield  {title} {\bibinfo {title} {Fast behavioral veriloga compact model for stochastic mtj},\ }in\ \href@noop {} {\emph {\bibinfo {booktitle} {ESSDERC 2021-IEEE 51st European Solid-State Device Research Conference (ESSDERC)}}}\ (\bibinfo {organization} {IEEE},\ \bibinfo {year} {2021})\ pp.\ \bibinfo {pages} {259--262}\BibitemShut {NoStop}%
\bibitem [{\citenamefont {Talatchian}\ \emph {et~al.}(2021)\citenamefont {Talatchian}, \citenamefont {Daniels}, \citenamefont {Madhavan}, \citenamefont {Pufall}, \citenamefont {Ju{\'e}}, \citenamefont {Rippard}, \citenamefont {McClelland},\ and\ \citenamefont {Stiles}}]{talatchianMutualControlStochastic2021}%
  \BibitemOpen
  \bibfield  {author} {\bibinfo {author} {\bibfnamefont {P.}~\bibnamefont {Talatchian}}, \bibinfo {author} {\bibfnamefont {M.~W.}\ \bibnamefont {Daniels}}, \bibinfo {author} {\bibfnamefont {A.}~\bibnamefont {Madhavan}}, \bibinfo {author} {\bibfnamefont {M.~R.}\ \bibnamefont {Pufall}}, \bibinfo {author} {\bibfnamefont {E.}~\bibnamefont {Ju{\'e}}}, \bibinfo {author} {\bibfnamefont {W.~H.}\ \bibnamefont {Rippard}}, \bibinfo {author} {\bibfnamefont {J.~J.}\ \bibnamefont {McClelland}},\ and\ \bibinfo {author} {\bibfnamefont {M.~D.}\ \bibnamefont {Stiles}},\ }\bibfield  {title} {\bibinfo {title} {Mutual control of stochastic switching for two electrically coupled superparamagnetic tunnel junctions},\ }\href {https://doi.org/10.1103/PhysRevB.104.054427} {\bibfield  {journal} {\bibinfo  {journal} {Physical Review B}\ }\textbf {\bibinfo {volume} {104}},\ \bibinfo {pages} {054427} (\bibinfo {year} {2021})}\BibitemShut {NoStop}%
\bibitem [{\citenamefont {Carboni}\ \emph {et~al.}(2019)\citenamefont {Carboni}, \citenamefont {Vernocchi}, \citenamefont {Siddik}, \citenamefont {Harms}, \citenamefont {Lyle}, \citenamefont {Sandhu},\ and\ \citenamefont {Ielmini}}]{carboniPhysicsBasedCompactModel2019}%
  \BibitemOpen
  \bibfield  {author} {\bibinfo {author} {\bibfnamefont {R.}~\bibnamefont {Carboni}}, \bibinfo {author} {\bibfnamefont {E.}~\bibnamefont {Vernocchi}}, \bibinfo {author} {\bibfnamefont {M.}~\bibnamefont {Siddik}}, \bibinfo {author} {\bibfnamefont {J.}~\bibnamefont {Harms}}, \bibinfo {author} {\bibfnamefont {A.}~\bibnamefont {Lyle}}, \bibinfo {author} {\bibfnamefont {G.}~\bibnamefont {Sandhu}},\ and\ \bibinfo {author} {\bibfnamefont {D.}~\bibnamefont {Ielmini}},\ }\bibfield  {title} {\bibinfo {title} {A {{Physics-Based Compact Model}} of {{Stochastic Switching}} in {{Spin-Transfer Torque Magnetic Memory}}},\ }\href {https://doi.org/10.1109/TED.2019.2933315} {\bibfield  {journal} {\bibinfo  {journal} {IEEE Transactions on Electron Devices}\ }\textbf {\bibinfo {volume} {66}},\ \bibinfo {pages} {4176} (\bibinfo {year} {2019})}\BibitemShut {NoStop}%
\bibitem [{\citenamefont {Yang}\ \emph {et~al.}(2022)\citenamefont {Yang}, \citenamefont {Zhang}, \citenamefont {Zhang},\ and\ \citenamefont {Wang}}]{yangUniversalCompactModel2022}%
  \BibitemOpen
  \bibfield  {author} {\bibinfo {author} {\bibfnamefont {X.}~\bibnamefont {Yang}}, \bibinfo {author} {\bibfnamefont {Y.}~\bibnamefont {Zhang}}, \bibinfo {author} {\bibfnamefont {Y.}~\bibnamefont {Zhang}},\ and\ \bibinfo {author} {\bibfnamefont {P.}~\bibnamefont {Wang}},\ }\bibfield  {title} {\bibinfo {title} {A {{Universal Compact Model}} for {{Spin-Transfer Torque-Driven Magnetization Switching}} in {{Magnetic Tunnel Junction}}},\ }\href {https://doi.org/10.1109/TED.2022.3207976} {\bibfield  {journal} {\bibinfo  {journal} {IEEE Transactions on Electron Devices}\ }\textbf {\bibinfo {volume} {69}},\ \bibinfo {pages} {6453} (\bibinfo {year} {2022})}\BibitemShut {NoStop}%
\bibitem [{\citenamefont {Schnitzspan}\ \emph {et~al.}(2023)\citenamefont {Schnitzspan}, \citenamefont {Kl{\"a}ui},\ and\ \citenamefont {Jakob}}]{schnitzspanNanosecondSuperparamagneticTunnel}%
  \BibitemOpen
  \bibfield  {author} {\bibinfo {author} {\bibfnamefont {L.}~\bibnamefont {Schnitzspan}}, \bibinfo {author} {\bibfnamefont {M.}~\bibnamefont {Kl{\"a}ui}},\ and\ \bibinfo {author} {\bibfnamefont {G.}~\bibnamefont {Jakob}},\ }\href@noop {} {\bibinfo {title} {Nanosecond {{True Random Number Generation}} with {{Superparamagnetic Tunnel Junctions}} -- {{Identification}} of {{Joule Heating}} and {{Spin-Transfer-Torque}} effects}} (\bibinfo {year} {2023}),\ \Eprint {https://arxiv.org/abs/2301.05694} {arxiv:2301.05694 [cond-mat]} \BibitemShut {NoStop}%
\bibitem [{\citenamefont {Safranski}\ \emph {et~al.}(2021)\citenamefont {Safranski}, \citenamefont {Kaiser}, \citenamefont {Trouilloud}, \citenamefont {Hashemi}, \citenamefont {Hu},\ and\ \citenamefont {Sun}}]{safranskiDemonstrationNanosecondOperation2021}%
  \BibitemOpen
  \bibfield  {author} {\bibinfo {author} {\bibfnamefont {C.}~\bibnamefont {Safranski}}, \bibinfo {author} {\bibfnamefont {J.}~\bibnamefont {Kaiser}}, \bibinfo {author} {\bibfnamefont {P.}~\bibnamefont {Trouilloud}}, \bibinfo {author} {\bibfnamefont {P.}~\bibnamefont {Hashemi}}, \bibinfo {author} {\bibfnamefont {G.}~\bibnamefont {Hu}},\ and\ \bibinfo {author} {\bibfnamefont {J.~Z.}\ \bibnamefont {Sun}},\ }\bibfield  {title} {\bibinfo {title} {Demonstration of {{Nanosecond Operation}} in {{Stochastic Magnetic Tunnel Junctions}}},\ }\href {https://doi.org/10.1021/acs.nanolett.0c04652} {\bibfield  {journal} {\bibinfo  {journal} {Nano Letters}\ }\textbf {\bibinfo {volume} {21}},\ \bibinfo {pages} {2040} (\bibinfo {year} {2021})}\BibitemShut {NoStop}%
\bibitem [{\citenamefont {Bapna}\ and\ \citenamefont {Majetich}(2017)}]{bapna2017current}%
  \BibitemOpen
  \bibfield  {author} {\bibinfo {author} {\bibfnamefont {M.}~\bibnamefont {Bapna}}\ and\ \bibinfo {author} {\bibfnamefont {S.~A.}\ \bibnamefont {Majetich}},\ }\bibfield  {title} {\bibinfo {title} {Current control of time-averaged magnetization in superparamagnetic tunnel junctions},\ }\href@noop {} {\bibfield  {journal} {\bibinfo  {journal} {Applied Physics Letters}\ }\textbf {\bibinfo {volume} {111}} (\bibinfo {year} {2017})}\BibitemShut {NoStop}%
\bibitem [{\citenamefont {Hayakawa}\ \emph {et~al.}(2021)\citenamefont {Hayakawa}, \citenamefont {Kanai}, \citenamefont {Funatsu}, \citenamefont {Igarashi}, \citenamefont {Jinnai}, \citenamefont {Borders}, \citenamefont {Ohno},\ and\ \citenamefont {Fukami}}]{hayakawa2021nanosecond}%
  \BibitemOpen
  \bibfield  {author} {\bibinfo {author} {\bibfnamefont {K.}~\bibnamefont {Hayakawa}}, \bibinfo {author} {\bibfnamefont {S.}~\bibnamefont {Kanai}}, \bibinfo {author} {\bibfnamefont {T.}~\bibnamefont {Funatsu}}, \bibinfo {author} {\bibfnamefont {J.}~\bibnamefont {Igarashi}}, \bibinfo {author} {\bibfnamefont {B.}~\bibnamefont {Jinnai}}, \bibinfo {author} {\bibfnamefont {W.}~\bibnamefont {Borders}}, \bibinfo {author} {\bibfnamefont {H.}~\bibnamefont {Ohno}},\ and\ \bibinfo {author} {\bibfnamefont {S.}~\bibnamefont {Fukami}},\ }\bibfield  {title} {\bibinfo {title} {Nanosecond random telegraph noise in in-plane magnetic tunnel junctions},\ }\href@noop {} {\bibfield  {journal} {\bibinfo  {journal} {Physical review letters}\ }\textbf {\bibinfo {volume} {126}},\ \bibinfo {pages} {117202} (\bibinfo {year} {2021})}\BibitemShut {NoStop}%
\bibitem [{\citenamefont {Leliaert}\ \emph {et~al.}(2017)\citenamefont {Leliaert}, \citenamefont {Mulkers}, \citenamefont {De~Clercq}, \citenamefont {Coene}, \citenamefont {Dvornik},\ and\ \citenamefont {Van~Waeyenberge}}]{leliaertAdaptivelyTimeStepping2017}%
  \BibitemOpen
  \bibfield  {author} {\bibinfo {author} {\bibfnamefont {J.}~\bibnamefont {Leliaert}}, \bibinfo {author} {\bibfnamefont {J.}~\bibnamefont {Mulkers}}, \bibinfo {author} {\bibfnamefont {J.}~\bibnamefont {De~Clercq}}, \bibinfo {author} {\bibfnamefont {A.}~\bibnamefont {Coene}}, \bibinfo {author} {\bibfnamefont {M.}~\bibnamefont {Dvornik}},\ and\ \bibinfo {author} {\bibfnamefont {B.}~\bibnamefont {Van~Waeyenberge}},\ }\bibfield  {title} {\bibinfo {title} {Adaptively time stepping the stochastic {{Landau-Lifshitz-Gilbert}} equation at nonzero temperature: {{Implementation}} and validation in {{MuMax3}}},\ }\href {https://doi.org/10.1063/1.5003957} {\bibfield  {journal} {\bibinfo  {journal} {AIP Advances}\ }\textbf {\bibinfo {volume} {7}},\ \bibinfo {pages} {125010} (\bibinfo {year} {2017})}\BibitemShut {NoStop}%
\bibitem [{\citenamefont {Endean}\ \emph {et~al.}(2014)\citenamefont {Endean}, \citenamefont {Weigelt}, \citenamefont {Victora},\ and\ \citenamefont {Dan~Dahlberg}}]{endean2014tunable}%
  \BibitemOpen
  \bibfield  {author} {\bibinfo {author} {\bibfnamefont {D.~E.}\ \bibnamefont {Endean}}, \bibinfo {author} {\bibfnamefont {C.}~\bibnamefont {Weigelt}}, \bibinfo {author} {\bibfnamefont {R.}~\bibnamefont {Victora}},\ and\ \bibinfo {author} {\bibfnamefont {E.}~\bibnamefont {Dan~Dahlberg}},\ }\bibfield  {title} {\bibinfo {title} {Tunable random telegraph noise in individual square permalloy dots},\ }\href@noop {} {\bibfield  {journal} {\bibinfo  {journal} {Applied Physics Letters}\ }\textbf {\bibinfo {volume} {104}} (\bibinfo {year} {2014})}\BibitemShut {NoStop}%
\bibitem [{\citenamefont {Shin}\ \emph {et~al.}(2010)\citenamefont {Shin}, \citenamefont {Kim},\ and\ \citenamefont {Kang}}]{shin2010compact}%
  \BibitemOpen
  \bibfield  {author} {\bibinfo {author} {\bibfnamefont {S.}~\bibnamefont {Shin}}, \bibinfo {author} {\bibfnamefont {K.}~\bibnamefont {Kim}},\ and\ \bibinfo {author} {\bibfnamefont {S.-M.}\ \bibnamefont {Kang}},\ }\bibfield  {title} {\bibinfo {title} {Compact models for memristors based on charge-flux constitutive relationships},\ }\href@noop {} {\bibfield  {journal} {\bibinfo  {journal} {IEEE Transactions on Computer-Aided Design of Integrated Circuits and Systems}\ }\textbf {\bibinfo {volume} {29}},\ \bibinfo {pages} {590} (\bibinfo {year} {2010})}\BibitemShut {NoStop}%
\bibitem [{\citenamefont {Ding}\ \emph {et~al.}(2022)\citenamefont {Ding}, \citenamefont {Peng}, \citenamefont {Li}, \citenamefont {Zhang}, \citenamefont {Wang}, \citenamefont {Song},\ and\ \citenamefont {Huang}}]{ding2022review}%
  \BibitemOpen
  \bibfield  {author} {\bibinfo {author} {\bibfnamefont {F.}~\bibnamefont {Ding}}, \bibinfo {author} {\bibfnamefont {B.}~\bibnamefont {Peng}}, \bibinfo {author} {\bibfnamefont {X.}~\bibnamefont {Li}}, \bibinfo {author} {\bibfnamefont {L.}~\bibnamefont {Zhang}}, \bibinfo {author} {\bibfnamefont {R.}~\bibnamefont {Wang}}, \bibinfo {author} {\bibfnamefont {Z.}~\bibnamefont {Song}},\ and\ \bibinfo {author} {\bibfnamefont {R.}~\bibnamefont {Huang}},\ }\bibfield  {title} {\bibinfo {title} {A review of compact modeling for phase change memory},\ }\href@noop {} {\bibfield  {journal} {\bibinfo  {journal} {Journal of Semiconductors}\ }\textbf {\bibinfo {volume} {43}},\ \bibinfo {pages} {023101} (\bibinfo {year} {2022})}\BibitemShut {NoStop}%
\bibitem [{\citenamefont {Jim{\'e}nez-Le{\'o}n}\ \emph {et~al.}(2021)\citenamefont {Jim{\'e}nez-Le{\'o}n}, \citenamefont {Sarmiento-Reyes},\ and\ \citenamefont {Rosales-Quintero}}]{jimenez2021compact}%
  \BibitemOpen
  \bibfield  {author} {\bibinfo {author} {\bibfnamefont {J.}~\bibnamefont {Jim{\'e}nez-Le{\'o}n}}, \bibinfo {author} {\bibfnamefont {L.~A.}\ \bibnamefont {Sarmiento-Reyes}},\ and\ \bibinfo {author} {\bibfnamefont {P.}~\bibnamefont {Rosales-Quintero}},\ }\bibfield  {title} {\bibinfo {title} {A compact modeling methodology for experimental memristive devices},\ }\href@noop {} {\bibfield  {journal} {\bibinfo  {journal} {IEEE Transactions on Computer-Aided Design of Integrated Circuits and Systems}\ }\textbf {\bibinfo {volume} {41}},\ \bibinfo {pages} {4851} (\bibinfo {year} {2021})}\BibitemShut {NoStop}%
\bibitem [{\citenamefont {Araujo}\ \emph {et~al.}(2022)\citenamefont {Araujo}, \citenamefont {Chopin},\ and\ \citenamefont {De~Wergifosse}}]{araujo2022data}%
  \BibitemOpen
  \bibfield  {author} {\bibinfo {author} {\bibfnamefont {F.~A.}\ \bibnamefont {Araujo}}, \bibinfo {author} {\bibfnamefont {C.}~\bibnamefont {Chopin}},\ and\ \bibinfo {author} {\bibfnamefont {S.}~\bibnamefont {De~Wergifosse}},\ }\bibfield  {title} {\bibinfo {title} {Data-driven thiele equation approach for solving the full nonlinear spin-torque vortex oscillator dynamics},\ }\href@noop {} {\bibfield  {journal} {\bibinfo  {journal} {arXiv preprint arXiv:2206.13596}\ } (\bibinfo {year} {2022})}\BibitemShut {NoStop}%
\bibitem [{\citenamefont {Faber}\ \emph {et~al.}(2009)\citenamefont {Faber}, \citenamefont {Zhao}, \citenamefont {Klein}, \citenamefont {Devolder},\ and\ \citenamefont {Chappert}}]{faber2009dynamic}%
  \BibitemOpen
  \bibfield  {author} {\bibinfo {author} {\bibfnamefont {L.-B.}\ \bibnamefont {Faber}}, \bibinfo {author} {\bibfnamefont {W.}~\bibnamefont {Zhao}}, \bibinfo {author} {\bibfnamefont {J.-O.}\ \bibnamefont {Klein}}, \bibinfo {author} {\bibfnamefont {T.}~\bibnamefont {Devolder}},\ and\ \bibinfo {author} {\bibfnamefont {C.}~\bibnamefont {Chappert}},\ }\bibfield  {title} {\bibinfo {title} {Dynamic compact model of spin-transfer torque based magnetic tunnel junction (mtj)},\ }in\ \href@noop {} {\emph {\bibinfo {booktitle} {2009 4th International Conference on Design \& Technology of Integrated Systems in Nanoscal Era}}}\ (\bibinfo {organization} {IEEE},\ \bibinfo {year} {2009})\ pp.\ \bibinfo {pages} {130--135}\BibitemShut {NoStop}%
\bibitem [{\citenamefont {Guo}\ \emph {et~al.}(2010)\citenamefont {Guo}, \citenamefont {Prenat}, \citenamefont {Javerliac}, \citenamefont {El~Baraji}, \citenamefont {De~Mestier}, \citenamefont {Baraduc},\ and\ \citenamefont {Dieny}}]{guo2010spice}%
  \BibitemOpen
  \bibfield  {author} {\bibinfo {author} {\bibfnamefont {W.}~\bibnamefont {Guo}}, \bibinfo {author} {\bibfnamefont {G.}~\bibnamefont {Prenat}}, \bibinfo {author} {\bibfnamefont {V.}~\bibnamefont {Javerliac}}, \bibinfo {author} {\bibfnamefont {M.}~\bibnamefont {El~Baraji}}, \bibinfo {author} {\bibfnamefont {N.}~\bibnamefont {De~Mestier}}, \bibinfo {author} {\bibfnamefont {C.}~\bibnamefont {Baraduc}},\ and\ \bibinfo {author} {\bibfnamefont {B.}~\bibnamefont {Dieny}},\ }\bibfield  {title} {\bibinfo {title} {Spice modelling of magnetic tunnel junctions written by spin-transfer torque},\ }\href@noop {} {\bibfield  {journal} {\bibinfo  {journal} {Journal of Physics D: Applied Physics}\ }\textbf {\bibinfo {volume} {43}},\ \bibinfo {pages} {215001} (\bibinfo {year} {2010})}\BibitemShut {NoStop}%
\bibitem [{\citenamefont {Torunbalci}\ \emph {et~al.}(2018)\citenamefont {Torunbalci}, \citenamefont {Upadhyaya}, \citenamefont {Bhave},\ and\ \citenamefont {Camsari}}]{torunbalciModularCompactModeling2018}%
  \BibitemOpen
  \bibfield  {author} {\bibinfo {author} {\bibfnamefont {M.~M.}\ \bibnamefont {Torunbalci}}, \bibinfo {author} {\bibfnamefont {P.}~\bibnamefont {Upadhyaya}}, \bibinfo {author} {\bibfnamefont {S.~A.}\ \bibnamefont {Bhave}},\ and\ \bibinfo {author} {\bibfnamefont {K.~Y.}\ \bibnamefont {Camsari}},\ }\bibfield  {title} {\bibinfo {title} {Modular {{Compact Modeling}} of {{MTJ Devices}}},\ }\href {https://doi.org/10.1109/TED.2018.2863538} {\bibfield  {journal} {\bibinfo  {journal} {IEEE Transactions on Electron Devices}\ }\textbf {\bibinfo {volume} {65}},\ \bibinfo {pages} {4628} (\bibinfo {year} {2018})}\BibitemShut {NoStop}%
\bibitem [{\citenamefont {Chowdhury}\ \emph {et~al.}(2023)\citenamefont {Chowdhury}, \citenamefont {Grimaldi}, \citenamefont {Aadit}, \citenamefont {Niazi}, \citenamefont {Mohseni}, \citenamefont {Kanai}, \citenamefont {Ohno}, \citenamefont {Fukami}, \citenamefont {Theogarajan}, \citenamefont {Finocchio}, \citenamefont {Datta},\ and\ \citenamefont {Camsari}}]{chowdhuryFullstackViewProbabilistic2023}%
  \BibitemOpen
  \bibfield  {author} {\bibinfo {author} {\bibfnamefont {S.}~\bibnamefont {Chowdhury}}, \bibinfo {author} {\bibfnamefont {A.}~\bibnamefont {Grimaldi}}, \bibinfo {author} {\bibfnamefont {N.~A.}\ \bibnamefont {Aadit}}, \bibinfo {author} {\bibfnamefont {S.}~\bibnamefont {Niazi}}, \bibinfo {author} {\bibfnamefont {M.}~\bibnamefont {Mohseni}}, \bibinfo {author} {\bibfnamefont {S.}~\bibnamefont {Kanai}}, \bibinfo {author} {\bibfnamefont {H.}~\bibnamefont {Ohno}}, \bibinfo {author} {\bibfnamefont {S.}~\bibnamefont {Fukami}}, \bibinfo {author} {\bibfnamefont {L.}~\bibnamefont {Theogarajan}}, \bibinfo {author} {\bibfnamefont {G.}~\bibnamefont {Finocchio}}, \bibinfo {author} {\bibfnamefont {S.}~\bibnamefont {Datta}},\ and\ \bibinfo {author} {\bibfnamefont {K.~Y.}\ \bibnamefont {Camsari}},\ }\href@noop {} {\bibinfo {title} {A full-stack view of probabilistic computing with p-bits: Devices, architectures and algorithms}} (\bibinfo {year} {2023}),\ \Eprint {https://arxiv.org/abs/2302.06457} {arxiv:2302.06457 [physics]}
  \BibitemShut {NoStop}%
\bibitem [{\citenamefont {Bunaiyan}\ \emph {et~al.}(2023)\citenamefont {Bunaiyan}, \citenamefont {Datta},\ and\ \citenamefont {Camsari}}]{bunaiyan2023heisenberg}%
  \BibitemOpen
  \bibfield  {author} {\bibinfo {author} {\bibfnamefont {S.}~\bibnamefont {Bunaiyan}}, \bibinfo {author} {\bibfnamefont {S.}~\bibnamefont {Datta}},\ and\ \bibinfo {author} {\bibfnamefont {K.~Y.}\ \bibnamefont {Camsari}},\ }\bibfield  {title} {\bibinfo {title} {Heisenberg machines with programmable spin-circuits},\ }\href@noop {} {\bibfield  {journal} {\bibinfo  {journal} {arXiv preprint arXiv:2312.01477}\ } (\bibinfo {year} {2023})}\BibitemShut {NoStop}%
\bibitem [{\citenamefont {Jung}\ \emph {et~al.}(2022)\citenamefont {Jung}, \citenamefont {Lee}, \citenamefont {Myung}, \citenamefont {Kim}, \citenamefont {Yoon}, \citenamefont {Kwon}, \citenamefont {Ju}, \citenamefont {Kim}, \citenamefont {Yi}, \citenamefont {Han}, \citenamefont {Kwon}, \citenamefont {Seo}, \citenamefont {Lee}, \citenamefont {Koh}, \citenamefont {Lee}, \citenamefont {Song}, \citenamefont {Choi}, \citenamefont {Ham},\ and\ \citenamefont {Kim}}]{jungCrossbarArrayMagnetoresistive2022}%
  \BibitemOpen
  \bibfield  {author} {\bibinfo {author} {\bibfnamefont {S.}~\bibnamefont {Jung}}, \bibinfo {author} {\bibfnamefont {H.}~\bibnamefont {Lee}}, \bibinfo {author} {\bibfnamefont {S.}~\bibnamefont {Myung}}, \bibinfo {author} {\bibfnamefont {H.}~\bibnamefont {Kim}}, \bibinfo {author} {\bibfnamefont {S.~K.}\ \bibnamefont {Yoon}}, \bibinfo {author} {\bibfnamefont {S.-W.}\ \bibnamefont {Kwon}}, \bibinfo {author} {\bibfnamefont {Y.}~\bibnamefont {Ju}}, \bibinfo {author} {\bibfnamefont {M.}~\bibnamefont {Kim}}, \bibinfo {author} {\bibfnamefont {W.}~\bibnamefont {Yi}}, \bibinfo {author} {\bibfnamefont {S.}~\bibnamefont {Han}}, \bibinfo {author} {\bibfnamefont {B.}~\bibnamefont {Kwon}}, \bibinfo {author} {\bibfnamefont {B.}~\bibnamefont {Seo}}, \bibinfo {author} {\bibfnamefont {K.}~\bibnamefont {Lee}}, \bibinfo {author} {\bibfnamefont {G.-H.}\ \bibnamefont {Koh}}, \bibinfo {author} {\bibfnamefont {K.}~\bibnamefont {Lee}}, \bibinfo {author} {\bibfnamefont {Y.}~\bibnamefont {Song}}, \bibinfo {author} {\bibfnamefont
  {C.}~\bibnamefont {Choi}}, \bibinfo {author} {\bibfnamefont {D.}~\bibnamefont {Ham}},\ and\ \bibinfo {author} {\bibfnamefont {S.~J.}\ \bibnamefont {Kim}},\ }\bibfield  {title} {\bibinfo {title} {A crossbar array of magnetoresistive memory devices for in-memory computing},\ }\href {https://doi.org/10.1038/s41586-021-04196-6} {\bibfield  {journal} {\bibinfo  {journal} {Nature}\ }\textbf {\bibinfo {volume} {601}},\ \bibinfo {pages} {211} (\bibinfo {year} {2022})}\BibitemShut {NoStop}%
\bibitem [{\citenamefont {Camsari}\ \emph {et~al.}(2017{\natexlab{b}})\citenamefont {Camsari}, \citenamefont {Salahuddin},\ and\ \citenamefont {Datta}}]{camsariImplementingPbitsEmbedded2017}%
  \BibitemOpen
  \bibfield  {author} {\bibinfo {author} {\bibfnamefont {K.~Y.}\ \bibnamefont {Camsari}}, \bibinfo {author} {\bibfnamefont {S.}~\bibnamefont {Salahuddin}},\ and\ \bibinfo {author} {\bibfnamefont {S.}~\bibnamefont {Datta}},\ }\bibfield  {title} {\bibinfo {title} {Implementing p-bits {{With Embedded MTJ}}},\ }\href {https://doi.org/10.1109/LED.2017.2768321} {\bibfield  {journal} {\bibinfo  {journal} {IEEE Electron Device Letters}\ }\textbf {\bibinfo {volume} {38}},\ \bibinfo {pages} {1767} (\bibinfo {year} {2017}{\natexlab{b}})}\BibitemShut {NoStop}%
\bibitem [{\citenamefont {Xiao}\ \emph {et~al.}(2005)\citenamefont {Xiao}, \citenamefont {Zangwill},\ and\ \citenamefont {Stiles}}]{xiao2005macrospin}%
  \BibitemOpen
  \bibfield  {author} {\bibinfo {author} {\bibfnamefont {J.}~\bibnamefont {Xiao}}, \bibinfo {author} {\bibfnamefont {A.}~\bibnamefont {Zangwill}},\ and\ \bibinfo {author} {\bibfnamefont {M.~D.}\ \bibnamefont {Stiles}},\ }\bibfield  {title} {\bibinfo {title} {Macrospin models of spin transfer dynamics},\ }\href@noop {} {\bibfield  {journal} {\bibinfo  {journal} {Physical Review B}\ }\textbf {\bibinfo {volume} {72}},\ \bibinfo {pages} {014446} (\bibinfo {year} {2005})}\BibitemShut {NoStop}%
\bibitem [{\citenamefont {Kanai}\ \emph {et~al.}()\citenamefont {Kanai}, \citenamefont {Hayakawa}, \citenamefont {Elyasi}, \citenamefont {Kobayashi}, \citenamefont {Igarashi}, \citenamefont {Jinnai}, \citenamefont {Borders}, \citenamefont {Bauer}, \citenamefont {Ohno},\ and\ \citenamefont {Fukami}}]{KanaiTBP}%
  \BibitemOpen
  \bibfield  {author} {\bibinfo {author} {\bibfnamefont {S.}~\bibnamefont {Kanai}}, \bibinfo {author} {\bibfnamefont {K.}~\bibnamefont {Hayakawa}}, \bibinfo {author} {\bibfnamefont {M.}~\bibnamefont {Elyasi}}, \bibinfo {author} {\bibfnamefont {K.}~\bibnamefont {Kobayashi}}, \bibinfo {author} {\bibfnamefont {J.}~\bibnamefont {Igarashi}}, \bibinfo {author} {\bibfnamefont {B.}~\bibnamefont {Jinnai}}, \bibinfo {author} {\bibfnamefont {W.~A.}\ \bibnamefont {Borders}}, \bibinfo {author} {\bibfnamefont {G.~E.~W.}\ \bibnamefont {Bauer}}, \bibinfo {author} {\bibfnamefont {H.}~\bibnamefont {Ohno}},\ and\ \bibinfo {author} {\bibfnamefont {S.}~\bibnamefont {Fukami}},\ }\bibfield  {title} {\bibinfo {title} {Stochastic switching time constant and instability in nanomagnets},\ }\href@noop {} {\bibinfo  {journal} {(to be published)}\ }\BibitemShut {NoStop}%
\bibitem [{\citenamefont {Berkov}\ and\ \citenamefont {Gorn}(2002)}]{berkov2002thermally}%
  \BibitemOpen
\bibfield  {journal} {  }\bibfield  {author} {\bibinfo {author} {\bibfnamefont {D.}~\bibnamefont {Berkov}}\ and\ \bibinfo {author} {\bibfnamefont {N.}~\bibnamefont {Gorn}},\ }\bibfield  {title} {\bibinfo {title} {Thermally activated processes in magnetic systems consisting of rigid dipoles: equivalence of the {Ito} and {Stratonovich} stochastic calculus},\ }\href@noop {} {\bibfield  {journal} {\bibinfo  {journal} {Journal of Physics: Condensed Matter}\ }\textbf {\bibinfo {volume} {14}},\ \bibinfo {pages} {L281} (\bibinfo {year} {2002})}\BibitemShut {NoStop}%
\bibitem [{\citenamefont {Desplat}\ and\ \citenamefont {Kim}(2020)}]{desplat_2020}%
  \BibitemOpen
  \bibfield  {author} {\bibinfo {author} {\bibfnamefont {L.}~\bibnamefont {Desplat}}\ and\ \bibinfo {author} {\bibfnamefont {J.-V.}\ \bibnamefont {Kim}},\ }\bibfield  {title} {\bibinfo {title} {Entropy-reduced retention times in magnetic memory elements: A case of the meyer-neldel compensation rule},\ }\href {https://doi.org/10.1103/PhysRevLett.125.107201} {\bibfield  {journal} {\bibinfo  {journal} {Phys. Rev. Lett.}\ }\textbf {\bibinfo {volume} {125}},\ \bibinfo {pages} {107201} (\bibinfo {year} {2020})}\BibitemShut {NoStop}%
\bibitem [{\citenamefont {Cimr{\'a}k}(2007)}]{cimrak2007survey}%
  \BibitemOpen
  \bibfield  {author} {\bibinfo {author} {\bibfnamefont {I.}~\bibnamefont {Cimr{\'a}k}},\ }\bibfield  {title} {\bibinfo {title} {A survey on the numerics and computations for the {Landau-Lifshitz} equation of micromagnetism},\ }\href@noop {} {\bibfield  {journal} {\bibinfo  {journal} {Archives of Computational Methods in Engineering}\ }\textbf {\bibinfo {volume} {15}},\ \bibinfo {pages} {1} (\bibinfo {year} {2007})}\BibitemShut {NoStop}%
\bibitem [{\citenamefont {Aron}\ \emph {et~al.}(2014)\citenamefont {Aron}, \citenamefont {Barci}, \citenamefont {Cugliandolo}, \citenamefont {Arenas},\ and\ \citenamefont {Lozano}}]{aron2014magnetization}%
  \BibitemOpen
  \bibfield  {author} {\bibinfo {author} {\bibfnamefont {C.}~\bibnamefont {Aron}}, \bibinfo {author} {\bibfnamefont {D.~G.}\ \bibnamefont {Barci}}, \bibinfo {author} {\bibfnamefont {L.~F.}\ \bibnamefont {Cugliandolo}}, \bibinfo {author} {\bibfnamefont {Z.~G.}\ \bibnamefont {Arenas}},\ and\ \bibinfo {author} {\bibfnamefont {G.~S.}\ \bibnamefont {Lozano}},\ }\bibfield  {title} {\bibinfo {title} {Magnetization dynamics: path-integral formalism for the stochastic {Landau--Lifshitz--Gilbert} equation},\ }\href@noop {} {\bibfield  {journal} {\bibinfo  {journal} {Journal of Statistical Mechanics: Theory and Experiment}\ }\textbf {\bibinfo {volume} {2014}},\ \bibinfo {pages} {P09008} (\bibinfo {year} {2014})}\BibitemShut {NoStop}%
\bibitem [{\citenamefont {Rom\'a}\ \emph {et~al.}(2014)\citenamefont {Rom\'a}, \citenamefont {Cugliandolo},\ and\ \citenamefont {Lozano}}]{PhysRevE.90.023203}%
  \BibitemOpen
  \bibfield  {author} {\bibinfo {author} {\bibfnamefont {F.}~\bibnamefont {Rom\'a}}, \bibinfo {author} {\bibfnamefont {L.~F.}\ \bibnamefont {Cugliandolo}},\ and\ \bibinfo {author} {\bibfnamefont {G.~S.}\ \bibnamefont {Lozano}},\ }\bibfield  {title} {\bibinfo {title} {Numerical integration of the stochastic {Landau-Lifshitz-Gilbert} equation in generic time-discretization schemes},\ }\href {https://doi.org/10.1103/PhysRevE.90.023203} {\bibfield  {journal} {\bibinfo  {journal} {Phys. Rev. E}\ }\textbf {\bibinfo {volume} {90}},\ \bibinfo {pages} {023203} (\bibinfo {year} {2014})}\BibitemShut {NoStop}%
\bibitem [{\citenamefont {Garc\'{\i}a-Palacios}\ and\ \citenamefont {L\'azaro}(1998)}]{PhysRevB.58.14937}%
  \BibitemOpen
  \bibfield  {author} {\bibinfo {author} {\bibfnamefont {J.~L.}\ \bibnamefont {Garc\'{\i}a-Palacios}}\ and\ \bibinfo {author} {\bibfnamefont {F.~J.}\ \bibnamefont {L\'azaro}},\ }\bibfield  {title} {\bibinfo {title} {Langevin-dynamics study of the dynamical properties of small magnetic particles},\ }\href {https://doi.org/10.1103/PhysRevB.58.14937} {\bibfield  {journal} {\bibinfo  {journal} {Phys. Rev. B}\ }\textbf {\bibinfo {volume} {58}},\ \bibinfo {pages} {14937} (\bibinfo {year} {1998})}\BibitemShut {NoStop}%
\bibitem [{\citenamefont {Ament}\ \emph {et~al.}(2016)\citenamefont {Ament}, \citenamefont {Rangarajan}, \citenamefont {Parthasarathy},\ and\ \citenamefont {Rakheja}}]{ament2016solving}%
  \BibitemOpen
  \bibfield  {author} {\bibinfo {author} {\bibfnamefont {S.}~\bibnamefont {Ament}}, \bibinfo {author} {\bibfnamefont {N.}~\bibnamefont {Rangarajan}}, \bibinfo {author} {\bibfnamefont {A.}~\bibnamefont {Parthasarathy}},\ and\ \bibinfo {author} {\bibfnamefont {S.}~\bibnamefont {Rakheja}},\ }\bibfield  {title} {\bibinfo {title} {Solving the stochastic landau-lifshitz-gilbert-slonczewski equation for monodomain nanomagnets: A survey and analysis of numerical techniques},\ }\href@noop {} {\bibfield  {journal} {\bibinfo  {journal} {arXiv preprint arXiv:1607.04596}\ } (\bibinfo {year} {2016})}\BibitemShut {NoStop}%
\bibitem [{\citenamefont {Boninsegna}\ \emph {et~al.}(2018)\citenamefont {Boninsegna}, \citenamefont {N{\"u}ske},\ and\ \citenamefont {Clementi}}]{boninsegna2018sparse}%
  \BibitemOpen
  \bibfield  {author} {\bibinfo {author} {\bibfnamefont {L.}~\bibnamefont {Boninsegna}}, \bibinfo {author} {\bibfnamefont {F.}~\bibnamefont {N{\"u}ske}},\ and\ \bibinfo {author} {\bibfnamefont {C.}~\bibnamefont {Clementi}},\ }\bibfield  {title} {\bibinfo {title} {Sparse learning of stochastic dynamical equations},\ }\href@noop {} {\bibfield  {journal} {\bibinfo  {journal} {The Journal of Chemical Physics}\ }\textbf {\bibinfo {volume} {148}} (\bibinfo {year} {2018})}\BibitemShut {NoStop}%
\bibitem [{\citenamefont {Callaham}\ \emph {et~al.}(2021)\citenamefont {Callaham}, \citenamefont {Loiseau}, \citenamefont {Rigas},\ and\ \citenamefont {Brunton}}]{callaham2021nonlinear}%
  \BibitemOpen
  \bibfield  {author} {\bibinfo {author} {\bibfnamefont {J.~L.}\ \bibnamefont {Callaham}}, \bibinfo {author} {\bibfnamefont {J.-C.}\ \bibnamefont {Loiseau}}, \bibinfo {author} {\bibfnamefont {G.}~\bibnamefont {Rigas}},\ and\ \bibinfo {author} {\bibfnamefont {S.~L.}\ \bibnamefont {Brunton}},\ }\bibfield  {title} {\bibinfo {title} {Nonlinear stochastic modelling with {L}angevin regression},\ }\href@noop {} {\bibfield  {journal} {\bibinfo  {journal} {Proceedings of the Royal Society A}\ }\textbf {\bibinfo {volume} {477}},\ \bibinfo {pages} {20210092} (\bibinfo {year} {2021})}\BibitemShut {NoStop}%
\bibitem [{\citenamefont {Gorj{\~a}o}\ \emph {et~al.}(2019)\citenamefont {Gorj{\~a}o}, \citenamefont {Heysel}, \citenamefont {Lehnertz},\ and\ \citenamefont {Tabar}}]{gorjao2019analysis}%
  \BibitemOpen
  \bibfield  {author} {\bibinfo {author} {\bibfnamefont {L.~R.}\ \bibnamefont {Gorj{\~a}o}}, \bibinfo {author} {\bibfnamefont {J.}~\bibnamefont {Heysel}}, \bibinfo {author} {\bibfnamefont {K.}~\bibnamefont {Lehnertz}},\ and\ \bibinfo {author} {\bibfnamefont {M.~R.~R.}\ \bibnamefont {Tabar}},\ }\bibfield  {title} {\bibinfo {title} {Analysis and data-driven reconstruction of bivariate jump-diffusion processes},\ }\href@noop {} {\bibfield  {journal} {\bibinfo  {journal} {Physical Review E}\ }\textbf {\bibinfo {volume} {100}},\ \bibinfo {pages} {062127} (\bibinfo {year} {2019})}\BibitemShut {NoStop}%
\bibitem [{Note1()}]{Note1}%
  \BibitemOpen
  \bibinfo {note} {{We note that in reality the driving noise terms likely possess additional structure beyond mere white noise. However, as we will show in Sec.~\ref {sec:results}, the white noise assumption does very well in reproducing the statistics observed in experiment. Physically, the choice of white noise is the unique choice that ensures the dissipative mechanisms of the system are local in time~\cite {kubo1966fluctuation}.}}\BibitemShut {Stop}%
\bibitem [{Note2()}]{Note2}%
  \BibitemOpen
  \bibinfo {note} {These conditional moments are reminiscent of Kolmogorov's structure functions in turbulence. However, structure functions are conditional moments that use a spatial separation and spatial ensemble average to obtain a scalar value for fluid velocity, whereas the Kramers-Moyal coefficients here are evaluated from a voltage-time trace (which by the ergodic hypothesis is equivalent to an ensemble average) that integrates solely over the probability distribution.}\BibitemShut {Stop}%
\bibitem [{\citenamefont {Risken}\ and\ \citenamefont {Risken}(1996)}]{risken1996fokker}%
  \BibitemOpen
  \bibfield  {author} {\bibinfo {author} {\bibfnamefont {H.}~\bibnamefont {Risken}}\ and\ \bibinfo {author} {\bibfnamefont {H.}~\bibnamefont {Risken}},\ }\href@noop {} {\emph {\bibinfo {title} {The {F}okker-{P}lanck {E}quation}}}\ (\bibinfo  {publisher} {Springer},\ \bibinfo {year} {1996})\BibitemShut {NoStop}%
\bibitem [{\citenamefont {Honisch}\ and\ \citenamefont {Friedrich}(2011)}]{honisch2011estimation}%
  \BibitemOpen
  \bibfield  {author} {\bibinfo {author} {\bibfnamefont {C.}~\bibnamefont {Honisch}}\ and\ \bibinfo {author} {\bibfnamefont {R.}~\bibnamefont {Friedrich}},\ }\bibfield  {title} {\bibinfo {title} {Estimation of {K}ramers-{M}oyal coefficients at low sampling rates},\ }\href@noop {} {\bibfield  {journal} {\bibinfo  {journal} {Physical {R}eview {E}}\ }\textbf {\bibinfo {volume} {83}},\ \bibinfo {pages} {066701} (\bibinfo {year} {2011})}\BibitemShut {NoStop}%
\bibitem [{\citenamefont {Friedrich}\ \emph {et~al.}(2000)\citenamefont {Friedrich}, \citenamefont {Peinke},\ and\ \citenamefont {Renner}}]{friedrich2000quantify}%
  \BibitemOpen
  \bibfield  {author} {\bibinfo {author} {\bibfnamefont {R.}~\bibnamefont {Friedrich}}, \bibinfo {author} {\bibfnamefont {J.}~\bibnamefont {Peinke}},\ and\ \bibinfo {author} {\bibfnamefont {C.}~\bibnamefont {Renner}},\ }\bibfield  {title} {\bibinfo {title} {How to {Q}uantify {D}eterministic and {R}andom {I}nfluences on the {S}tatistics of the {F}oreign {E}xchange {M}arket},\ }\href@noop {} {\bibfield  {journal} {\bibinfo  {journal} {Physical Review Letters}\ }\textbf {\bibinfo {volume} {84}},\ \bibinfo {pages} {5224} (\bibinfo {year} {2000})}\BibitemShut {NoStop}%
\bibitem [{\citenamefont {Pawula}(1967)}]{PhysRev.162.186}%
  \BibitemOpen
  \bibfield  {author} {\bibinfo {author} {\bibfnamefont {R.~F.}\ \bibnamefont {Pawula}},\ }\bibfield  {title} {\bibinfo {title} {Approximation of the linear {Boltzmann} equation by the {Fokker-Planck} equation},\ }\href {https://doi.org/10.1103/PhysRev.162.186} {\bibfield  {journal} {\bibinfo  {journal} {Phys. Rev.}\ }\textbf {\bibinfo {volume} {162}},\ \bibinfo {pages} {186} (\bibinfo {year} {1967})}\BibitemShut {NoStop}%
\bibitem [{\citenamefont {Gottschall}\ and\ \citenamefont {Peinke}(2008)}]{gottschall2008definition}%
  \BibitemOpen
  \bibfield  {author} {\bibinfo {author} {\bibfnamefont {J.}~\bibnamefont {Gottschall}}\ and\ \bibinfo {author} {\bibfnamefont {J.}~\bibnamefont {Peinke}},\ }\bibfield  {title} {\bibinfo {title} {On the definition and handling of different drift and diffusion estimates},\ }\href@noop {} {\bibfield  {journal} {\bibinfo  {journal} {New Journal of Physics}\ }\textbf {\bibinfo {volume} {10}},\ \bibinfo {pages} {083034} (\bibinfo {year} {2008})}\BibitemShut {NoStop}%
\bibitem [{\citenamefont {Rydin~Gorj{\~a}o}\ \emph {et~al.}(2021)\citenamefont {Rydin~Gorj{\~a}o}, \citenamefont {Witthaut}, \citenamefont {Lehnertz},\ and\ \citenamefont {Lind}}]{rydin2021arbitrary}%
  \BibitemOpen
  \bibfield  {author} {\bibinfo {author} {\bibfnamefont {L.}~\bibnamefont {Rydin~Gorj{\~a}o}}, \bibinfo {author} {\bibfnamefont {D.}~\bibnamefont {Witthaut}}, \bibinfo {author} {\bibfnamefont {K.}~\bibnamefont {Lehnertz}},\ and\ \bibinfo {author} {\bibfnamefont {P.~G.}\ \bibnamefont {Lind}},\ }\bibfield  {title} {\bibinfo {title} {Arbitrary-{O}rder {F}inite-{T}ime {C}orrections for the {K}ramers-{M}oyal {O}perator},\ }\href@noop {} {\bibfield  {journal} {\bibinfo  {journal} {Entropy}\ }\textbf {\bibinfo {volume} {23}},\ \bibinfo {pages} {517} (\bibinfo {year} {2021})}\BibitemShut {NoStop}%
\bibitem [{\citenamefont {Sicard}\ \emph {et~al.}(2021)\citenamefont {Sicard}, \citenamefont {Koskin}, \citenamefont {Annibale},\ and\ \citenamefont {Rosta}}]{sicard2021position}%
  \BibitemOpen
  \bibfield  {author} {\bibinfo {author} {\bibfnamefont {F.}~\bibnamefont {Sicard}}, \bibinfo {author} {\bibfnamefont {V.}~\bibnamefont {Koskin}}, \bibinfo {author} {\bibfnamefont {A.}~\bibnamefont {Annibale}},\ and\ \bibinfo {author} {\bibfnamefont {E.}~\bibnamefont {Rosta}},\ }\bibfield  {title} {\bibinfo {title} {Position {D}ependent {D}iffusion from {B}iased {S}imulations and {M}arkov {S}tate {M}odel {A}nalysis},\ }\href@noop {} {\bibfield  {journal} {\bibinfo  {journal} {Journal of Chemical Theory and Computation}\ }\textbf {\bibinfo {volume} {17}},\ \bibinfo {pages} {2022} (\bibinfo {year} {2021})}\BibitemShut {NoStop}%
\bibitem [{\citenamefont {Vercauteren}\ and\ \citenamefont {Hartmann}(2005)}]{vercauteren2005numerical}%
  \BibitemOpen
  \bibfield  {author} {\bibinfo {author} {\bibfnamefont {N.}~\bibnamefont {Vercauteren}}\ and\ \bibinfo {author} {\bibfnamefont {C.}~\bibnamefont {Hartmann}},\ }\emph {\bibinfo {title} {Numerical investigation of solutions of {L}angevin equations}},\ \href@noop {} {Ph.D. thesis},\ \bibinfo  {school} {\'{E}cole Polytechnique F\'{e}d\'{e}rale de Lausanne} (\bibinfo {year} {2005})\BibitemShut {NoStop}%
\bibitem [{Note3()}]{Note3}%
  \BibitemOpen
  \bibinfo {note} {Legrendre polynomials were also considered because of their similar properties, but we chose Chebyshev polynomials because they obtained more accurate results with fewer total coefficients.}\BibitemShut {Stop}%
\bibitem [{\citenamefont {Van~Kampen}(1981)}]{van1981ito}%
  \BibitemOpen
  \bibfield  {author} {\bibinfo {author} {\bibfnamefont {N.~G.}\ \bibnamefont {Van~Kampen}},\ }\bibfield  {title} {\bibinfo {title} {It{\^o} versus {S}tratonovich},\ }\href@noop {} {\bibfield  {journal} {\bibinfo  {journal} {Journal of Statistical Physics}\ }\textbf {\bibinfo {volume} {24}},\ \bibinfo {pages} {175} (\bibinfo {year} {1981})}\BibitemShut {NoStop}%
\bibitem [{\citenamefont {Daniels}\ \emph {et~al.}(2023)\citenamefont {Daniels}, \citenamefont {Borders}, \citenamefont {Prasad}, \citenamefont {Madhavan}, \citenamefont {Gibeault}, \citenamefont {Adeyeye}, \citenamefont {Pocher}, \citenamefont {Wan}, \citenamefont {Tran}, \citenamefont {Katine} \emph {et~al.}}]{daniels2023neural}%
  \BibitemOpen
  \bibfield  {author} {\bibinfo {author} {\bibfnamefont {M.~W.}\ \bibnamefont {Daniels}}, \bibinfo {author} {\bibfnamefont {W.~A.}\ \bibnamefont {Borders}}, \bibinfo {author} {\bibfnamefont {N.}~\bibnamefont {Prasad}}, \bibinfo {author} {\bibfnamefont {A.}~\bibnamefont {Madhavan}}, \bibinfo {author} {\bibfnamefont {S.}~\bibnamefont {Gibeault}}, \bibinfo {author} {\bibfnamefont {T.}~\bibnamefont {Adeyeye}}, \bibinfo {author} {\bibfnamefont {L.}~\bibnamefont {Pocher}}, \bibinfo {author} {\bibfnamefont {L.}~\bibnamefont {Wan}}, \bibinfo {author} {\bibfnamefont {M.}~\bibnamefont {Tran}}, \bibinfo {author} {\bibfnamefont {J.~A.}\ \bibnamefont {Katine}}, \emph {et~al.},\ }\bibfield  {title} {\bibinfo {title} {Neural networks three ways: unlocking novel computing schemes using magnetic tunnel junction stochasticity},\ }in\ \href@noop {} {\emph {\bibinfo {booktitle} {Spintronics XVI}}},\ Vol.\ \bibinfo {volume} {12656}\ (\bibinfo {organization} {SPIE},\ \bibinfo {year} {2023})\ pp.\ \bibinfo {pages}
  {84--94}\BibitemShut {NoStop}%
\bibitem [{\citenamefont {Kostinski}\ and\ \citenamefont {Amir}(2016)}]{kostinski2016elementary}%
  \BibitemOpen
  \bibfield  {author} {\bibinfo {author} {\bibfnamefont {S.}~\bibnamefont {Kostinski}}\ and\ \bibinfo {author} {\bibfnamefont {A.}~\bibnamefont {Amir}},\ }\bibfield  {title} {\bibinfo {title} {An elementary derivation of first and last return times of 1d random walks},\ }\href@noop {} {\bibfield  {journal} {\bibinfo  {journal} {American journal of physics}\ }\textbf {\bibinfo {volume} {84}},\ \bibinfo {pages} {57} (\bibinfo {year} {2016})}\BibitemShut {NoStop}%
\bibitem [{\citenamefont {Fitzhugh}(1983)}]{fitzhughStatisticalPropertiesAsymmetric1983}%
  \BibitemOpen
  \bibfield  {author} {\bibinfo {author} {\bibfnamefont {R.}~\bibnamefont {Fitzhugh}},\ }\bibfield  {title} {\bibinfo {title} {Statistical properties of the asymmetric random telegraph signal, with applications to single-channel analysis},\ }\href {https://doi.org/10.1016/0025-5564(83)90028-7} {\bibfield  {journal} {\bibinfo  {journal} {Mathematical Biosciences}\ }\textbf {\bibinfo {volume} {64}},\ \bibinfo {pages} {75} (\bibinfo {year} {1983})}\BibitemShut {NoStop}%
\bibitem [{\citenamefont {Lemons}\ and\ \citenamefont {Langevin}(2002)}]{lemons2002introduction}%
  \BibitemOpen
  \bibfield  {author} {\bibinfo {author} {\bibfnamefont {D.~S.}\ \bibnamefont {Lemons}}\ and\ \bibinfo {author} {\bibfnamefont {P.}~\bibnamefont {Langevin}},\ }\href@noop {} {\emph {\bibinfo {title} {An introduction to stochastic processes in physics}}}\ (\bibinfo  {publisher} {JHU Press},\ \bibinfo {year} {2002})\BibitemShut {NoStop}%
\bibitem [{Note4()}]{Note4}%
  \BibitemOpen
  \bibinfo {note} {We remark that, unlike what we would expect in a real magnetic tunnel junction, there is no physically imposed ultraviolet cutoff for macrospin dynamics. This is not particularly problematic, though, since any good circuit simulation involving such a device would impose its own high-frequency cutoffs through the finite bandwidth of realistic circuit components.}\BibitemShut {Stop}%
\bibitem [{\citenamefont {Gilmore}\ \emph {et~al.}(2010)\citenamefont {Gilmore}, \citenamefont {Stiles}, \citenamefont {Seib}, \citenamefont {Steiauf},\ and\ \citenamefont {F\"ahnle}}]{PhysRevB.81.174414}%
  \BibitemOpen
  \bibfield  {author} {\bibinfo {author} {\bibfnamefont {K.}~\bibnamefont {Gilmore}}, \bibinfo {author} {\bibfnamefont {M.~D.}\ \bibnamefont {Stiles}}, \bibinfo {author} {\bibfnamefont {J.}~\bibnamefont {Seib}}, \bibinfo {author} {\bibfnamefont {D.}~\bibnamefont {Steiauf}},\ and\ \bibinfo {author} {\bibfnamefont {M.}~\bibnamefont {F\"ahnle}},\ }\bibfield  {title} {\bibinfo {title} {Anisotropic damping of the magnetization dynamics in {Ni}, {Co}, and {Fe}},\ }\href {https://doi.org/10.1103/PhysRevB.81.174414} {\bibfield  {journal} {\bibinfo  {journal} {Phys. Rev. B}\ }\textbf {\bibinfo {volume} {81}},\ \bibinfo {pages} {174414} (\bibinfo {year} {2010})}\BibitemShut {NoStop}%
\bibitem [{\citenamefont {Soumah}\ \emph {et~al.}(2024)\citenamefont {Soumah}, \citenamefont {Desplat}, \citenamefont {Phan}, \citenamefont {Valli}, \citenamefont {Madhavan}, \citenamefont {Disdier}, \citenamefont {Auffret}, \citenamefont {Sousa}, \citenamefont {Ebels},\ and\ \citenamefont {Talatchian}}]{soumah2024nanosecond}%
  \BibitemOpen
  \bibfield  {author} {\bibinfo {author} {\bibfnamefont {L.}~\bibnamefont {Soumah}}, \bibinfo {author} {\bibfnamefont {L.}~\bibnamefont {Desplat}}, \bibinfo {author} {\bibfnamefont {N.-T.}\ \bibnamefont {Phan}}, \bibinfo {author} {\bibfnamefont {A.~S.~E.}\ \bibnamefont {Valli}}, \bibinfo {author} {\bibfnamefont {A.}~\bibnamefont {Madhavan}}, \bibinfo {author} {\bibfnamefont {F.}~\bibnamefont {Disdier}}, \bibinfo {author} {\bibfnamefont {S.}~\bibnamefont {Auffret}}, \bibinfo {author} {\bibfnamefont {R.}~\bibnamefont {Sousa}}, \bibinfo {author} {\bibfnamefont {U.}~\bibnamefont {Ebels}},\ and\ \bibinfo {author} {\bibfnamefont {P.}~\bibnamefont {Talatchian}},\ }\bibfield  {title} {\bibinfo {title} {Nanosecond stochastic operation in perpendicular superparamagnetic tunnel junctions},\ }\href@noop {} {\bibfield  {journal} {\bibinfo  {journal} {arXiv preprint arXiv:2402.03452}\ } (\bibinfo {year} {2024})}\BibitemShut {NoStop}%
\bibitem [{\citenamefont {Chavent}\ \emph {et~al.}(2020)\citenamefont {Chavent}, \citenamefont {Iurchuk}, \citenamefont {Tillie}, \citenamefont {Bel}, \citenamefont {Lamard}, \citenamefont {Vila}, \citenamefont {Ebels}, \citenamefont {Sousa}, \citenamefont {Dieny}, \citenamefont {Di~Pendina} \emph {et~al.}}]{chavent2020multifunctional}%
  \BibitemOpen
  \bibfield  {author} {\bibinfo {author} {\bibfnamefont {A.}~\bibnamefont {Chavent}}, \bibinfo {author} {\bibfnamefont {V.}~\bibnamefont {Iurchuk}}, \bibinfo {author} {\bibfnamefont {L.}~\bibnamefont {Tillie}}, \bibinfo {author} {\bibfnamefont {Y.}~\bibnamefont {Bel}}, \bibinfo {author} {\bibfnamefont {N.}~\bibnamefont {Lamard}}, \bibinfo {author} {\bibfnamefont {L.}~\bibnamefont {Vila}}, \bibinfo {author} {\bibfnamefont {U.}~\bibnamefont {Ebels}}, \bibinfo {author} {\bibfnamefont {R.~C.}\ \bibnamefont {Sousa}}, \bibinfo {author} {\bibfnamefont {B.}~\bibnamefont {Dieny}}, \bibinfo {author} {\bibfnamefont {G.}~\bibnamefont {Di~Pendina}}, \emph {et~al.},\ }\bibfield  {title} {\bibinfo {title} {A multifunctional standardized magnetic tunnel junction stack embedding sensor, memory and oscillator functionality},\ }\href@noop {} {\bibfield  {journal} {\bibinfo  {journal} {Journal of Magnetism and Magnetic Materials}\ }\textbf {\bibinfo {volume} {505}},\ \bibinfo {pages} {166647} (\bibinfo {year}
  {2020})}\BibitemShut {NoStop}%
\bibitem [{\citenamefont {Ma}\ \emph {et~al.}(2021)\citenamefont {Ma}, \citenamefont {Sidi El~Valli}, \citenamefont {Krei{\ss}ig}, \citenamefont {Di~Pendina}, \citenamefont {Protze}, \citenamefont {Ebels}, \citenamefont {Prenat}, \citenamefont {Chavent}, \citenamefont {Iurchuk}, \citenamefont {Sousa} \emph {et~al.}}]{ma2021microwave}%
  \BibitemOpen
  \bibfield  {author} {\bibinfo {author} {\bibfnamefont {R.}~\bibnamefont {Ma}}, \bibinfo {author} {\bibfnamefont {A.}~\bibnamefont {Sidi El~Valli}}, \bibinfo {author} {\bibfnamefont {M.}~\bibnamefont {Krei{\ss}ig}}, \bibinfo {author} {\bibfnamefont {G.}~\bibnamefont {Di~Pendina}}, \bibinfo {author} {\bibfnamefont {F.}~\bibnamefont {Protze}}, \bibinfo {author} {\bibfnamefont {U.}~\bibnamefont {Ebels}}, \bibinfo {author} {\bibfnamefont {G.}~\bibnamefont {Prenat}}, \bibinfo {author} {\bibfnamefont {A.}~\bibnamefont {Chavent}}, \bibinfo {author} {\bibfnamefont {V.}~\bibnamefont {Iurchuk}}, \bibinfo {author} {\bibfnamefont {R.}~\bibnamefont {Sousa}}, \emph {et~al.},\ }\bibfield  {title} {\bibinfo {title} {Microwave functionality of spintronic devices implemented in a hybrid complementary metal oxide semiconductor and magnetic tunnel junction technology},\ }\href@noop {} {\bibfield  {journal} {\bibinfo  {journal} {Electronics Letters}\ }\textbf {\bibinfo {volume} {57}},\ \bibinfo {pages} {264} (\bibinfo {year}
  {2021})}\BibitemShut {NoStop}%
\bibitem [{\citenamefont {Dieny}\ and\ \citenamefont {Chshiev}(2017)}]{dieny2017perpendicular}%
  \BibitemOpen
  \bibfield  {author} {\bibinfo {author} {\bibfnamefont {B.}~\bibnamefont {Dieny}}\ and\ \bibinfo {author} {\bibfnamefont {M.}~\bibnamefont {Chshiev}},\ }\bibfield  {title} {\bibinfo {title} {Perpendicular magnetic anisotropy at transition metal/oxide interfaces and applications},\ }\href@noop {} {\bibfield  {journal} {\bibinfo  {journal} {Reviews of Modern Physics}\ }\textbf {\bibinfo {volume} {89}},\ \bibinfo {pages} {025008} (\bibinfo {year} {2017})}\BibitemShut {NoStop}%
\bibitem [{\citenamefont {Strelkov}\ \emph {et~al.}(2018)\citenamefont {Strelkov}, \citenamefont {Chavent}, \citenamefont {Timopheev}, \citenamefont {Sousa}, \citenamefont {Prejbeanu}, \citenamefont {Buda-Prejbeanu},\ and\ \citenamefont {Dieny}}]{strelkov2018impact}%
  \BibitemOpen
  \bibfield  {author} {\bibinfo {author} {\bibfnamefont {N.}~\bibnamefont {Strelkov}}, \bibinfo {author} {\bibfnamefont {A.}~\bibnamefont {Chavent}}, \bibinfo {author} {\bibfnamefont {A.}~\bibnamefont {Timopheev}}, \bibinfo {author} {\bibfnamefont {R.}~\bibnamefont {Sousa}}, \bibinfo {author} {\bibfnamefont {I.}~\bibnamefont {Prejbeanu}}, \bibinfo {author} {\bibfnamefont {L.}~\bibnamefont {Buda-Prejbeanu}},\ and\ \bibinfo {author} {\bibfnamefont {B.}~\bibnamefont {Dieny}},\ }\bibfield  {title} {\bibinfo {title} {Impact of joule heating on the stability phase diagrams of perpendicular magnetic tunnel junctions},\ }\href {https://journals.aps.org/prb/abstract/10.1103/PhysRevB.98.214410} {\bibfield  {journal} {\bibinfo  {journal} {Physical Review B}\ }\textbf {\bibinfo {volume} {98}},\ \bibinfo {pages} {214410} (\bibinfo {year} {2018})}\BibitemShut {NoStop}%
\bibitem [{Note5()}]{Note5}%
  \BibitemOpen
  \bibinfo {note} {Numerical experimental power spectral densities were estimated using Welch's method~\cite {welch1967use}, with a Hann window.}\BibitemShut {Stop}%
\bibitem [{\citenamefont {Kubo}(1966)}]{kubo1966fluctuation}%
  \BibitemOpen
  \bibfield  {author} {\bibinfo {author} {\bibfnamefont {R.}~\bibnamefont {Kubo}},\ }\bibfield  {title} {\bibinfo {title} {The fluctuation-dissipation theorem},\ }\href@noop {} {\bibfield  {journal} {\bibinfo  {journal} {Reports on progress in physics}\ }\textbf {\bibinfo {volume} {29}},\ \bibinfo {pages} {255} (\bibinfo {year} {1966})}\BibitemShut {NoStop}%
\bibitem [{\citenamefont {Welch}(1967)}]{welch1967use}%
  \BibitemOpen
  \bibfield  {author} {\bibinfo {author} {\bibfnamefont {P.}~\bibnamefont {Welch}},\ }\bibfield  {title} {\bibinfo {title} {The use of fast fourier transform for the estimation of power spectra: a method based on time averaging over short, modified periodograms},\ }\href@noop {} {\bibfield  {journal} {\bibinfo  {journal} {IEEE Transactions on audio and electroacoustics}\ }\textbf {\bibinfo {volume} {15}},\ \bibinfo {pages} {70} (\bibinfo {year} {1967})}\BibitemShut {NoStop}%
\end{thebibliography}%

\end{document}